%% file: WhitePaper-SKBePol.tex
\documentclass[12pt]{article}
\pdfoutput=1
\usepackage{amsmath,amssymb}
\usepackage{xspace}
\usepackage{hyperref}
\hypersetup{
colorlinks,
linkcolor={blue},
linktocpage=true,
citecolor={blue},
urlcolor={blue}
}

\usepackage{graphicx}

\usepackage{booktabs}
\usepackage{caption}
\usepackage{subcaption}
\usepackage{slashed}

\usepackage{float}
\usepackage{authblk}

\usepackage[utf8]{inputenc}
\usepackage[numbers,sort&compress]{natbib}
\usepackage{graphicx}
\usepackage{verbatim} 
\usepackage{booktabs}
\usepackage[percent]{overpic}
\def\epem       {\ensuremath{e^+e^-}\xspace}
\def\mumu       {\ensuremath{\mu^+\mu^-}\xspace}
\def\uubar {\ensuremath{u\overline u}\xspace}
\def\ddbar {\ensuremath{d\overline d}\xspace}
\def\ssbar {\ensuremath{s\overline s}\xspace}
\def\ccbar {\ensuremath{c\overline c}\xspace}
\def\bbbar {\ensuremath{b\overline b}\xspace}
\def\tautau     {\ensuremath{\tau^+\tau^-}\xspace}
\usepackage{relsize}
\def\babar{\mbox{\slshape B\kern-0.1em{\smaller A}\kern-0.1em B\kern-0.1em{\smaller A\kern-0.2em R}}}
\def\piz   {\ensuremath{\pi^0}\xspace}
\newcommand{\taupi}{$\tau^\pm\rightarrow\pi^\pm\nu_\tau$~}
\newcommand{\taue}{$\tau^\pm\rightarrow e^\pm\overline{\nu}_e\nu_\tau$~}
\newcommand{\taumu}{$\tau^\pm\rightarrow \mu^\pm\overline{\nu}_\mu\nu_\tau$~}
\newcommand{\taurho}{$\tau^\pm\rightarrow(\rho^\pm\rightarrow\pi^\pm\pi^0)\nu_\tau$~}
\newcommand{\taua}{$\tau^\pm\rightarrow(a_1^\pm\rightarrow\pi^\pm\pi^0\pi^0)\nu_\tau$~}
\newcommand{\fb}{fb$^{-1}$}
\newcommand{\ab}{ab$^{-1}$}
\newcommand{\stw}{$\sin^2\theta_{W}$}

\newcommand\snowmass{\begin{center}\rule[-0.2in]{\hsize}{0.01in}\\\rule{\hsize}{0.01in}\\
\vskip 0.1in Submitted to the  Proceedings of the US Community Study\\ 
on the Future of Particle Physics (Snowmass 2021)\\ 
\rule{\hsize}{0.01in}\\\rule[+0.2in]{\hsize}{0.01in} \end{center}}

\graphicspath{{Plots/}}

\renewcommand\Affilfont{\itshape\small}

\begin{document}
\renewcommand\Affilfont{\itshape\small}

\begin{titlepage}
\snowmass
\begin{center}
\begin{large}
 {\bf Snowmass 2021 White Paper\\ Upgrading SuperKEKB with a Polarized Electron Beam:\\ Discovery Potential and Proposed Implementation}\\ 
 \end{large}
 
 \vspace*{.1in}
 April 13, 2022
\vspace{.5cm}

{\large US Belle~II Group} 
\footnote{
\input{USAuthors} 
}
\\ {\large and} \\
{\large  Belle~II/SuperKEKB e- Polarization Upgrade Working Group } 
\footnote{
U.~Wienands,
M.~Hoferichter,
C.~Hearty,
I.~A.~Koop, A.~V.~Otboev, Yu.~M.~Shatunov,
B.~Parker,
A.~Schwartz,
A.~Vossen,
A.~Accardi,
M.~Kuriki, Z.~Liptak,
F.~R.~Le~Diberder, A.~Martens, K.~Trabelsi, F.~Zomer,
D.~Zhou,
Sw.~Banerjee,
W.~Deconinck, M.~Gericke, J.~Mammei,
A.~Signori,
R.~Baartman, T.~Planche,
A.~Beaubien, T.~Junginger, C.~Miller, K.~Moorthy, Y.~Peng, J.~M.~Roney 
}

\end{center}

\noindent {\bf Corresponding Author(s):}\\
\noindent {Swagato Banerjee  (swagato.banerjee@louisville.edu)}\\
\noindent {J. Michael Roney  (mroney@uvic.ca)}\\

\noindent {\bf Thematic Area(s):}\\
\noindent $\blacksquare$ (RF0) Frontier for Rare Processes and Precision Measurements\\
\noindent $\blacksquare$ (RF01) Weak Decays of b and c \\
\noindent $\blacksquare$ (RF02) Strange \& Light Quarks \\
\noindent $\blacksquare$ (RF03) Fundamental Physics in Small Experiments \\
\noindent $\blacksquare$ (RF05) Charged Lepton Flavor Violation (electrons, muons and taus) \\
\noindent $\blacksquare$ (RF06) Dark Sector at Low Energies \\
\noindent $\blacksquare$ (AF05) Accelerators for rare processes and precision measurements \\
\noindent $\blacksquare$ (EF04) EW Physics: EW Precision Physics and constraining new physics \\


\newpage
\begin{center}
{\bf\large Executive Summary}\\
\end{center}
\input{ExecSummary}



\end{titlepage}
 
\tableofcontents
\cleardoublepage
\pagestyle{plain} 
\setcounter{page}{1}
\pagenumbering{arabic}




\section{Introduction}
\IfFileExists{Introduction.tex}{\input{Introduction.tex}}{}

\section{Precision Electroweak Program}
\IfFileExists{ChiralBellePhysics.tex}{\input{ChiralBellePhysics.tex}}{}

\subsection{Muon Pair $A_{LR}$}
\IfFileExists{MuPair.tex}{\input{MuPair.tex}}{}

\subsection{Tau Pair $A_{LR}$}
\IfFileExists{TauPair.tex}{\input{TauPair.tex}}{}

\subsection{Charm and Beauty $A_{LR}$}
\IfFileExists{CharmandBeauty.tex}{\input{CharmandBeauty.tex}}{}

\IfFileExists{Charm_ALR.tex}{\input{Charm_ALR.tex}}{}

\IfFileExists{Beauty.tex}{\input{Beauty.tex}}{}

\subsection{Bhabha $A_{LR}$}
\IfFileExists{Bhabha.tex}{\input{Bhabha.tex}}{}

\section{Tau $\boldsymbol{g-2}$}
\IfFileExists{Taug-2.tex}{\input{Taug-2.tex}}{}

\section{Tau EDM }
\IfFileExists{TauEDM.tex}{\input{TauEDM.tex}}{}

\section{Tau LFV}
\IfFileExists{TauLFV.tex}{\input{TauLFV.tex}}{}

\section{QCD: Dynamical mass generation studies with polarized beams }
\IfFileExists{QCD.tex}{\input{QCD.tex}}{}

\pagebreak

\section{Polarized Source}
\IfFileExists{PolarizedSource.tex}{\input{PolarizedSource.tex}}{}

\section{Beam-Beam Effects on Polarization}
\IfFileExists{BeamBeamEffects.tex}{\input{BeamBeamEffects.tex}}{}

\section{Spin Rotator}
Three approaches for implementing  spin rotators into the HER are presented. Two approaches are described in the ``BINP Spin Rotator Concepts" subsection, in which a modest conversion of  the geometry of electron beam bends in the experimental section of the HER storage ring to provide drift gaps with a length of about 10~m for installing spin rotators, are proposed. One BINP approach uses a conventional spin rotator with separate solenoid and quadrupole magnets  and the second makes use of combined function solenoid-quadrupole magnets.  The third approach is described in the subsection entitled ``Dipole-Solenoid-Quadrupoles Combined Function Magnets Concept", in which four existing dipoles on either side of the IP are replaced with dipole-solenoid-quadrupoles combined function spin rotator magnets that can be constructed using the direct wind technology used at BNL with the intention of minimizing changes to the HER lattice. So far, these approaches indicate that implementing spin rotators in the HER is feasible and the next step in this R\&D program is to perform long-term particle tracking studies to determine whether or not the dynamic aperture with spin rotators has worsened
in comparison with the version of the ring optics without spin rotators.

\subsection{BINP Spin Rotator Concepts}
\IfFileExists{SpinRotator-BINPDesign.tex}{\input{SpinRotator-BINPDesign.tex}}{}

\clearpage
\subsection{Dipole-Solenoid-Quadrupoles Combined Function Magnets Concept}
\IfFileExists{SpinRotator-ANLDesign.tex}{\input{SpinRotator-ANLDesign.tex}}{}
\clearpage

\subsection{Direct Wind Magnet for Spin Rotators} 
\IfFileExists{DirectWindSpinRotator.tex}{\input{DirectWindSpinRotator.tex}}{}
\clearpage

\section{Compton Polarimetry}
\IfFileExists{ComptonPolarimetry.tex}{\input{ComptonPolarimetry.tex}}{}

\section{Tau Polarimetry}
\IfFileExists{TauPolarimetry.tex}{\input{TauPolarimetry.tex}}{}

\IfFileExists{Schedule.tex}{\input{Schedule.tex}}{}

\IfFileExists{WorkPackages.tex}{\input{WorkPackages.tex}}{}

\IfFileExists{CostEstimates.tex}{\input{CostEstimates.tex}}{}

\section{Summary and Next Steps}
\IfFileExists{SummaryConclusions.tex}{\input{SummaryConclusions.tex}}{}
Unique and powerful sensitivities to new physics via precision neutral current measurements at 10~GeV are enabled by an upgrade of SuperKEKB to have polarized electron beams. With  a  measurement of $\sin^2\theta_W$ having a precision comparable to the current world-average $Z^0$-pole value, but at 10~GeV, Chiral Belle will be uniquely sensitive to the presence of dark sector parity violating bosons with masses below the Z$^0$. Moreover, it would be the only facility able to probe neutral current universality relations between charm and beauty quarks and all three charged leptons at energies below  the $Z^0$ pole and would measure ratios of neutral current couplings with unprecedentedly high precision.  The $\tau$-pairs produced with polarized beams will also provide the only means to measure the third-generation $g-2$ at a precision that can begin to approach the ${\cal O}(10^{-6})$ level,  the equivalent of the muon $g-2$ anomaly scaled by $(m_{\tau}/m{_\mu})^2$ as expected in Minimal Flavor Violation scenarios. With 40~ab$^{-1}$, a measurement at the $10^{-5}$ is possible and to reach ${\cal O}(10^{-6})$ will require more statistics as well as improved measurements of $m_{\tau}$ and $M_{\Upsilon(1S)}$.  Other physics will also benefit from the polarization, including lepton flavor violation searches in $\tau$ decays,  $\tau$ Michel parameter measurements, $\tau$ EDM measurements and QCD hadronization studies. 

It has been demonstrated that the challenging problem of providing a spin rotator that is transparent to the rest of the SuperKEKB lattice can be solved, with some options being presented in this paper. Developments of a polarized source are underway, as are solutions for  implementing Compton polarimeters. Unique to the Chiral Belle program is the  additional means of  measuring the beam polarization at the IP using $\tau$ pair events with a relative systematic uncertainty of 0.4\%, as described in this paper. 
 
The next steps involve developing the full Conceptual Design after which a Technical Design wold commence and cost estimates produced.
The capital costs for such an upgrade are expected to be substantially less than half of the annual  power costs of operating the SuperKEKB accelerator. It can also be expected that a significant fraction of those capital costs will be provided by non-Japanese groups on Belle~II. It is possible to plan for the upgrade to commence during a long shutdown at the end of this decade and completed over a number of summer shutdowns following such a long shutdown. In such a scenario, the polarization program could begin
while SuperKEKB completes its program of delivering 50~$ab^{-1}$ of data to Belle~II and continued beyond that program.
 
\vspace*{1cm}
\bibliographystyle{apsrev4-1_mod.bst}
\bibliography{references}

\end{document}

%% file: USAuthors.tex
 D.M. Asner, H. Atmacan, Sw. Banerjee, J.V. Bennett, M. Bertemes, M. Bessner, D. Biswas, G. Bonvicini, N. Brenny, R.A. Briere, T.E. Browder, C. Chen, S. Choudhury, D. Cinabro, J. Cochran, L. M. Cremaldi, A. Di Canto, S. Dubey, K. Flood, B. G. Fulsom, V. Gaur, R. Godang, T. Gu, Y. Guan, J. Guilliams, C. Hadjivasiliou, O. Hartbrich, W.W. Jacobs, D.E. Jaffe, S. Kang, L. Kapitánová, C. Ketter, A. Khatri, K. Kinoshita, S. Kohani, H. Korandla, I. Koseoglu Sari, R. Kroeger, J. Kumar, K.J. Kumara, T. Lam, P.J. Laycock, L. Li, D. Liventsev, F. Meier, S. Mitra, A. Natochii, N. Nellikunnummel, K.A. Nishimura, E.R. Oxford, A. Panta, K. Parham, T.K. Pedlar, R. Peschke, L.E. Piilonen, S. Pokharel, S. Prell, H. Purwar, D.E. Ricalde Herrmann, C. Rosenfeld, D. Sahoo, D.A. Sanders, A. Sangal, V. Savinov, S. Schneider, J. Schueler, A.J. Schwartz, V. Shebalin, A. Sibidanov, Z.S. Stottler, J. Strube, S. Tripathi, S.E. Vahsen, G.S. Varner, A. Vossen, D. Wang, E. Wang, L. Wood, J. Yelton, Y. Zhai, B. Zhang

%% file: ExecSummary.tex
 Upgrading the SuperKEKB electron-positron collider with polarized electron
beams opens a new program of precision physics at a center-of-mass energy
of 10.58~GeV, the mass of the $\Upsilon(4S)$ meson.
This white paper describes the physics potential of this `Chiral Belle' program. It includes projections for precision measurements of $\sin^2\theta_W$ that can be obtained from independent left-right asymmetry measurements of
 \epem transitions to pairs of electrons, muons, taus, charm and b-quarks.
 The $\sin^2\theta_W$ precision obtainable at SuperKEKB will match that of the LEP/SLC world average but at the centre-of-mass energy of 10.58~GeV. Measurements of the couplings for muons, charm, and $b$-quarks will be substantially
  improved and the existing $3\sigma$ discrepancy between the SLC $A_{LR}$ and LEP $A_{FB}^b$
 measurements will be addressed. Precision measurements of neutral current universality will be more than an order of magnitude more precise than 
 currently available. As the energy scale is  well away from the $Z^0$-pole, the precision measurements will have sensitivity to the presence of a  parity-violating dark sector gauge boson, $Z_{\rm dark}$.
 The  program also enables the measurement of the 
 anomalous magnetic moment $g-2$ form factor of the $\tau$ to be made at an unprecedented level of precision. A precision of $10^{-5}$ level is accessible with 40~ab$^{-1}$ and with more data it would start to approach the $10^{-6}$ level.
 This technique would provide the most precise information from the third generation about potential new physics explanations of the muon $g-2$ $4\sigma$ anomaly.  
 Additional  $\tau$ and QCD physics programs  enabled or enhanced with having polarized electron beams are also discussed in this White Paper. 

In order to implement $e^-$ beam polarization in the SuperKEKB high energy ring (HER),  three hardware upgrades are required: 1) introduction of a low-emittance polarized source that supplies SuperKEKB with  transversely polarized electrons that provide separate data sets with opposite polarization states; 2) a system of spin rotator magnets that rotate the spin of the electrons in the beam to be longitudinal before the interaction point (IP) where the Belle~II detector is located, and then back to transversely polarized after the IP; and 3) a  Compton polarimeter that provides  online measurements of the beam polarization at a location between the first spin rotator  and the IP. A precision measurement of the polarization is also made at the IP by analysing the spin-dependent decay kinematics of $\tau$ leptons produced in  a  $e^+e^- \rightarrow \tau^+\tau^-$ data set. This White Paper will review the current status of the R\&D associated with the three hardware projects and describes the $\tau$ polarimetery analysis of 0.4ab$^{-1}$ of $e^+e^-$ data collected at the $\Upsilon(4S)$  with the \babar~experiment that shows the high precision that can achieved. This paper includes a summary of the path forward in R\&D and next steps required to implement this  upgrade and access its exciting discovery potential. 

%% file: Introduction.tex
 
 The SuperKEKB \epem collider, operating at a center-of-mass energy of 10.58~GeV
with its  high luminosity,
 can access new windows for discovery with the Belle~II 
experiment if it is upgraded to have a longitudinally polarized electron beam.
Upgrading SuperKEKB to provide large samples of left-handed and right-handed initial-state electrons with approximately 70\% polarization at the Belle~II interaction point creates   a unique and versatile facility for probing new physics with precision  measurements that no other experiments, current or planned,   can achieve. This upgrade is being considered and the associated R\&D program is underway. This paper describes the physics potential of this ``Chiral Belle'' program and how the upgrade can be implemented.

 The measurements in this 'Chiral Belle' program include precision measurements of $\sin^2\theta_W$ obtained from independent left-right asymmetry measurements of
 \epem transitions to pairs of electrons, muons, taus, charm and b-quarks.
 The $\sin^2\theta_W$ precision obtainable at SuperKEKB will match that of the LEP/SLC world average but at the center-of-mass energy of 10.58~GeV,
 thereby probing the neutral current couplings with unprecedented precision at a new energy scale, which is sensitive to the running of the couplings. 
 World average measurements of the individual neutral current vector coupling constants to b- and c-quarks and muons in particular will be
 substantially improved, and the residual $3\sigma$ discrepancy between the SLC $A_{LR}$ and LEP $A_{FB}^b$
 measurements will be addressed. Precision measurements of neutral current universality will be more than an order of magnitude more precise than 
 currently available, with measurements sensitive to parity-violating dark sector gauge bosons.  If there is a dark-sector parity-violating equivalent of the Standard Model (SM) $Z^0$, a $Z_{dark}$, with a mass below that of the SM $Z^0$, then precision parity-violation measurements at 10~GeV provide a unique sensitivity to its presence, particularly if a $Z_{dark}$ has different couplings to different SM fermions.
 The Chiral Belle physics program also enables the measurement of the tau-lepton g-2 at an unprecedented and unrivaled level of precision. It is the only method for measuring tau g-2 that can approach a level of precision of interest in a Minimal Flavor Violation third-generation analog of the muon g-2 $4\sigma$ anomaly.
 Other  physics uniquely enabled with polarized electron beams includes precision measurements of the tau EDM and tau Michel parameters. In addition, searches for lepton flavor violation in tau decays and dynamical mass generation hadronization studies are enhanced will be enhanced with polarized beams.

 The upgrade to SuperKEKB involves three hardware projects:\\
1)  A low-emittance polarized electron source in which 
  the electron beams would be produced via a polarized laser illuminating a ``strained lattice'' GaAs photocathode
 as was done for SLD~\cite{ALEPH:2005ab}. 
 The source would produce a sample of left-handed and a sample of right-handed longitudinally polarized electron bunches whose spin would be rotated to be transversely polarized (with spins vertically up or vertically down) before encountering any dipole magnets, thereby ensuring spin stability in the 
  SuperKEKB 7~GeV electron storage ring - the High Energy Ring (HER);\\
2) A pair of spin-rotators, one positioned before and the other after the interaction region, to rotate the spin to longitudinal
 prior to collisions and back to transverse following collisions. 
 The challenge is to design the rotators to be transparent to the rest of the HER lattice, which requires   minimizing couplings between vertical and horizontal planes and addressing higher order and chromatic effects  in the design to ensure the luminosity is not degraded; \\
3) A Compton polarimeter that measures the beam polarization before the beam enters the interaction region. The beam polarization is also measured directly at the IP via the analysis of decays of $\tau$ leptons produced in $e^+e^-\rightarrow\tau^+\tau^-$ events. 

 The rest of this paper provides more details of the physics discovery potential and how electron polarization can be implemented in an
 upgrade to SuperKEKB while maintaining the high luminosity.

%% file: ChiralBellePhysics.tex
 A data sample of 20~ab$^{-1}$ with a polarized electron beam enables Belle~II to measure the weak neutral current vector coupling 
 constants of the $b$-quark, $c$-quark and muon at significantly higher precision than any previous experiment. 
 With 40~ab$^{-1}$ of polarized beam data, the precision of
 the vector couplings to the tau and electron can be measured at a level comparable to current world averages, which are dominated by
 LEP and SLD measurements at the Z$^0$-pole.

 Within the framework of the Standard Model (SM) these measurements of $g_V^f$, the neutral current vector coupling for fermion $f$,
 can be used to determine the weak mixing angle, $\theta_W$, 
through the relation: $g_V^f = T_3^f - 2Q_f \sin^2\theta_W$, where $T_3^f$ is the $3^{rd}$ component of weak isospin of fermion $f$,
 $Q_f$ is its electric charge in units of electron charge and
the notational conventions of Reference~\cite{ALEPH:2005ab} are used.

As described in Reference~\cite{SuperB:2013cxb},
with  polarized electron beams an \epem collider at 10.58~GeV enables the
 determination of $g_V^f$  by measuring the left-right asymmetry, $A_{LR}^f$, for each identified final-state fermion-pair in the process
$e^+e^- \rightarrow f \overline{f}$. With SM Born cross sections for 100\% left-handed ($\sigma_{\mathbf L}$) and 100\% right-handed ($\sigma_{\mathbf R}$)  initial state electrons in s-channel processes, such as  \epem$ \rightarrow \mu^+\mu^-$,
\begin{equation}
 A_{LR}^{f}({\rm SM}) = \frac{ \sigma_{\mathbf L}^f - \sigma_{\mathbf R}^f}{\sigma_{\mathbf L}^f + \sigma_{\mathbf R}^f} = \frac{s G_F }{\sqrt{2} \pi \alpha Q_f} g_A^e g_V^f 
\label{eq:ALRBorn}
\end{equation}
where $g_A^e = T_3^e = -\frac{1}{2}$ is the neutral current axial coupling of the electron, $G_F$ is the Fermi coupling constant, and
$s$ is the square of the center-of-mass energy.
The Bhabha channel,  $\epem \rightarrow \epem$, with a significant t-channel contribution, has a different dependence on the SM couplings than presented in Equation~\ref{eq:ALRBorn}, as discussed in section~\ref{Bhabha}. These left-right asymmetries arise from $\gamma-Z$ interference and, although the
 SM asymmetries are small (approximately $-6\times 10^{-4}$ for the $\mu$ and $\tau$ leptons, $+1.5\times 10^{-4}$ for electrons, $-5\times 10^{-3}$ for charm and  $-2\%$ for the $b$-quarks), unprecedented precisions can be achieved 
 because of the combination of both the high luminosity of SuperKEKB and a $70\%$ beam polarization measured with precision of better than $\pm 0.5\%$.
Note that, because of the small asymmetries, the denominator  is dominated by the parity-conserving components of the cross-section. Nonetheless, the measurements include these  parity-violating components in both numerator and denominator. 

Independent measurements of $A^f_{LR}$ for the different final state fermions ($f=e, \mu, \tau$, c-quark, b-quark) are performed by selecting  pure samples of each event type and counting the numbers of such events when
the beam longitudinal polarization is left-handed (${\mathbf L}$) and separately when it is right-handed (${\mathbf R}$), so that:
\begin{equation}
A_{LR}^f({\rm measured}) = \frac{N_{\mathbf L}^f - N_{\mathbf R}^f}{N_{\mathbf L}^f+N_{\mathbf R}^f}\langle Pol \rangle 
\end{equation}
$\langle Pol \rangle$ is the average electron beam polarization for the  sample under consideration:
\begin{equation}
\langle Pol \rangle =\frac{1}{2} \left[ \left(\frac{N_{eR}-N_{eL}}{N_{eR}+N_{eL}}\right)_{\mathbf R} -
 \left(\frac{N_{eR}-N_{eL}}{N_{eR}+N_{eL}}\right)_{\mathbf L} \right]
\end{equation}
where  $N_{eR}$ is the actual number of right-handed electrons and $N_{eL}$
 the actual number of left-handed electrons in the event samples where the electron beam bunch is nominally left polarized or right polarized,
 as indicated by the `${\mathbf L}$' and `${\mathbf R}$' subscripts.

 High precision measurements of $A_{LR}^f$, and consequently of $\sin^2\theta_W^f$, are possible at such an upgraded SuperKEKB
 because with 20~ab$^{-1}$ of data Belle~II can identify between $10^9$ and $10^{10}$ final-state pairs of b-quarks, c-quarks, taus, muons and electrons
 with high purity and reasonable signal efficiency,
 and because all detector-related systematic errors can be made to cancel by flipping
 the laser polarization at the source from ${\mathbf R}$ to ${\mathbf L}$ in a random, but known, pattern.
 $\langle Pol \rangle$ would be measured in two ways. The first method uses a Compton polarimeter, which can be expected to have an absolute
 uncertainty at the Belle~II interaction point of less than $1\%$ and
 provides a `bunch-by-bunch' measurement of $\left(\frac{N_{eR}-N_{eL}}{N_{eR}+N_{eL}}\right)_{\mathbf R}$
 and $\left(\frac{N_{eR}-N_{eL}}{N_{eR}+N_{eL}}\right)_{\mathbf L}$. The uncertainty will be dominated by the need to predict the polarization loss
 from the Compton polarimeter to the interaction point. The second method 
 measures the polar angle dependence of the polarization of $\tau$-leptons produced in $e^+e^- \rightarrow \tau^+ \tau^-$ events using the kinematic distributions of the
 decay products of the $\tau$ separately for the ${\mathbf R}$ and ${\mathbf L}$ data samples.
 The forward-backward asymmetry of the tau-pair polarization is linearly dependent on $\langle Pol \rangle$ and therefore
 can be used to determine $\langle Pol \rangle$ to better than $0.5\%$ with a  10~ab$^{-1}$ ${\mathbf R}$ sample and 10~ab$^{-1}$ ${\mathbf L}$ sample at the Belle~II interaction point in a manner entirely independent of the Compton polarimeter, as discussed in section~\ref{TauPolarimetery}.
 The $\tau$ polarimetry method avoids the uncertainties associated with tracking the polarization losses to the interaction point, and also
 automatically accounts for any residual positron polarization that might be present. In addition, it automatically provides a luminosity-weighted beam
 polarization measurement.

 Table~\ref{Table-measurements} provides the sensitivities to electroweak parameters expected with polarized electron beams in an upgraded SuperKEKB from
$e^+e^- \rightarrow b \bar{b}$, $e^+e^- \rightarrow c \bar{c}$, $e^+e^- \rightarrow \tau^+ \tau^-$, $e^+e^- \rightarrow \mu^+ \mu^-$, 
and $e^+e^- \rightarrow e^+ e^-$ events selected by Belle~II. From this information the precision on the b-quark, c-quark and muon neutral current vector couplings 
will improve by a factor of four, seven and five, respectively, over the current world average values\cite{ALEPH:2005ab} with 20~ab$^{-1}$ of polarized data.

This is of particular importance for $g^b_V$, where the measurement of -0.3220$\pm$0.0077 is  2.8$\sigma$ higher than the SM value of -0.3437~\cite{ALEPH:2005ab}.
That discrepancy arises from the 3$\sigma$ difference between the SLC $A_{LR}$ measurements and LEP $A^b_{FB}$ measurements of $\sin^2\theta_W^{eff}$. 
A measurement of $g^b_V$ at an upgraded SuperKEKB that is four times more precise and which avoids the hadronization uncertainties that are a significant component
 of the uncertainties of the measurement of the forward-backward asymmetry at LEP, or any other forward-backward asymmetry measurement
 using on-shell $Z^0$ bosons, will be able to definitively resolve whether or not this is
a statistical fluctuation or a first hint of a genuine breakdown of the SM. 

\begin{figure}
\centering
\includegraphics[width=.6\textwidth]{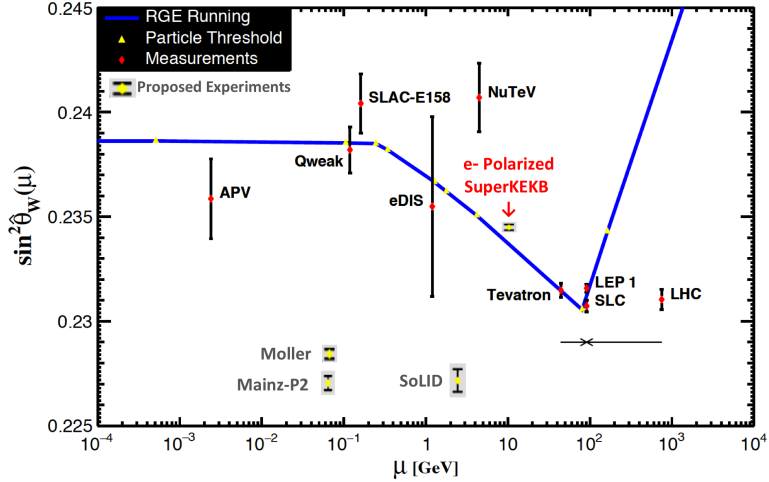}
\caption{ Determination of $\sin^2\theta_W$ at present and future experimental facilities as a function of energy scale,
 adapted from \cite{ErlerPDG2018,Becker:2018ggl,MOLLER:2014iki}.}
 \label{fig:sin2thetaW}
\end{figure}

\begin{figure}
\centering
\raisebox{0.0\height}{\includegraphics[width=0.55\textwidth]{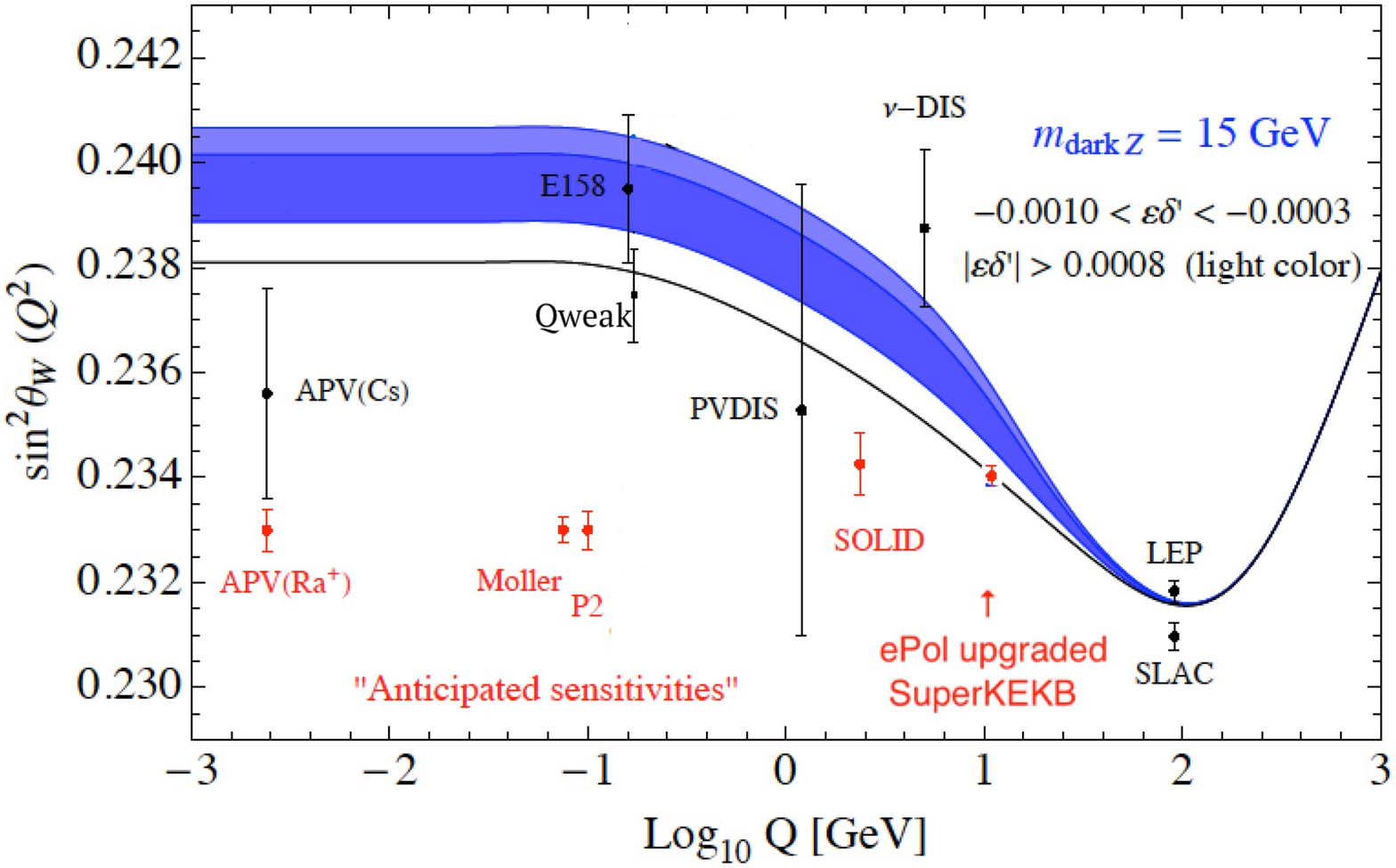}}
\raisebox{0.0\height}{\includegraphics[width=0.44\textwidth]{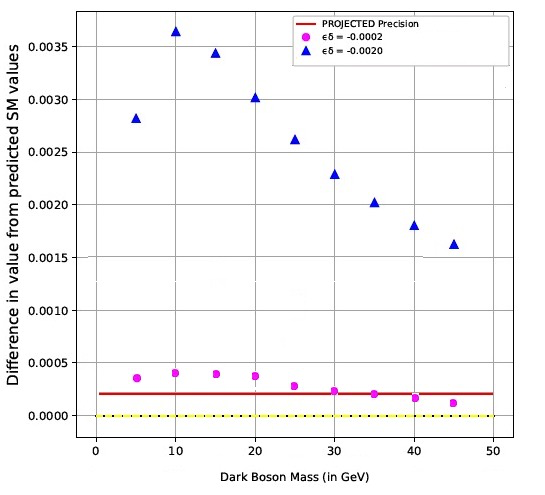}}
\caption{ Left: Dark blue band shows Q$^2$-dependent shift in $\sin^2\theta_W$ caused by a 15~GeV mass dark Z,
 adapted from \cite{Davoudiasl:2015bua}. Right:
 Comparison of the difference between  $\sin^2\theta_W$ at $Q^2=10.58^2$GeV$^2$ for the SM and in a model with two  values of the product of the model-dependent parameters, $\epsilon\delta$, as a function of $m_{dark Z}$.
 The projected precision of the measurement of
  $\sin^2\theta_W$ with 40~ab$^{-1}$ at Chiral Belle is indicated by the red line. The magnitude of the SM theory error on  $\sin^2\theta_W$ is shown by the yellow band.}
  \label{fig:sin2thetaW-NP}
\end{figure}

\begin{table*}[h]
\begin{center}
\begin{tabular}{|l|c|c|r|c|c|c|}
\hline
Final State & $A_{LR}^{SM}$ &  Relative $A_{LR}$  &$g^f_V$  & $\sigma(g^f_V)$  & $\sigma(g^f_V)$  & $\sigma(s^2 \theta_W)$ \\
Fermion     &               &   Error (\%)& W.A.\cite{ALEPH:2005ab}  & (20~ab$^{-1}$) & (40~ab$^{-1}$) & (40~ab$^{-1}$)             \\ 
\hline 
 b-quark     &  -0.020      &  0.4                &  -0.3220               &   0.002          &   0.002          & 0.003 \\
 (eff.=0.3)  &              &                     & $\pm 0.0077$           &improves x4        &                  &       \\
 \hline
 c-quark     &  -0.005      &  0.5                &  +0.1873               &   0.001          &   0.001          & 0.0008 \\
 (eff.=0.15)  &              &                     &  $\pm 0.0070$          &  improves x7      &                  &        \\
 \hline
 tau         & -0.0006      &  2.4                &  -0.0366               &   0.001         &   0.0008         &  0.0004 \\
 (eff.=0.25) &              &                     &  $\pm 0.0010$          &                  &                  &       \\ 
 \hline
 muon        & -0.0006      &  1.5                &  -0.03667              &   0.0007        &   0.0005         & 0.0003 \\
 (eff.=0.5)  &              &                     &  $\pm 0.0023$          &   improves x3     &                  &       \\
 \hline
 electron (17nb   & +0.00015      &  2.0                &  -0.3816               &   0.0009         &   0.0006         & 0.0003 \\
 acceptance, eff=.36)&          &                     &  $\pm 0.00047$         &                  &                  &       \\
 \hline
\end{tabular}
\caption{For each fermion pair cleanly identifiable in Belle~II for the given efficiency in column~1: column~2 gives
the SM value of $A_{LR}$; column~3, the expected 
relative error on $A_{LR}$ based on based on 40~ab$^{-1}$
and  a beam polarization at Belle~II of $0.700\pm0.003$ with an error of $\pm0.003$;
column~4, the current world average value of its neutral current vector coupling;
column~5, the projected error on $g_V^f$ with 20~ab$^{-1}$ of data; 
column~6, the projected error on $g_V^f$ with 40~ab$^{-1}$ of data; and 
column~7, the projected SuperKEKB/Belle~II error on $\sin^2\theta_W^{eff}$ with 40~ab$^{-1}$ of polarized $e^-$ beam data.
}
\label{Table-measurements}
\end{center}
\end{table*}

Table~\ref{Table-measurements} also indicates the uncertainties on $\sin^2\theta_W^{eff}$ that can be achieved with
 40~ab$^{-1}$ of polarized beam data - the combined uncertainty at Chiral  Belle would be comparable to the Z$^0$ world average measured uncertainty
of $\pm 0.00016$ from LEP and SLD\cite{ALEPH:2005ab} but made away from the  Z$^0$-pole at an order of magnitude  lower energy scale. Assuming lepton universality, the uncertainty on $\sin^2\theta_W^{eff}$ from the three Chiral Belle lepton measurements, including the common systematic uncertainty on the beam polarization measurement, is projected to be $\pm 0.00018$.
Figure~\ref{fig:sin2thetaW} shows the determinations of $\sin^2\theta_W$ as a function of energy scale at present and future experimental
facilities including SuperKEKB upgraded with a polarized electron beam delivering 40~ab$^{-1}$ of data to Belle~II.

This electroweak program with polarized electron beams in SuperKEKB would  also provide
the most precise tests of 
neutral current vector coupling universality for all available final-state fermions. The ratio of $A_{LR}^{f_1}/A_{LR}^{f_2}$, which provides a measure of the ratio $g_{V}^{f_1}/g_{V}^{f_2}$, does not depend on the beam polarization, which cancels in the ratio. Because of this cancellation, the universality ratio is measured with an uncertainty dominated by statistics.  For example, $g_V^b/g_V^c$ would be measured with a relative uncertainty below 0.3\%, an order of magnitude lower than the current uncertainty on this ratio.
In addition, right-handed $b$-quark couplings to the $Z$ can  be experimentally
probed with high precision at Belle~II with polarized beams.  Also, measurements with the
projected precision will enable Belle~II to probe parity violation
induced by the exchange of heavy particles such as a
hypothetical TeV-scale $Z^\prime$ boson(s). If such bosons only couple
to leptons they will not be produced at the LHC.  Moreover, the
SuperKEKB machine will have a unique possibility to probe parity
violation in the lepton sector mediated by light and very weakly
coupled particles often referred to as ``Dark Forces''.
Such forces have been entertained as a possible connecting link between normal and dark
matter \cite{Pospelov:2008jd,ArkaniHamed:2008qn}. 
SuperKEKB with polarization would be uniquely sensitivity to ``Dark Sector'' parity violating light neutral gauge
bosons, especially when $Z_{dark}$ is
 off-shell and with a mass between roughly 10 and 35~GeV \cite{Davoudiasl:2015bua} or even up to the Z$^0$ pole,
 or couples more to the 3rd generation (see Figure~\ref{fig:sin2thetaW-NP}).

The enhancement of
parity violation in the muon sector has been an automatic consequence
of some models \cite{Batell:2011qq} that aim at explaining the
unexpected result for the recent Lamb shift measurement in muonic
hydrogen \cite{Pohl:2010zza}. The left-right asymmetry of the $e^-e^+
\to \mu^-\mu^+$ in such models is expected to be enhanced at
low-to-intermediate energies, and SuperKEKB with polarized beams may provide a
conclusive test of such models, as well as impose new constraints on a
parity-violating dark sector.




%% file: Mupair.tex
\label{MuALR}
A  study with  polarized $e^+e^+ \rightarrow \mu^+\mu^-$ KKMC Monte Carlo events processed with the Belle~II GEANT4 detector response software and Belle~II event reconstruction software, validated with  Belle~II $\mu$-pair data, demonstrates that a high purity selection with an efficiency of 50\% for a measurement of $A^{\mu}_{LR}$ is quite feasible~\cite{CHeartyMuonALR}.  The selection of muon pairs starts from a set of two track events. The two tracks are required to be back-to-back, to have $\theta_{cms}>30^\circ$, and to have an invariant mass $M_{\mu\mu}$ in the range $8.7 < M_{\mu\mu} < 11$~GeV/c$^2$. The angular region corresponds to that covered by the muon identification system. To suppress Bhabha events, which make up the vast majority of the two-track sample, both tracks are required to be identified as muons. The fraction of Bhabha  events in the final selected sample is 0.014\%. The efficiency of this selection can be derived from data, and exceeds  $96\%$ within that geometrical acceptance. The trigger efficiency, also derived from Belle~II collision data, is essentially 100\%. 
The overall selection efficiency is 50.4\%, which for the 1147~pb$^{-1}$ production cross section, corresponds to an effective cross-section for selected events of 578~pb$^{-1}$.  For a 10~ab$^{-1}$ sample of each polarization state (20~ab$^{-1}$ in total) having 70\% polarization, and assuming $A^{\mu}_{LR}({\rm measured}) = A^{\mu}_{LR}({\rm SM}) \times \langle Pol \rangle = -0.00064$~\cite{Aleksejevs} $\times 0.70  = -0.00045$, the statistical uncertainty on $A_{LR}$ is 2.1\%. With a 40~ab$^{-1}$ sample, the relative statistical uncertainty is 1.5\%.

With this selection the decays of $\Upsilon(1S)$, $\Upsilon(2S)$, and $\Upsilon(3S)$ to $\mu^+ \mu^-$ make up approximately 0.1\% of selected events. 
These are produced via initial state radiation to the lower mass $\Upsilon$ states and such `radiative return' processes have been extensively studied (see e.g. \cite{BaBar:2003waz}). If one assumes the asymmetry for these events is that of
$b \bar b$ events, $A^b_{LR}({\rm SM})=-0.020$, then they would have an asymmetry 32 times larger than $A^{\mu}_{LR}$. As the Chiral Belle program calls for a precision measurement of $A^b_{LR}$ (see section \ref{charmandbeauty}), and the contributions of the  $\Upsilon$  resonances to the sample can be measured independently of the $A_{LR}$ measurement, this systematic effect can be controlled and is not expected to contribute significantly to the overall uncertainty.
The uncertainty on the measured polarization produces an uncertainty on $A_{LR}$ of 0.4\%, as discussed in section~\ref{TauPolarimetery}. Other uncertainties, including residual contamination from Bhabha, tau pair, and $q \bar q$ events, are expected to contribute a small amount to the systematic uncertainty (less than 0.2\%).  With a 40~ab$^{-1}$ sample having 20~ab$^{-1}$ with 70\% left-handed beam electrons and 20~ab$^{-1}$ with 70\% right-handed, the total uncertainty would be 1.5\%.

%% file: TauPair.tex
\label{TauALR}

A high-purity ($> 98\%$ purity) sample of $e^+e^- \rightarrow \tau^+\tau^-$ events with reasonable efficiency can be obtained with  a selection in which
cross-feeds between $\tau$ decay modes are ignored in the optimization process, since the $A^{\tau}_{LR}$ measurement is insensitive to such cross-feeds. Note that this is not the approach taken in many published $\tau$ analyses, as well as the selection in the Tau Polarimetry section of this white paper,  where backgrounds from cross-feeds are required to be minimized. 

Such a selection approach that ignores cross-feeds has been developed by considering the following $\tau$ decay modes:
 $\tau^{\pm} \rightarrow e^{\pm} \nu \nu$, 
 $\tau \rightarrow \mu^{\pm} \nu \nu$,
  $\tau^{\pm} \rightarrow h^{\pm} \nu$,
 $\tau^{\pm} \rightarrow h^{\pm} \pi^0 \nu$,
 $\tau^{\pm} \rightarrow h 2\pi^0 \nu$, and
  $\tau^{\pm} \rightarrow 3h^{\pm} \nu$.
  Here, $h$ is a charge pion or kaon.  
 
 Considering the following mutually exclusive selections that require one of the $\tau$'s to decay leptonically:
 \begin{itemize}
      \item Events with $\tau^{\pm} \rightarrow e^{\pm} \nu \nu$ and $\tau^{\mp} \rightarrow \mu^{\mp} \nu \nu$. BaBar MC studies indicated this has a 58\% selection efficiency, excluding branching fractions,  and 99\% purity. Including the branching fractions factor, this selects 3.6\% of $e^+e^- \rightarrow \tau^+\tau^-$ events.
     \item $\tau^{\pm} \rightarrow e^{\pm} \nu \nu$ and $\tau^{\mp}$ decays to any of the above listed semileptonic decay modes (i.e. the above listed modes except $\tau^{\mp} \rightarrow e^{\mp} \nu \nu$ or $\tau^{\mp} \rightarrow \mu^{\mp} \nu \nu$). Based on a  similar inclusive $\tau^+\tau^-$ event selection described in reference~\cite{BaBar:2020nlq}, this is expected to have an efficiency (excluding the branching fractions factor), of 63\% with a 99\% purity. It would select 13\% of $\tau$-pair events.
     \item $\tau^{\pm} \rightarrow \mu^{\pm} \nu \nu$ and $\tau^{\mp}$ decays to any of the above listed semileptonic modes.  With an efficiency of 50\% (99\% purity),  it selects 10\% of $\tau$-pairs.
 \end{itemize}
Combining these three selections alone provide a high purity selection of at least 25\% of $\tau$-pair events. 
Early studies that include  a $\tau \rightarrow \rho \nu$ tag, in addition to the leptonic tagged event selection described above, indicate that this can add substantially to the fraction of $\tau$-pair events that can be selected with high purity and that, consequently,  the 25\% should be considered a lower limit to the fraction of  the overall fraction of $\tau$-pair events that can be selected with high purity. The relative statistical uncertainty on $A^{\tau}_{LP}$(measured) with 40~ab$^{-1}$ of data (20~ab$^{-1}$ left-handed and 20~ab$^{-1}$ right-handed) would be 2.3\%. 

As with the muon pair sample discussed in section~\ref{MuALR}, there will be a non-negligible effect on the $A^{\tau}_{LR}$(measured) from radiative returns to the $\Upsilon(1S,2S,3S)\rightarrow \tau^+\tau^-$ that have an asymmetry equal to $A^b_{LR}$, which will be measured with a 0.5\% relative uncertainty. The rate of production of these states can be determined from studies of the  muon pair sample independent of the $A^{\mu}_{LR}$ measurement and the known SM branching fractions of the ${\cal B}(\Upsilon(1S,2S,3S)\rightarrow \tau^+\tau^-)/
{\cal B}(\Upsilon(1S,2S,3S)\rightarrow \mu^+\mu^-$) under the assumption of lepton universality, or from direct measurements as in  reference~\cite{BaBar:2020nlq}. As with the muon pair study, this effect is expected to have a small contribution to the total  uncertainty on  $A^{\tau}_{LR}$(measured), projected to be 2.4\% with 40~ab$^{-1}$ of data, with the beam polarization uncertainty the dominant systematic uncertainty.

%% file: CharmandBeauty.tex
\label{charmandbeauty}
\subsubsection{Introduction}

In order to cleanly separate  $e^+e^- \rightarrow c\Bar{c}$ events from $e^+e^- \rightarrow  b\Bar{b}$ events with high efficiency for the  separate $A_{LR}$ measurements, we develop  $b\Bar{b}$ and $c\Bar{c}$ selectors that use event-shape variables. Although this approach is sufficient for selecting $b\Bar{b}$ events, additional requirements must be imposed in the $c\Bar{c}$ selector in order to suppress additional backgrounds arising from  non-$b\Bar{b}$ sources: light quarks ($uds$) and $\tau$-pairs. 

A machine learning tool is trained to classify between $c\Bar{c}$ and $b\Bar{b}$ events using the ten most important event-shape variables (as identified by the model). The model chosen for this training is a gradient boosted decision tree (GBDT) with depth of 5, trained over 100 iterations. The model was also trained on a larger number of variables (23), but negligible benefits were found to adding extra variables. 

Further selections are tested on the $c\Bar{c}$, $b\Bar{b}$ events and their backgrounds $u\Bar{u}$, $d\Bar{d}$, $s\Bar{s}$ and $\tau^+\tau^-$ ($uds\tau$) requiring events to have an identified lepton ($e^\pm$ or $\mu^\pm$) as a way to reduce background rates from $uds$ backgrounds in the $c\Bar{c}$ sample.

\subsubsection{Training and evaluation of the model}

Monte Carlo (MC) simulation samples of $b\Bar{b}$, $c\Bar{c}$ and $uds\tau$ events are collected. Table \ref{tab:cc_bb_numbers} shows the number of events per sample. The number of events per sample is normalized to the cross section of each process \cite{BelleIIPhysBook}.

\begin{table}[hbt!]
\centering
\begin{tabular}{ccccccccc}
\toprule
dataset       & charged & mixed & $b\Bar{b}$ & $c\Bar{c}$ & $u\Bar{u}$ & $u\Bar{u}$ & $s\Bar{s}$ & $\tau^+\tau^-$    \\ \midrule
MC number     & 268591  & 253669 & 522260 & 661033 & 798314 & 199454 & 190501 & 457103 \\ \midrule
Reconstructed & 268580  & 253659 & 522239 & 646343 & 788986 & 197119 & 180206 & 457102 \\ \bottomrule
\end{tabular}
\caption{Number of events generated and reconstructed in each simulation sample. The $b\Bar{b}$ events are separately simulated as charged ($B^+B^-$) and mixed ($B^0\Bar{B}^0$) and are shown separately next to the total number of $b\Bar{b}$.}\label{tab:cc_bb_numbers}
\end{table}

The events are reconstructed without any decay requirements, and event-shape variables are calculated based on every candidate particle available in the event. Table \ref{tab:cc_bb_feature_importance} shows the ten event-shape variables used for training with their relative importance.

\begin{table}[hbt!]
\centering
\begin{tabular}{cc}
\toprule
Variable                   & Feature Importance \\ \midrule
foxWolframR2               & 0.590              \\ \midrule
thrust                     & 0.184              \\ \midrule
foxWolframR1               & 0.081              \\ \midrule
harmonicMomentThrust0      & 0.060              \\ \midrule
thrustAxisCosTheta         & 0.039              \\ \midrule
harmonicMomentCollision2   & 0.020              \\ \midrule
foxWolframR3               & 0.010              \\ \midrule
aplanarity                 & 0.006              \\ \midrule
harmonicMomentThrust2      & 0.006              \\ \midrule
sphericity                 & 0.004              \\ \bottomrule
\end{tabular}
\caption{Importance of the variables as calculated by the GBDT, ordered by decreasing importance. The importance metric has arbitrary units and is used to quantify the decision weight of the variable during GBDT decisions.}
\label{tab:cc_bb_feature_importance}
\end{table}

\begin{figure}[hbt!]
    \centering
    \includegraphics[width=0.8\textwidth]{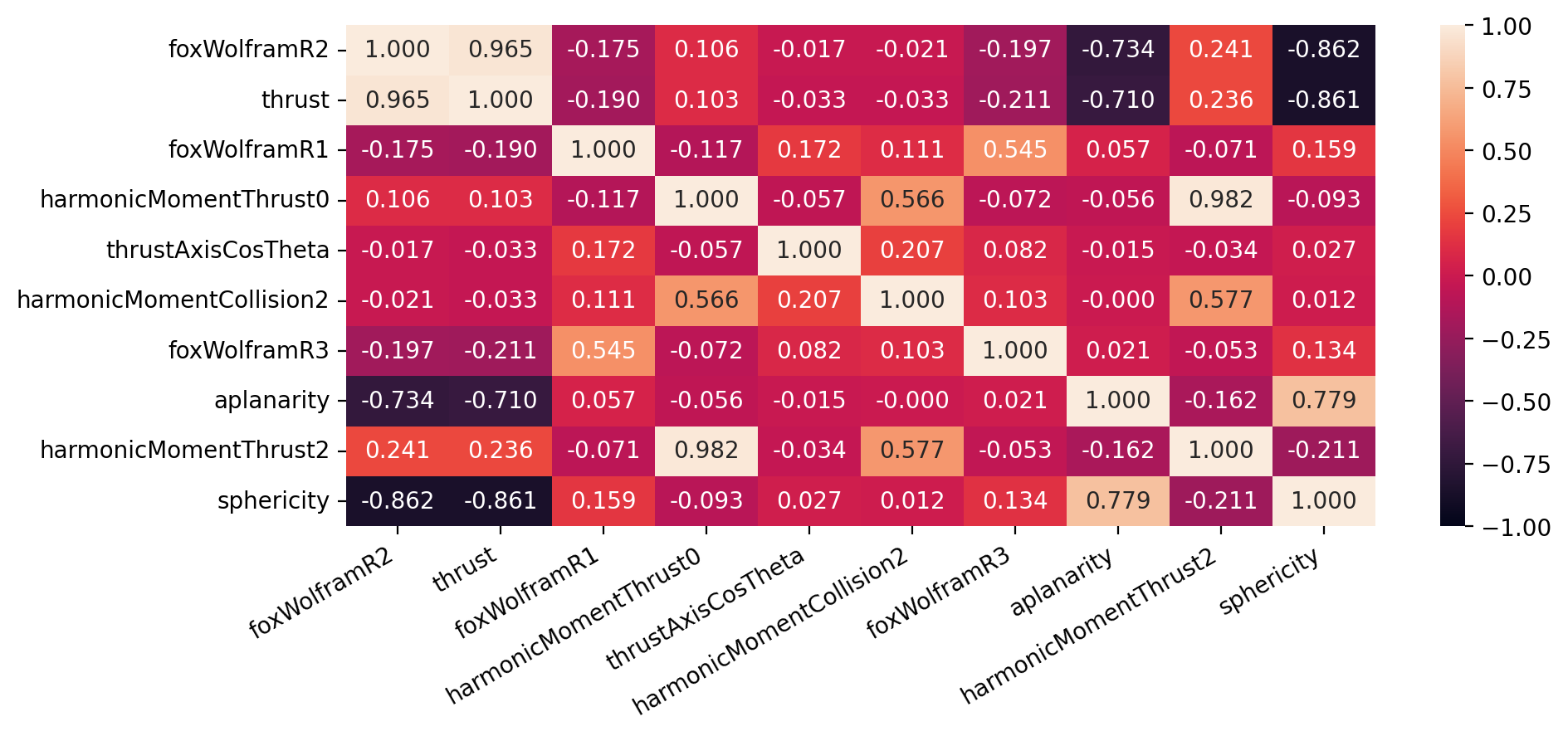}
    \caption{Correlation matrix for the event-shape variables of the $c\Bar{c}$ simulation.}
    \label{fig:corr_cc}
\end{figure}

\begin{figure}[hbt!]
    \centering
    \includegraphics[width=0.8\textwidth]{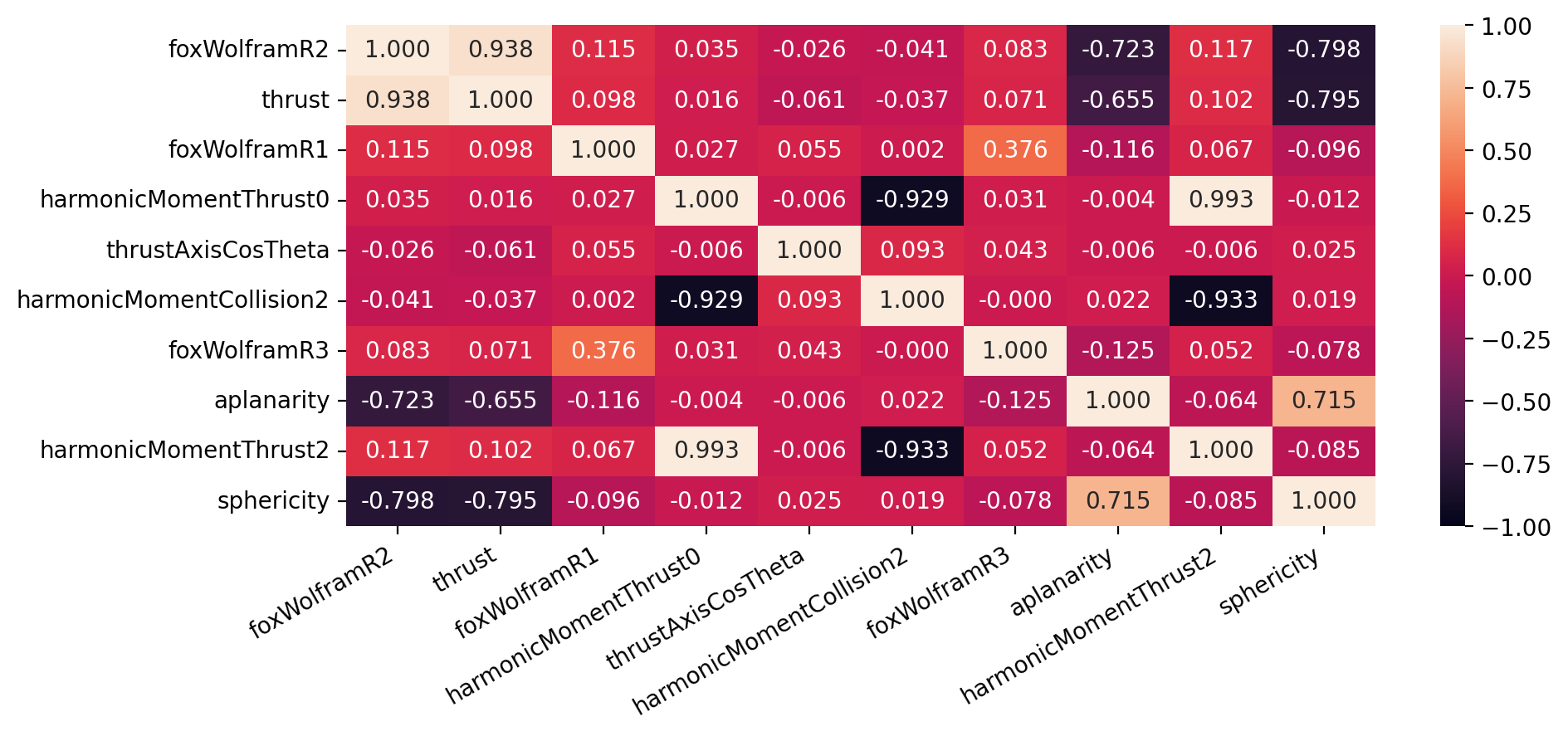}
    \caption{Correlation matrix for the event-shape variables of the $b\Bar{b}$ ($B^\pm$) simulation.}
    \label{fig:corr_charged}
\end{figure}

\begin{figure}[hbt!]
    \centering
    \includegraphics[width=0.8\textwidth]{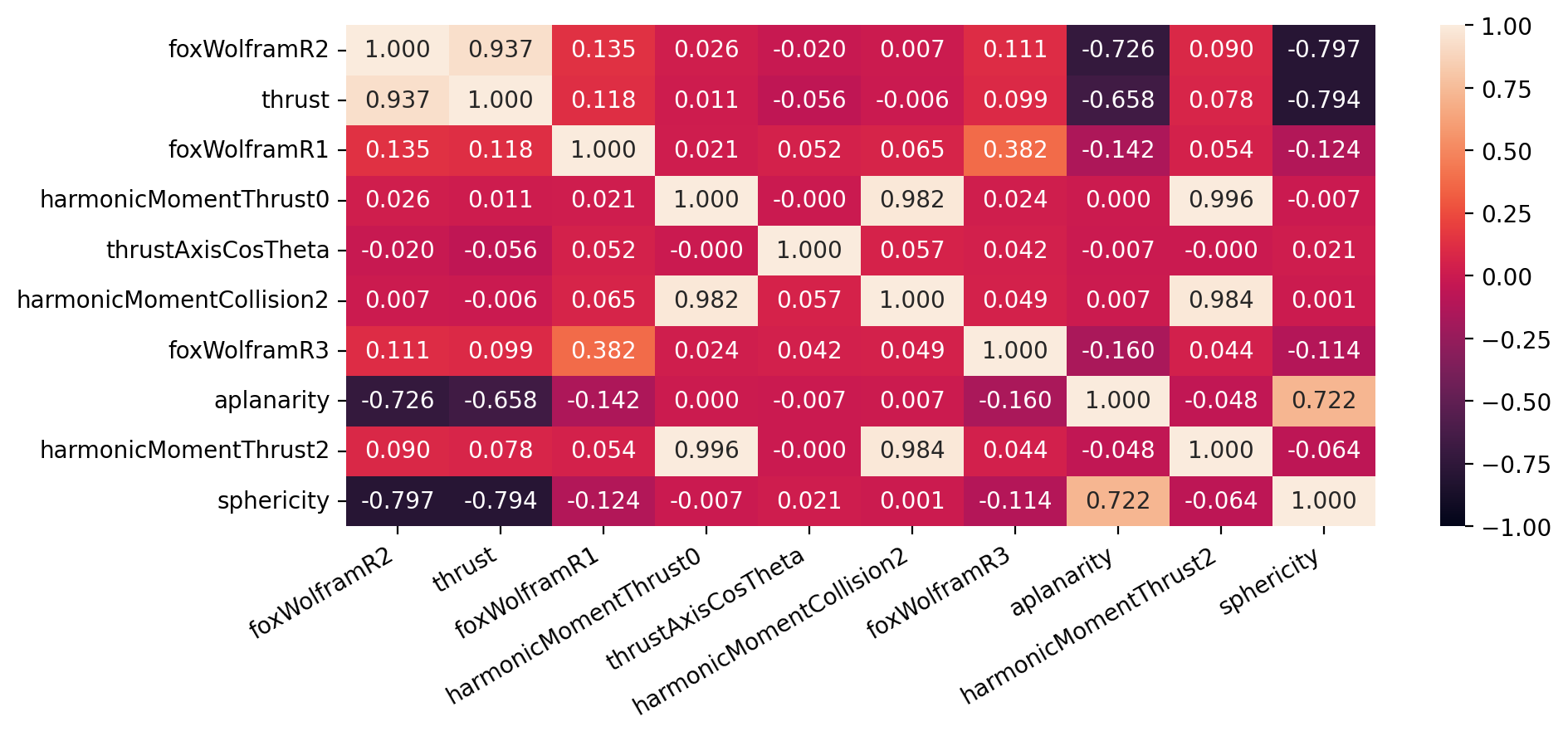}
    \caption{Correlation matrix for the event-shape variables of the $b\Bar{b}$ ($B^0$) simulation.}
    \label{fig:corr_mixed}
\end{figure}

The following figures show the correlation matrix of the different event-shape variables for the different simulation samples. The $c\Bar{c}$ simulation is shown in figure \ref{fig:corr_cc}, the $b\Bar{b}$ ($B^\pm$) simulation is shown in figure \ref{fig:corr_charged} and the $b\Bar{b}$ ($B^0$) simulation is shown in figure \ref{fig:corr_mixed}.

\begin{figure}[hbt!]
    \centering
    \includegraphics[width=0.45\textwidth]{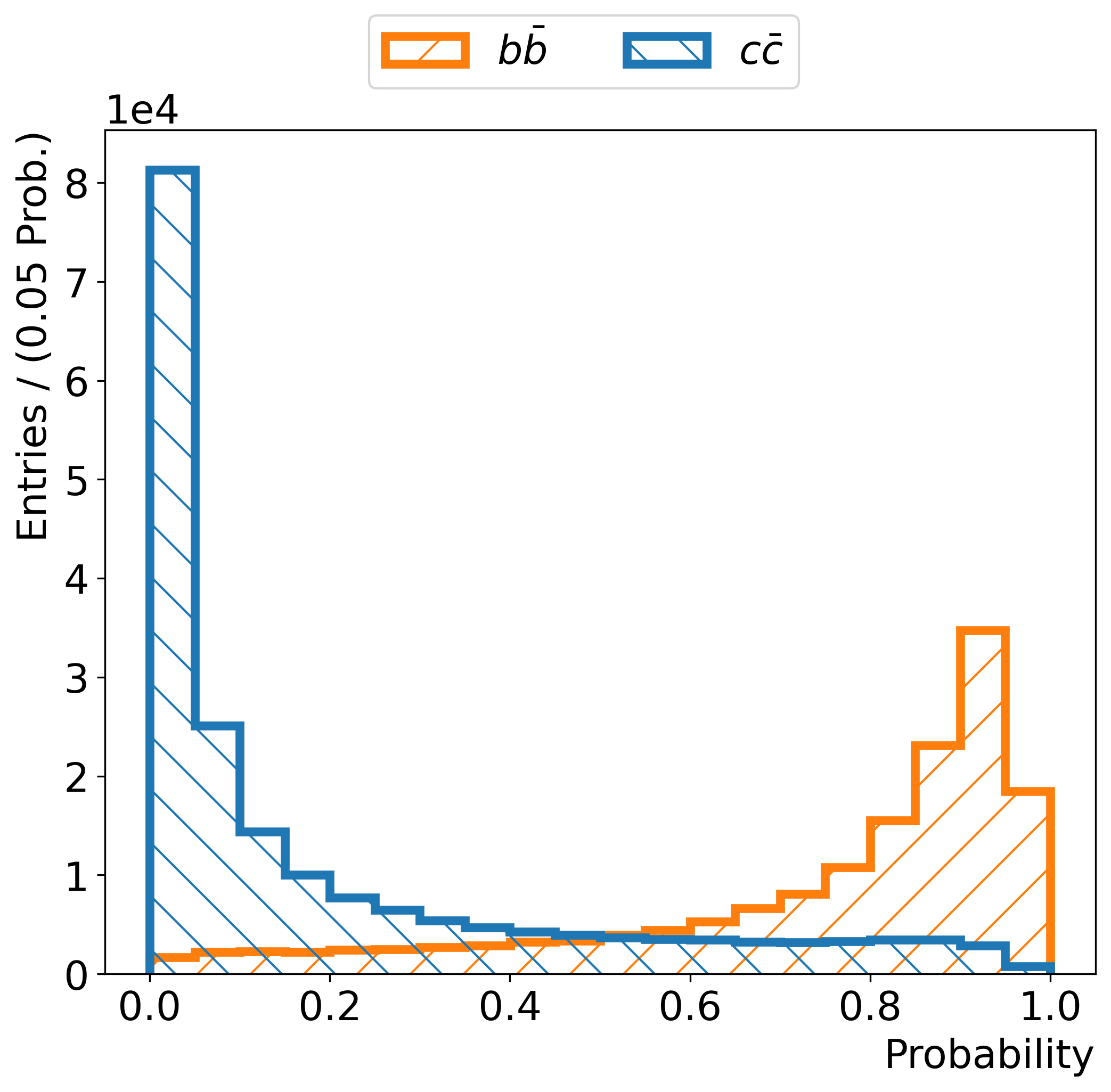}
    \caption{GBDT prediction of events, where the probability is the certainty of the model that the event is a $b\bar{b}$ event. Identified by color are the truth value of a given prediction ($c\Bar{c}$ or $b\Bar{b}$).}
    \label{fig:cc_bb_prob_distribution}
\end{figure}

Once trained, the GBDT is used to predict if an event is from a $c\Bar{c}$ event or a $b\Bar{b}$ event. Figure \ref{fig:cc_bb_prob_distribution} shows the output distribution of the GBDT model when classifying a validation sample. 

\subsubsection{Evaluation}

The performance of the GBDT model is evaluated on two distinct tasks: the identification of $c\Bar{c}$ and $b\Bar{b}$ events. The performance is obtained using two Receiver Operating Characteristic (ROC) curves and their associated Area Under the Curve (AUC). The ROC curve is expressed using efficiencies and fake rates as defined in equation \ref{eq:cc_bb_fake_rate}

\begin{align}\label{eq:cc_bb_fake_rate}
    b\bar{b} \text{ Efficiency} = \frac{\# \text{ of real }b\bar{b}\text{ predicted as }b\bar{b}}{\# \text{total number of MC generated }b\bar{b}} \\
    c\bar{c} \text{ Fake rate} = \frac{\# \text{ of real }c\bar{c}\text{ misidentified as }b\bar{b}}{\# \text{total number of MC generated }c\bar{c}}
\end{align}

\begin{figure}[hbt!]
    \centering
    \begin{subfigure}{0.45\textwidth}
    \includegraphics[width=\linewidth]{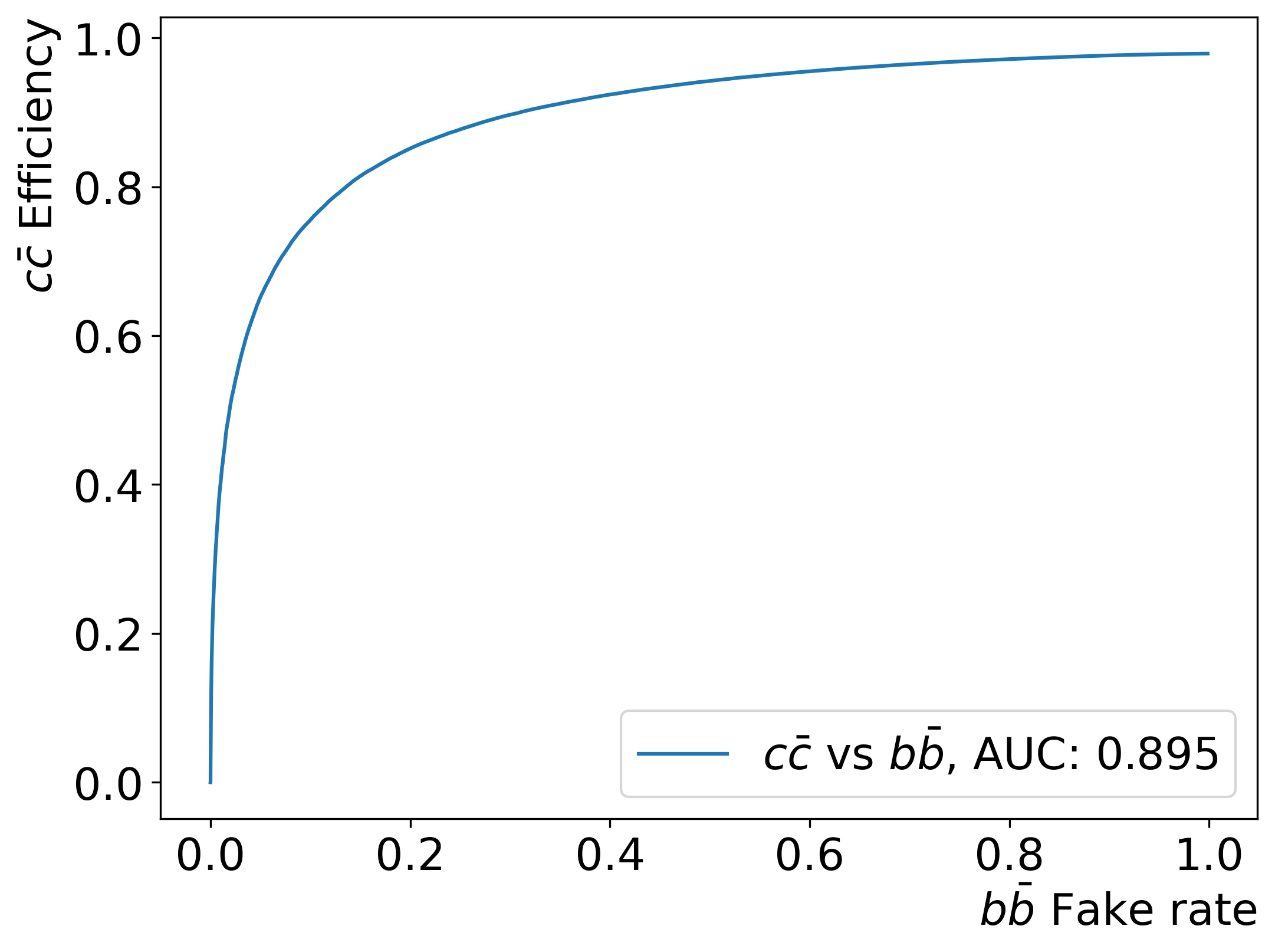}
    \caption{$c\bar{c}$ classification.}\label{fig:ROC_cc}
    \end{subfigure}
    \begin{subfigure}{0.45\textwidth}
    \includegraphics[width=\linewidth]{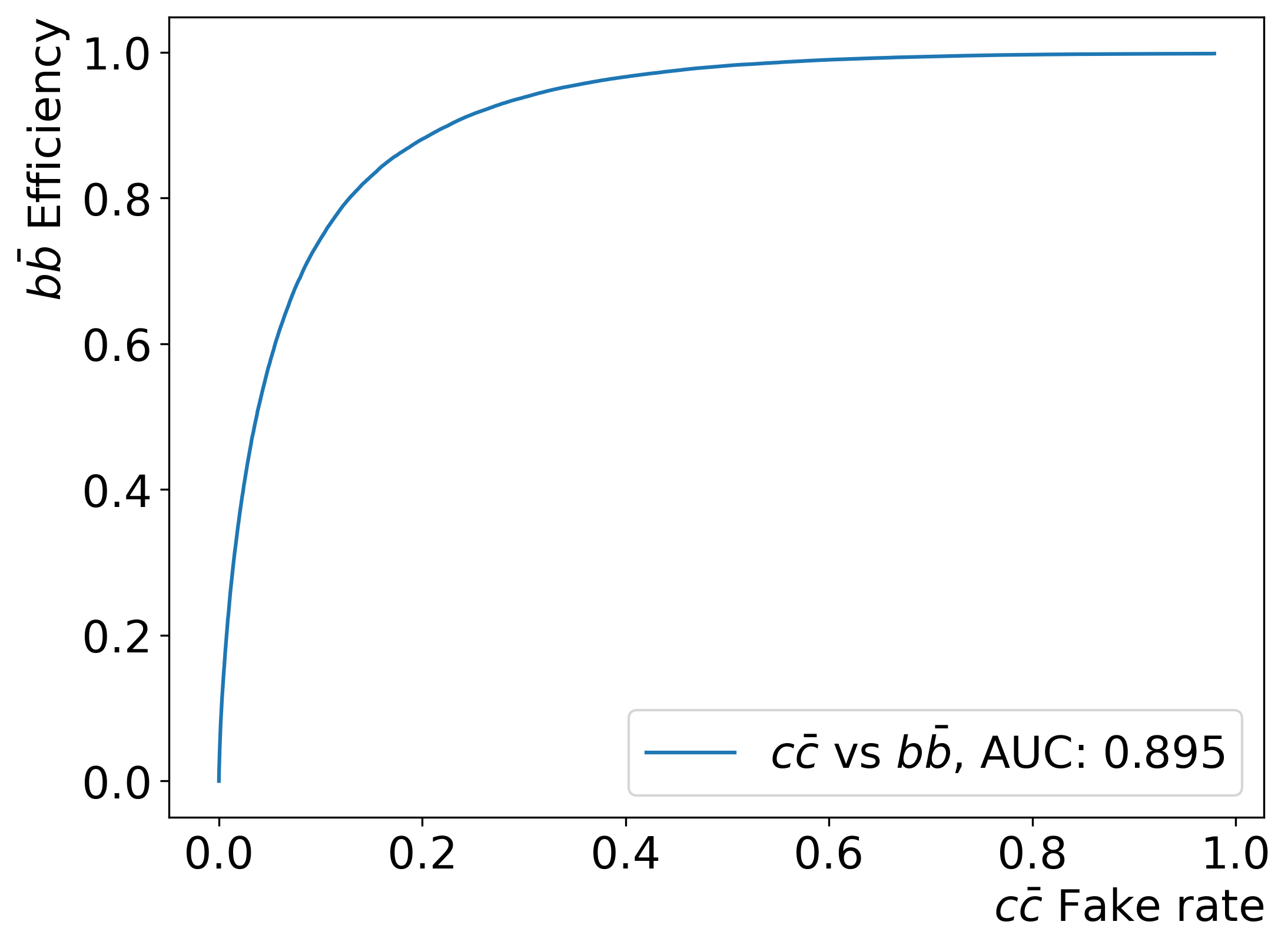}
    \caption{$b\bar{b}$ classification.}\label{fig:ROC_bb}
    \end{subfigure}
    \caption{ROC curves of the $c\Bar{c}$, $b\Bar{b}$ classifier. The definition of fake rates and efficiencies can be found in equation \ref{eq:cc_bb_fake_rate}.}
    \label{fig:cc_bb_ROC_curves}
\end{figure}

Figure \ref{fig:cc_bb_ROC_curves} shows the ROC curves calculated on a validation sample of $c\bar{c}$ and $b\bar{b}$. The figures show that the model can discriminate between $c\Bar{c}$ and $b\Bar{b}$ events with high purity.

\subsubsection{Classification of $c\Bar{c}$}

Table \ref{tab:cc_eff} shows the fake rate of $c\Bar{c}$ at different efficiencies for figure \ref{fig:ROC_cc}. Only the fake rates of $b\bar{b}$ is shown as $uds\tau$ backgrounds are classified as $c\bar{c}$ events by the GBDT model.
Table \ref{tab:cc_eff_nb} shows the rates of table \ref{tab:cc_eff} converted to number of events per nb$^{-1}$.



\begin{table}
\centering
\begin{subtable}[]{0.35\textwidth}
\centering
\begin{tabular}{cc}
\toprule
\multicolumn{2}{c}{Fraction}       \\ \midrule
$c\bar{c}$ efficiency & $b\bar{b}$ fake rate \\ \midrule
0.102            & 0.000           \\ \midrule
0.201            & 0.002           \\ \midrule
0.301            & 0.005           \\ \midrule
0.402            & 0.010           \\ \midrule
0.500            & 0.020           \\ \midrule
0.599            & 0.037           \\ \midrule
0.700            & 0.069           \\ \midrule
0.800            & 0.138           \\ \midrule
0.900            & 0.311           \\ \midrule
0.979            & 0.998           \\ \bottomrule
\end{tabular}
\caption{Efficiency and fake rates.}\label{tab:cc_eff}
\end{subtable}
\begin{subtable}[]{0.35\textwidth}
\centering
\begin{tabular}{cc}
\toprule
\multicolumn{2}{c}{Events per nb$^{-1}$}  \\ \midrule
$c\Bar{c}$ & $b\bar{b}$ \\ \midrule
0.133 & 0.000 \\ \midrule
0.261 & 0.002 \\ \midrule
0.391 & 0.006 \\ \midrule
0.523 & 0.011 \\ \midrule
0.650 & 0.022 \\ \midrule
0.779 & 0.041 \\ \midrule
0.910 & 0.077 \\ \midrule
1.040 & 0.153 \\ \midrule
1.170 & 0.345 \\ \midrule
1.273 & 1.108 \\ \bottomrule
\end{tabular}
\caption{Number of events selected as $c\bar{c}$..}\label{tab:cc_eff_nb}
\label{tab:table1_b}
\end{subtable}
\caption{Classification of $c\Bar{c}$ events against $b\bar{b}$ events. These tables show the results from figure \ref{fig:ROC_cc}.}
\end{table}

\subsubsection{Classification of $b\Bar{b}$}

Table \ref{tab:bb_eff} shows the fake rate of $b\Bar{b}$ at different efficiencies for figure \ref{fig:ROC_bb}. The fake rates of $uds\tau$ backgrounds are included.
Table \ref{tab:bb_eff_nb} shows the rates of Table \ref{tab:bb_eff} converted to number of events per nb$^{-1}$.

\begin{table}[hbt!]
\centering
\begin{tabular}{@{}ccccccc@{}}
\toprule
\multicolumn{7}{c}{Fraction}                                                                                          \\ \midrule
Efficiency & \multicolumn{1}{l}{Background} & \multicolumn{5}{c}{Fake Rate}                                      \\
$b\Bar{b}$ & Fraction                               & $c\Bar{c}$ & $u\Bar{u}$ & $d\Bar{d}$ & $s\Bar{s}$ & $\tau^+\tau^-$ \\ \midrule
0.098      & 0.088                               & 0.003      & 0.002      & 0.002      & 0.002      & 0.002          \\ \midrule
0.196      & 0.091                               & 0.008      & 0.004      & 0.004      & 0.004      & 0.002          \\ \midrule
0.301      & 0.110                               & 0.015      & 0.008      & 0.008      & 0.008      & 0.003          \\ \midrule
0.401      & 0.128                               & 0.024      & 0.013      & 0.013      & 0.014      & 0.003          \\ \midrule
0.501      & 0.153                               & 0.038      & 0.020      & 0.020      & 0.022      & 0.003          \\ \midrule
0.599      & 0.181                               & 0.056      & 0.029      & 0.030      & 0.032      & 0.004          \\ \midrule
0.699      & 0.221                               & 0.083      & 0.044      & 0.046      & 0.050      & 0.004          \\ \midrule
0.800      & 0.281                               & 0.130      & 0.070      & 0.073      & 0.079      & 0.006          \\ \midrule
0.900      & 0.380                               & 0.226      & 0.124      & 0.131      & 0.144      & 0.014          \\ \midrule
0.998      & 0.802                               & 0.979      & 0.966      & 0.966      & 0.941      & 0.996          \\ \bottomrule
\end{tabular}
\caption{Efficiency and fake rates for the classification of $b\Bar{b}$. This table shows the results from figure \ref{fig:ROC_bb}.}\label{tab:bb_eff}
\end{table}

\begin{table}[hbt!]
\centering
\begin{tabular}{cccccc}
\toprule
\multicolumn{6}{c}{Events selected as $b\bar{b}$ per nb$^{-1}$}\\ \midrule
\multicolumn{1}{c}{$b\Bar{b}$} & \multicolumn{1}{c}{$c\Bar{c}$} & \multicolumn{1}{c}{$u\Bar{u}$} & \multicolumn{1}{c}{$d\Bar{d}$} & \multicolumn{1}{c}{$s\Bar{s}$} & \multicolumn{1}{c}{$\tau^+\tau^-$} \\ \midrule
0.109                                & 0.004                               & 0.003                               & 0.001                               & 0.001                              & 0.002                                \\ \midrule
0.218                                & 0.010                               & 0.006                               & 0.002                               & 0.002                              & 0.002                                \\ \midrule
0.334                                & 0.020                               & 0.013                               & 0.003                               & 0.003                              & 0.003                                \\ \midrule
0.445                                & 0.031                               & 0.021                               & 0.005                               & 0.005                              & 0.003                                \\ \midrule
0.556                                & 0.049                               & 0.032                               & 0.008                               & 0.008                              & 0.003                                \\ \midrule
0.665                                & 0.073                               & 0.047                               & 0.012                               & 0.012                              & 0.004                                \\ \midrule
0.776                                & 0.108                               & 0.071                               & 0.018                               & 0.019                              & 0.004                                \\ \midrule
0.888                                & 0.169                               & 0.113                               & 0.029                               & 0.030                              & 0.006                                \\ \midrule
0.999                                & 0.294                               & 0.200                               & 0.052                               & 0.055                              & 0.013                                \\ \midrule
1.108                                & 1.273                               & 1.555                               & 0.386                               & 0.358                              & 0.915                                \\ \bottomrule
\end{tabular}
\caption{Events selected as $b\bar{b}$ per nb$^{-1}$ for the classification of $b\Bar{b}$.}\label{tab:bb_eff_nb}
\end{table}

\begin{table}[hbt!]
\centering
\begin{subtable}[]{0.49\textwidth}
\centering
\resizebox{\textwidth}{!}{%
\begin{tabular}{ccccccc}
\toprule
\multicolumn{7}{c}{Fraction}                                                                                                                                                   \\ \midrule
           & \multicolumn{3}{c}{MC Truth}                                                    & \multicolumn{3}{c}{ID Cuts}                                                     \\
Type       & \multicolumn{1}{c}{$\mu$} & \multicolumn{1}{c}{$e$} & \multicolumn{1}{c}{Total} & \multicolumn{1}{c}{$\mu$} & \multicolumn{1}{c}{$e$} & \multicolumn{1}{c}{Total} \\ \midrule
$b\Bar{b}$ & 0.26                      & 0.27                    & 0.40                      & 0.33                      & 0.22                    & 0.41                      \\ \midrule
$c\Bar{c}$ & 0.13                      & 0.16                    & 0.28                      & 0.25                      & 0.12                    & 0.34                      \\ \midrule
$u\Bar{u}$ & 0.02                      & 0.06                    & 0.08                      & 0.16                      & 0.05                    & 0.20                      \\ \midrule
$d\Bar{d}$ & 0.02                      & 0.06                    & 0.08                      & 0.16                      & 0.05                    & 0.20                      \\ \midrule
$s\Bar{s}$ & 0.02                      & 0.05                    & 0.07                      & 0.14                      & 0.04                    & 0.17                      \\ \midrule
$\tau^+\tau^-$                   & 0.25                     & 0.27                   & 0.47                     & 0.27                     & 0.21                   & 0.45                     \\ \bottomrule
\end{tabular}}
\caption{Fraction of events containing at least one lepton.}\label{tab:lepton_frac}
\end{subtable}
\hspace{\fill}
\begin{subtable}[]{0.49\textwidth}
\centering
\resizebox{\textwidth}{!}{%
\begin{tabular}{ccccccc}
\toprule
\multicolumn{7}{c}{Events per nb$^{-1}$} \\ \midrule
                      & \multicolumn{3}{c}{MC Truth}                                         & \multicolumn{3}{c}{ID Cuts}                                       \\
Type & \multicolumn{1}{c}{$\mu$} & \multicolumn{1}{c}{$e$} & \multicolumn{1}{c}{Total} & \multicolumn{1}{c}{$\mu$} & \multicolumn{1}{c}{$e$} & \multicolumn{1}{c}{Total} \\ \midrule
$b\Bar{b}$                    & 0.29                     & 0.30                   & 0.45                     & 0.36                     & 0.24                   & 0.46                     \\ \midrule
$c\Bar{c}$                    & 0.17                     & 0.21                   & 0.36                     & 0.32                     & 0.15                   & 0.44                     \\ \midrule
$u\Bar{u}$                    & 0.04                     & 0.10                   & 0.13                     & 0.25                     & 0.07                   & 0.31                     \\ \midrule
$d\Bar{d}$                    & 0.01                     & 0.02                   & 0.03                     & 0.06                     & 0.02                   & 0.08                     \\ \midrule
$s\Bar{s}$                    & 0.01                     & 0.02                   & 0.03                     & 0.05                     & 0.01                   & 0.07                     \\ \midrule
$\tau^+\tau^-$                   & 0.23                     & 0.24                   & 0.44                     & 0.25                     & 0.20                   & 0.41                     \\ \bottomrule
\end{tabular}}
\caption{Number of events containing at least one lepton.}\label{tab:lepton_nb}
\end{subtable}
\caption{Simulation sets containing at least one lepton. The sets are selected using cuts: truth is the number MC generated number of events with leptons. ID cuts uses a cut on the default particle identification tool ($>$0.95) and on $E/p$ ($>$0.85 electron).}
\end{table}

\subsubsection{Lepton requirement study}

Also explored was whether or not applying a requirement that a muon or electron be present in the event would provide an additional means of reducing the $uds$ backgrounds in the $c\bar{c}$ and $b\bar{b}$ selections. Such cuts can be used separately from the selection of $c\bar{c}$ from $b\bar{b}$ events with the GBDT. Table \ref{tab:lepton_frac} shows the fraction of events kept after requiring the event to have a lepton ($\mu^\pm$ or $e^\pm$). Table \ref{tab:lepton_nb} shows the fraction of table \ref{tab:lepton_frac} converted to number of events per nb$^{-1}$. 
Although it is evident that $uds$ backgrounds can be partially suppressed using a lepton requirement, higher purity against the $uds$ backgrounds will require additional measures to develop a sufficiently pure $c\bar{c}$ sample, such as requiring the presence of a charm meson. 

\subsubsection{Beauty $A_{LR}$}
For the $b\bar{b}$ sample, a GBDT requirement alone that provides a 30\% efficiency that has an 88\% purity, with half of the backgrounds coming from $c\bar{c}$. Studies with data below the $\Upsilon(4S)$ resonance  will enable these backgrounds to be measured precisely, and therefore they would introduce a $\cal{O}$(0.1\%) systematic uncertainty on $A_{LR}$ for $b\bar{b}$. The statistical error would be significantly below that and the dominant uncertainty would arise from the systematic error of the measurement of the beam polarization. As shown in the Tau Polarimetry section, the systematic component of the uncertainty that can be achieved is  0.3\%. Assuming this is 100\% correlated across running periods, it can be used as a reliable estimate of the overall systematic uncertainty that can be achieved. Concerning the statistical component of the polarization,  with a 20~ab$^{-1}$ sample, which consists of  10~ab$^{-1}$ of left-handed beam electrons and 10~ab$^{-1}$ of right-handed beam electrons, the statistical error would project to be less than 0.1\%. For a 70\% beam polarization, these translate into a 0.4\% relative uncertainty on the beam polarization, which will dominate the precision with which the $A_{LR}$ for $b\bar{b}$ can be measured.

%% file: Charm_ALR.tex
\subsubsection{Charm $A_{LR}$}
A  $c\bar{c}$ sample can be obtained that has an efficiency of 50\% and $b\bar{b}$ background that is only 3\% of the combined number of  $c\bar{c}$ and $b\bar{b}$ events selected.  However, additional requirements are required to suppress large $uds\tau$ backgrounds. For example, requiring the presence of a charm meson in the event would provide a high purity $c\bar{c}$ sample. If one assumes a 30\% efficiency to find any charm meson,  this would provide a 15\% overall efficiency for selecting  a  $c\bar{c}$ sample, and  with   40~ab$^{-1}$ of data leads to a 0.3\% relative statistical error on $A_{LR}$ for $c\bar{c}$. This would be less than the 0.4\% relative uncertainty coming from the systematic uncertainty from the beam polarization, described above. Under these assumptions, the total relative uncertainty would be 0.6\% with 20~ab$^{-1}$ of data and 0.5\% with 40~ab$^{-1}$. The development of an efficient charm meson finder forms part of the next steps in the R\&D program. 



\begin{figure}
    \centering
    \includegraphics[width=0.45\linewidth]{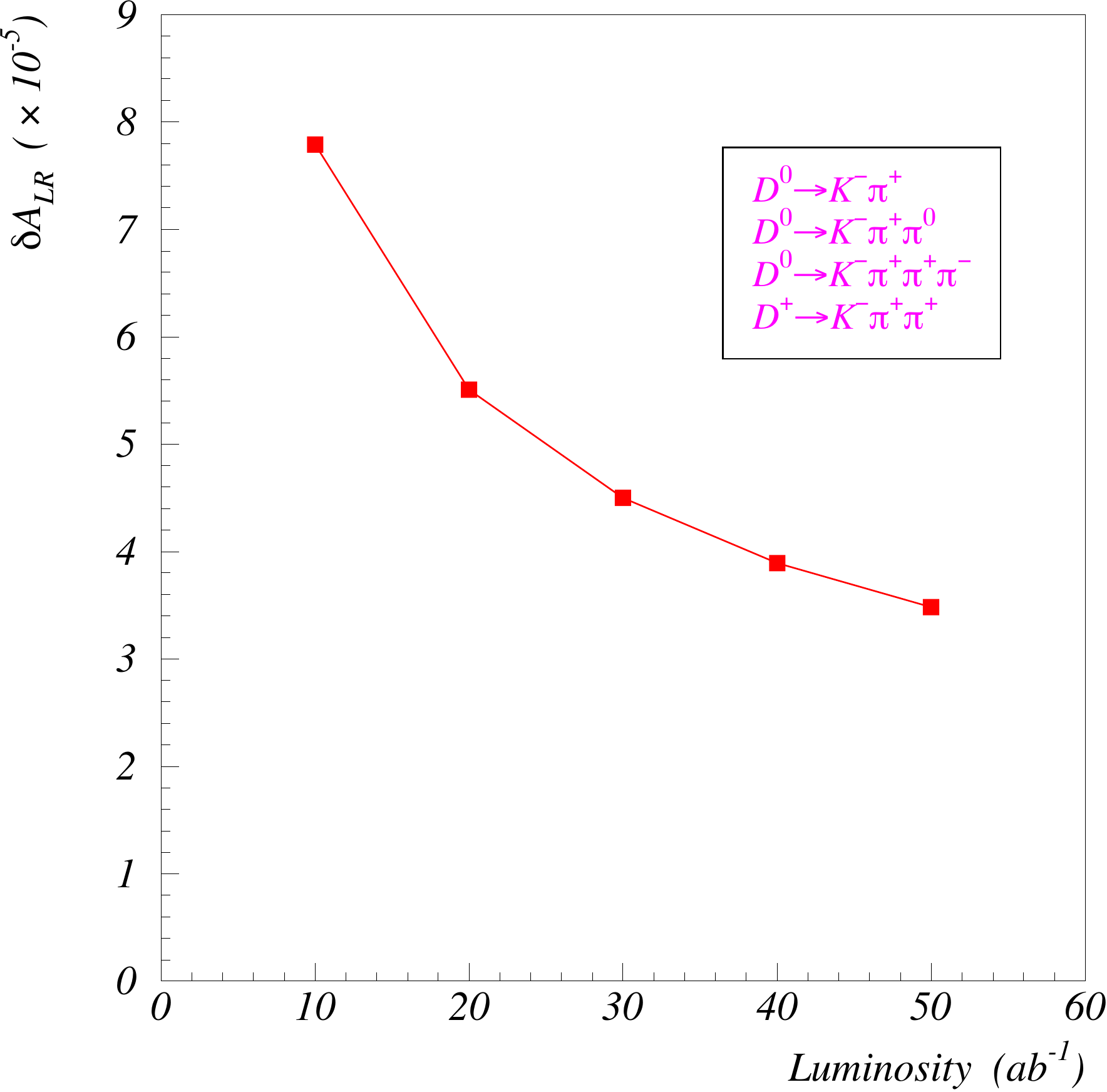}
    \hskip0.30in
    \includegraphics[width=0.45\linewidth]{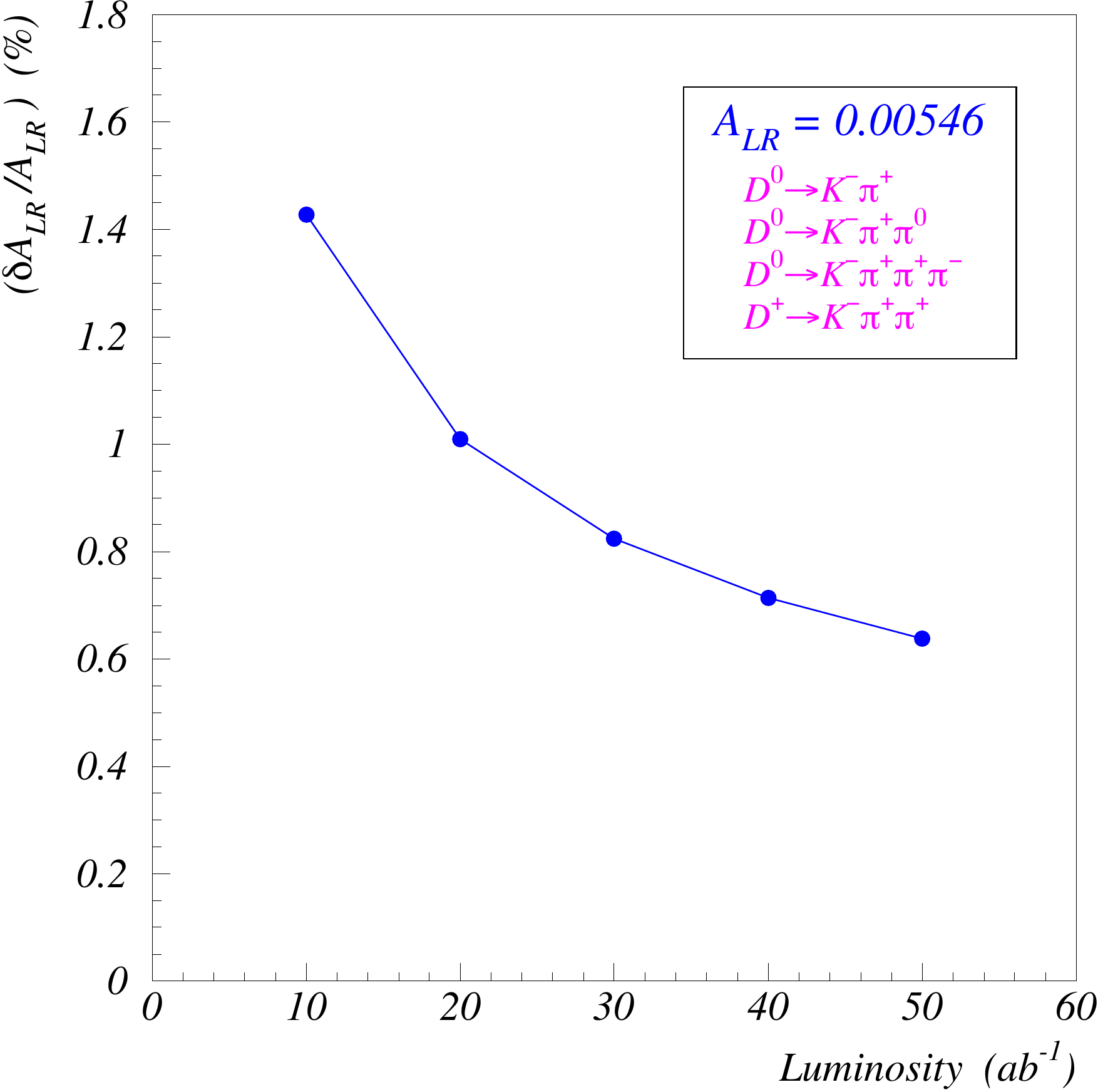}
    \caption{Left:~absolute uncertainty on $A^{}_{LR}$ obtained 
    by reconstructing Cabibbo-favored $D^{*+}\to D^0\pi^+$, 
    $D^0\to K^-\pi^+$, $K^-\pi^+\pi^0$, $K^-\pi^+\pi^+\pi^-$ decays,
    and $D^{*+}\to D^+\pi^0$,  $D^+\to K^-\pi^+\pi^+$ decays.
    Right:~the corresponding fractional uncertainty on $A^{}_{LR}$
    in percent, assuming $A^{}_{LR}=0.00546$.}
    \label{fig:charm_ALR}
\end{figure}

One can obtain a conservative estimate for the precision on
$A^{}_{LR}$ that can be obtained by scaling the signal yields 
of charm decays reconstructed at Belle and
Belle~II~\cite{Belle-II:2021cxx,Belle:2005xmv}. We consider 
the Cabibbo-favored decays $D^0\to K^-\pi^+$, $K^-\pi^+\pi^0$, $K^-\pi^+\pi^+\pi^-$, and $D^+\to K^-\pi^+\pi^+$. To reduce
backgrounds from non-$c\bar{c}$ sources, we require that 
the $D^0$ and $D^+$ mesons originate from $D^{*+}\to D^0\pi^+$ 
and $D^{*+}\to D^+\pi^0$ decays. For these decay modes, the 
purity of the Belle and Belle~II samples ranged from 91-99\%.
The resulting uncertainty on $A^{}_{LR}$ is plotted in 
Fig.~\ref{fig:charm_ALR}. The figure shows that a 
fractional uncertainty of below 1\% could be obtained 
with $>\!20$~ab$^{-1}$ of data for each beam polarization.


%% file: Bhabha.tex
\begin{figure}
    \centering
    \includegraphics[trim=60 205 60 210,clip,width=0.5\linewidth]{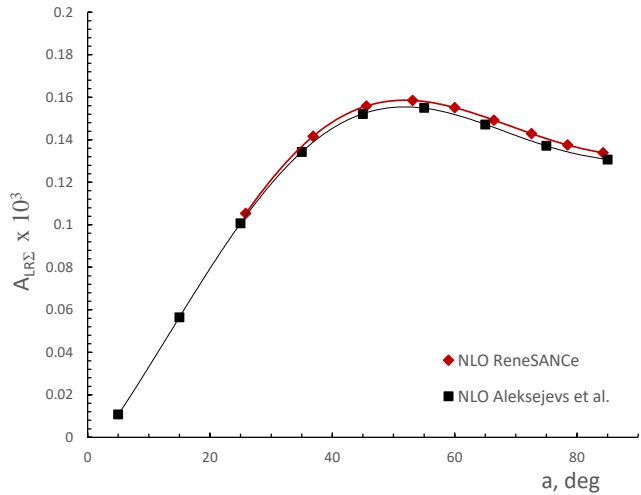}
    \caption{Distribution of A$_{LR}$ in Bhabhas with an angular acceptance of -$\cos a$ to $\cos a$}
    \label{fig:bhabhaALR}
\end{figure}

\label{Bhabha}
The $A_{LR}$ behaviour in the Bhabha events (\epem$\rightarrow$\epem) differs from the other final state fermions due to the presence of t-channel scattering. A detailed NLO calculation of A$_{LR}$ for Bhabhas has been carried out by Aleksejevs \textit{et. al.}\cite{AleksBhabha}. A luminosity paper has been published by the Belle II collaboration\cite{B2lumi}, and includes the detector efficiency for collecting Bhabhas. Belle II reports a high purity selection with an efficiency of 36\% for an angular acceptance corresponding to a cross section of 17.4 nb. 
From these results the expected value of A$^e_{LR}$ is +$1.5\times 10^{-4}$, and with a 40 \ab~data sample and 70\% beam polarization, a relative statistical uncertainty of 2\% is projected. This 2\% relative uncertainty is expected to be statistically dominated. This would translate into a uncertainty of 0.0003 for a measurement of \stw~ using only the Bhabhas, and similarly an uncertainty of 0.0006 on a measurement of $g_V^e$.
Initial studies comparing the NLO value of A$_{LR}$ to the ReneSANCe MC generator\cite{ReneSANCe} have been carried out. The preliminary results from the generator show good agreement on the behaviour of A$_{LR}$ over the angular acceptance as shown in Figure \ref{fig:bhabhaALR}. Further studies are being carried out to determine if the small difference between the two approaches is related to differences in variable definitions or reflects the level of uncertainty in the NLO calculations.

%% file: Taug-2.tex
So far, a test of physics beyond the SM (BSM) via a measurement of the anomalous magnetic moment of the $\tau$, $a_\tau$, has proven elusive, with current limits not precise enough to even resolve the Schwinger term in the QED expansion~\cite{DELPHI:2003nah,Gonzalez-Sprinberg:2000lzf}. Meanwhile, given the current tension in $a_\mu$~\cite{Abi:2021gix,Aoyama:2020ynm}, such a test 
would provide valuable complementary information  in case the muonic tension did signal contributions from new particles or interactions. However, for a meaningful BSM test, $a_\tau$ needs to be probed at the level of $10^{-6}$, a conclusion that derives from scaling the possible BSM contribution to $a_\mu$ with the square of the lepton masses as expected in Minimal Flavor Violation, from the size of the electroweak contribution~\cite{Eidelman:2007sb}, and from the study of concrete BSM scenarios~\cite{Crivellin:2021spu}.

Reaching such a level of precision is challenging, but could potentially be achieved at a SuperKEKB upgrade with polarized electrons in a precision study of $e^+e^-\to\tau^+\tau^-$ at or around the $\Upsilon$ resonances. Such measurements at $s\simeq (10\,\text{GeV})^2$ allow one to extract the Pauli form factor $F_2$ at the same energy, in such a way that the comparison to the SM prediction reveals a potential BSM contribution. If the associated BSM scale lies beyond the electroweak scale, a mismatch in $\text{Re}\,(F_2)$ can be directly interpreted as $a_\tau^\text{BSM}$, while bounds for light BSM degrees of freedom become model dependent.

In order to measure $\text{Re}\,(F^\text{eff}_2)$, the effective form factor that can be extracted directly in terms of observable asymmetries,  with this procedure requiring the selection of $\tau$-pair events in which both $\tau^+$ and $\tau^-$ decay semileptonically: $e^+e^- \rightarrow \tau^+\tau^- \rightarrow h^+{\bar \nu_{\tau}} h^-\nu_{\tau}$, which enables the reconstruction of the production plane and direction of flight, as described in Ref.~\cite{Kuhn:1993ra}.  Polarization is needed because, without it, one can only extract $F_2$  from the angular dependence of the cross section, which would require controlling the normalization at the $10^{-6}$ level, or by using the ``Normal Asymmetry'' as defined in Refs.~\cite{Bernabeu:2007rr,Bernabeu:2008ii}, which  is only sensitive to $\text{Im}\,(F_2)$~\cite{Bernabeu:2007rr,Bernabeu:2008ii}. Such approaches will be limited by systematic uncertainties associated with modeling the detector asymmetries that do not cleanly cancel. 

In contrast, with a polarized beam,  asymmetries between data taken with a left-polarized and right-polarized beam benefit from cancellations of systematic uncertainties associated with the detector asymmetries, since it is the beam that is changing polarization state under identical detector responses.
Two left-right beam polarization  asymmetries are used, a  transverse ($A_T$) and a longitudinal $(A_L)$ asymmetry, as  suggested in Ref.~\cite{Bernabeu:2007rr}, and described below. A particular linear combination of $A_T$ and $A_L$ cancels large contributions from $F_1$ and is proportional to the effective $\text{Re}\,(F_2)$:
\begin{equation}
{\rm Re} (F_2^\text{eff})=\mp \frac{8(3-\beta^2)}{3\pi\gamma\beta^2\alpha_\pm}\Big(A_T^{\pm} -\frac{\pi}{2\gamma}A_L^{\pm}\Big),
\label{RF2eff}
\end{equation}
where the $\pm$ refers to the charge on the $\tau$ being considered, $\alpha_{\pm} \equiv  
 (m^2_{\tau}-2m^2_{h^{\pm}})/(m^2_{\tau}+2m^2_{h^{\pm}})$ is the polarization analyzer~\cite{Bernabeu:1993er}, and $\gamma = 1/\sqrt{1-\beta^2} =  E_{\tau}/m_{\tau}$. 

\begin{figure}[t!] 
\centering
\includegraphics[width = 4 in]{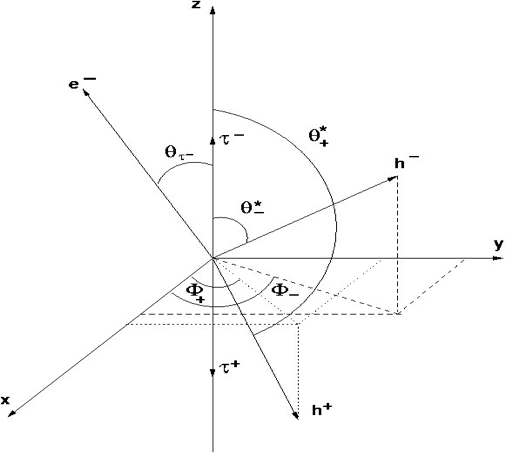}
\caption{
Coordinate system for e$^+$e$^-\rightarrow \tau^+\tau^-$; $\tau^+\rightarrow h^+{\bar \nu_{\tau}}$ and $\tau^- \rightarrow h^- \nu_{\tau}$ events used in $\tau$ $g-2$ and EDM measurements~\cite{Bernabeu:2006wf}. Here the $z$-axis is aligned with $\tau^-$ momentum, 
 $\theta_{\tau-}$ is the production angle of the $\tau$ with respect to the beam electron direction in the center-of-mass, and the azimuthal and polar 
angles   of the
produced hadrons, $h^{\pm}$, in $\tau^{\pm}$ rest frame, are
 $\phi_{\pm}$ and $\theta^*_{\pm}$, respectively.
The tau production plane and direction of flight are fully reconstructed using the technique described in Ref.~\cite{Kuhn:1993ra}.
}
\label{fig:taug-2-coordinates}
\end{figure}

Using the coordinate systems defined in Fig.~\ref{fig:taug-2-coordinates}, the transverse asymmetry for  the $\tau^+$ (and separately for the $\tau^-$) is measured by counting events with $\pi/2 < \phi_{\pm} < 3\pi/2$ when the beam is right-polarized ($R_e$) and also when the beam is left-polarized ($L_e$) and, taking their difference, then doing the same 
for events with $-\pi/2 < \phi_{\pm} < \pi/2$. 
 Subtracting the former from the latter gives $A_T^{\pm}$: 
 \begin{equation}
 A_T^{\pm} =\frac{1}{2\sigma} \left[ \int^{\pi/2}_{-\pi/2}  \left( \left(\frac{d\sigma^{R_e}}{d\phi_{\pm}}\right)-
\left(\frac{d\sigma^{L_e}}{d\phi_{\pm}}\right) \right) d\phi_{\pm}
- 
\int^{3\pi/2}_{\pi/2}\left( \left(\frac{d\sigma^{R_e}}{d\phi_{\pm}}\right)-
\left(\frac{d\sigma^{L_e}}{d\phi_{\pm}}\right) \right) d\phi_{\pm}
\right].
 \end{equation}
 
The longitudinal asymmetry measurement involves
 the $R_e$-$L_e$ asymmetries as well, along with the asymmetries associated with two angular observables ($z = \cos\theta_{\tau-}$ and $z^*_{\pm}=\cos\theta^*_{\pm}$), 
  after integrating over all other angles.
Defining 
\begin{equation}
A_{RL} =  \frac{d^2\sigma^{R_e}}{dz^*_{\pm}dz} - \frac{d^2\sigma^{L_e}}{dz^*_{\pm}dz},   
\end{equation}
the  longitudinal asymmetry is: 
  \begin{align}
 A_L^{\pm} = & \frac{1}{2\sigma} \left[ 
 \int^{1}_{0} dz^*_{\pm} \left( 
 \int^{1}_{0} dz \left( A_{RL} \right) - 
 \int^{0}_{-1} dz \left(  A_{RL}  \right) \right) 
 -\int^{0}_{-1} dz^*_{\pm} \left( 
 \int^{1}_{0} dz \left(  A_{RL} \right) - 
 \int^{0}_{-1} dz \left( A_{RL}  \right) \right) \right].
 \end{align}
In order to measure $A_L^{\pm}$, for each beam polarization state ($R_e$~or~$L_e$), the experiment counts the events in four bins separately for the identified $\tau^{+}$ and $\tau^{-}$: 
 \begin{itemize}
     \item $N_{\pm}^{FF}$($R_e$~or~$L_e$): number of $\tau^{\pm}$ decays with  $0<\cos\theta_{\tau-}<1$ and $h^{\pm}$ in $0<\cos\theta^*_{\pm}<1$;
     \item  $N_{\pm}^{FB}$($R_e$~or~$L_e$): number of $\tau^{\pm}$ decays with  $0<\cos\theta_{\tau-}<1$ and $h^{\pm}$ in $-1<\cos\theta^*_{\pm}<0$;
     \item  $N_{\pm}^{BF}$($R_e$~or~$L_e$): number of $\tau^{\pm}$ decays with $-1<\cos\theta_{\tau-}<0$ and $h^{\pm}$ in $0<\cos\theta^*_{\pm}<1$;
     \item  $N_{\pm}^{BB}$($R_e$~or~$L_e$): number of $\tau^{\pm}$ decays with $-1<\cos\theta_{\tau-}<0$ and $h^{\pm}$ in $-1<\cos\theta^*_{\pm}<0$.
 \end{itemize}
 Defining the $R_e$ and $L_e$ asymmetries as:
 \begin{align}
 A^{\pm}_{FF} &= \frac{N_{\pm}^{FF}(R_e) - N_{\pm}^{FF}(L_e)}
 {N_{\pm}^{FF}(R_e) + N_{\pm}^{FF}(L_e)}, 
 & A^{\pm}_{BF} &= \frac{N_{\pm}^{BF}(R_e) - N_{\pm}^{BF}(L_e)}
                  {N_{\pm}^{BF}(R_e) + N_{\pm}^{BF}(L_e)}, \notag\\
 A^{\pm}_{FB} &= \frac{N_{\pm}^{FB}(R_e) - N_{\pm}^{FB}(L_e)}
                 {N_{\pm}^{FB}(R_e) + N_{\pm}^{FB}(L_e)},
& A^{\pm}_{BB} &= \frac{N_{\pm}^{BB}(R_e) - N_{\pm}^{BB}(L_e)} 
                       {N_{\pm}^{BB}(R_e) + N_{\pm}^{BB}(L_e)},
                       \label{asymmetries}
\end{align} 
the longitudinal asymmetry for the $\tau^{\pm}$ is then: 
\begin{equation}
A^{\pm}_L = \frac{1}{2} \left[ (A^{\pm}_{FF} - A^{\pm}_{BF}) - (A^{\pm}_{FB} - A^{\pm}_{BB}) \right].
\end{equation}
Note that the unpolarized cross section, after integration over $\phi_{\pm}$, no longer depends on $\theta_\pm^*$ and is further symmetric in $z$, in such a way that in Eq.~\eqref{asymmetries} each term can indeed be normalized to the sum of events in each bin separately. The same holds true for $A_T^\pm$, as the unpolarized cross section does not distinguish between the two regions in $\phi_\pm$ either.   


In the SM at 10~GeV, Re($F_2^\text{eff}$)=$-268.77(50)\times 10^{-6}$~\cite{Crivellin:2021spu}. Consequently, the factor $\frac{8(3-\beta^2)}{3\pi\gamma\beta^2\alpha_\pm}$ in Eq.~\ref{RF2eff} only needs to be  controlled at the 0.5\% level in order to achieve a ppm measurement. However, the cancellation of the large contribution from $F_1$,  which motivates the subtraction $(A_T^{\pm} -\frac{\pi}{2\gamma}A_L^{\pm})$,
 requires the $\pi/(2\gamma)$ factor to be very well controlled. Since $\gamma = E_{\tau}/m_{\tau} = E_\text{cm}/(2m_{\tau})$, its uncertainty is determined by the uncertainties on the $\tau$ mass and mass of the $\Upsilon(1S)$, which is used to calibrate the center-of-mass energy in the machine. Current uncertainties on $m_{\tau}$ and  $M_{\Upsilon(1S)}$~\cite{Zyla:2020zbs} limit this precision to $1\times 10^{-5}$. This contribution to the overall uncertainty on the precision of 
 Re($F_2^\text{eff}$) will scale in proportion with the 
 uncertainties of the measurements on $m_{\tau}$ and  $M_{\Upsilon(1S)}$, and therefore motivates improved measurements of those quantities.  
 
With 40~ab$^{-1}$ of $e^+e^- \rightarrow \tau^+\tau^-$ data, requiring both the $\tau^+$ and $\tau^-$ to decay semileptonically and assuming a 60\% selection efficiency, the statistical error on 
Re($F_2^\text{eff}$) would be $1\times 10^{-5}$.
As both $A_T$ and $A_L$ involve differences in the polarization state of the beam,  the dominant detector systematic uncertainties cancel and one is left dealing with the detector related systematic   uncertainties in the residual differences. These can be accurately controlled in  analyses that do  not use the beam polarization information. 
Consequently, the measurements of $A_T$ and $A_L$ at a polarization-upgraded SuperKEKB   constitutes a promising
way to precisely measure $a_\tau$.
The path towards eventually constraining $a_\tau^\text{BSM}$ at the $10^{-6}$ level will require more statistics as well as higher precision measurements of $m_{\tau}$ and  $M_{\Upsilon(1S)}$. 

Key requirements for such a program are studied in Ref.~\cite{Crivellin:2021spu}: first, a precision of $10^{-6}$ necessitates the consideration of two-loop effects, as provided therein for the resonance-enhanced case of a measurement on the $\Upsilon(nS)$, $n=1,2,3$. However,  the typical spread in beam energies results in the continuum $\tau^+\tau^-$ pairs to dominate over resonant ones, and such a measurement would require significant dedicated runs on the lower $\Upsilon$ resonances to gather enough statistics. On the other hand, the broader physics program calls for most data to be collected at the center-of-mass energy of $\Upsilon(4S)$ mass, where $e^+e^- \rightarrow \tau^+ \tau^-$ events are produced non-resonantly, since the $\Upsilon(4S)$ has a negligible leptonic branching fraction. 
Given that this  non-resonant data would allow for a significant increase in statistics, and translate to improved limits on $a_\tau^\text{BSM}$, this strongly motivates  the full two-loop calculation in the coming years.

%% file: TauEDM.tex
\def\CP                {\ensuremath{C\!P}\xspace}
\def\fb   {\ensuremath{\mbox{\,fb}}\xspace}
\def\invfb   {\ensuremath{\mbox{\,fb}^{-1}}\xspace}
\def\invab   {\ensuremath{\mbox{\,ab}^{-1}}\xspace}

The electric dipole moment (EDM, $d_\tau$) of the $\tau$ lepton characterizes the time-reversal or charge-parity (\CP) violation properties at the $\gamma\tau\tau$ vertex. The SM predicts an extremely small value, $d_\tau \approx 10^{-37}$ $e$cm~\cite{Booth:1993af,Mahanta:1996er}, many orders of magnitude below any experimental sensitivity. Independent measurements of the electric dipole moments of $e$, $\mu$ and $\tau$ are necessary to determine the flavor dependence of \CP violating phases in the possible mixing between three generation of the charged lepton sector. In general, the strength of \CP violation may be different for different flavors and Belle II is uniquely suited to test a large class of new physics models which predict enhanced contributions in EDM of the $\tau$ lepton at observable levels of $10^{-19}$ $e$cm ~\cite{Bernreuther:1996dr,Huang:1996jr}.

Experimental studies of the $\tau$ EDM are very clean tools of discovery of new physics, because they rely on measurement of asymmetries with relatively small systematic uncertainties in the measurement. 
Current best results are from a recent Belle study~\cite{Belle:2021ybo}, where the squared spin-density matrix of the $\tau$ production vertex is extended to include contributions proportional to the real and imaginary parts of the $\tau$ EDM. The expectation values of the optimal observable were measured  yielding
Re$(d_\tau) = (-6.2 \pm 6.3) \times 10^{-18}~{e}\rm{cm}$ 
and 
Im$(d_\tau) = (-4.0 \pm 3.2) \times 10^{-18}~{e}\rm{cm}$. The results are obtained using 833~\invfb{} of data, with the dominant systematic uncertainty associated to mismatch of data and the simulated distributions of momentum and angular variables. The sensitivity studies on EDM at Belle II has shown that the agreement between data and Monte Carlo can be improved significantly. Conservative assumption on benchmark scenarios of  systematic uncertainties at Belle II show that we expect to probe $\tau$ EDM at the $10^{-19}$ level with the 50~\invab data set. 

The $\tau$ EDM not only influences the angular distributions, but also the polarization of the $\tau$ produced in electron-positron annihilation. The beam polarization substantially improves the experimental sensitivity for $\tau$ EDM by allowing measurements of the polarization of a single $\tau$, rather than measurements of correlations between two $\tau$ leptons produced in the same event~\cite{Gonzalez-Sprinberg:2007owj,Bernabeu:2006wf}. 
The proposed beam polarization upgrade at SuperKEKB will further experimental sensitivity, since the uncertainties in modeling the forward-background asymmetry in the detector response are independent of beam polarization and will largely cancel. Such an increase in experimental sensitivity will allow to unambiguously discriminate between the new physics contributions to the $\tau$ EDM at the level of $10^{-20}~e\rm{cm}$, which is two orders of magnitude below any other existing bounds~\cite{Ananthanarayan:1994af,Bernabeu:2006wf, Bernreuther:2021elu}.

%% file: TauLFV.tex
\begin{figure}[!htb]
  \begin{center}
   \includegraphics[width=.6\textwidth]{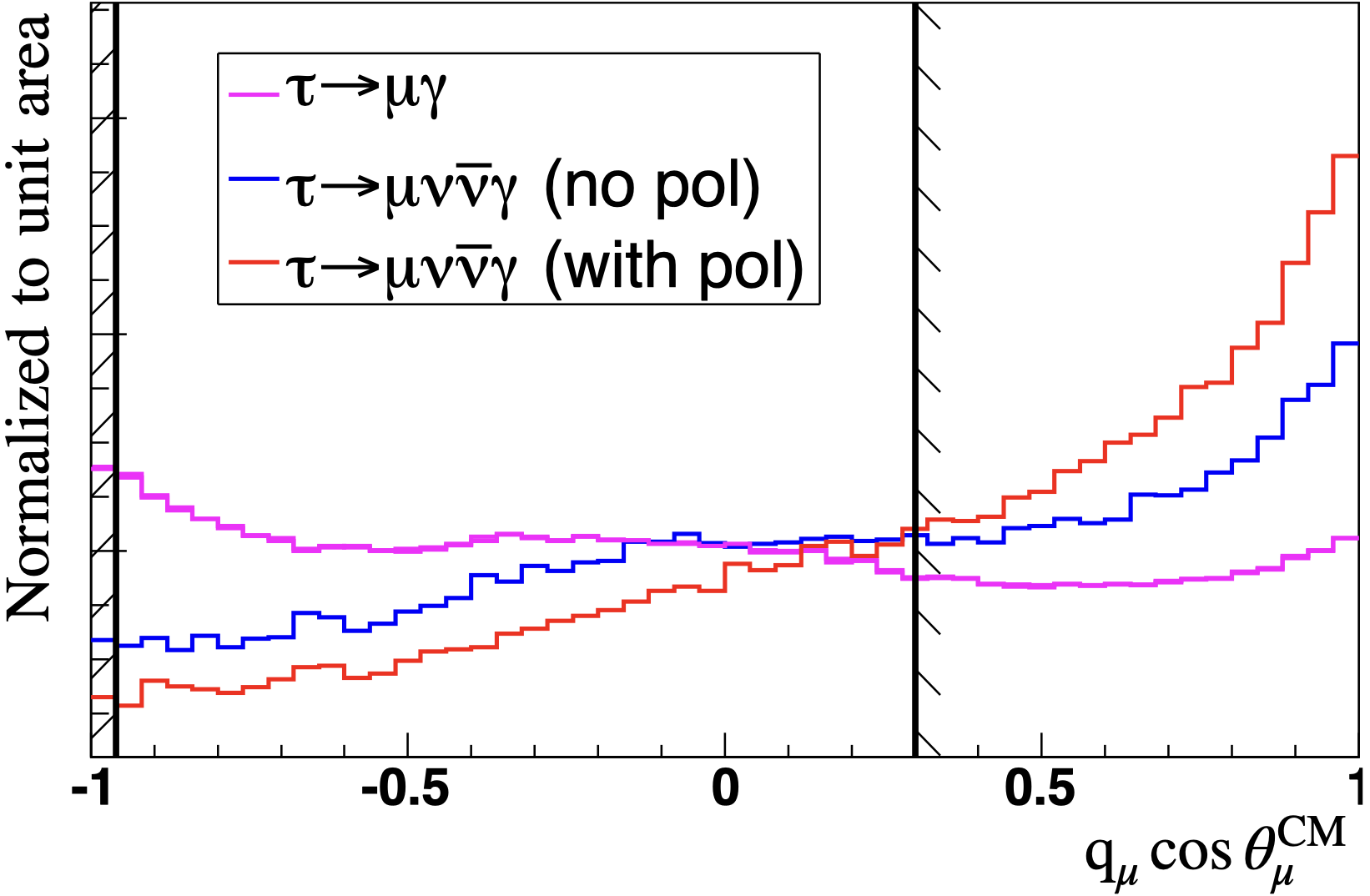}
  \end{center}
  \vspace{1ex}
  \caption{\label{fig:tau-lfv-exp}%
    Distribution of the cosine of the angle between the signal-side muon momentum and e$^-$ beam momentum in the center-of-mass frame,  multiplied by the
    muon charge for signal and background events with and without electron
    beam polarization in the $\tau\to\mu\nu\bar\nu\gamma$ search analysis.}
\end{figure}

Many models predict lepton flavor violation (LFV) in $\tau$ decays at $10^{-10}$--$10^{-8}$ levels, which will be probed by 
the huge data sample of $10^{11}$ single $\tau^-$ decays at Belle II.
Upper limits will improve current bounds by  an order of magnitude in the next decade, probing LFV in $\tau$ decays down to few parts in $10^{-10}-10^{-9}$~\cite{BelleIIPhysBook}.
Substantial gains are possible by re-optimizing the analysis for the
Belle II detector, exploiting beam polarization effects. The high
energy electrons beam at SuperKEKB are expected to be ${\sim}70\%$ longitudinally polarized, influencing the angular distribution of the $\tau$ decay products in a way that depends on the interaction that causes LFV.  With beam polarization, the helicity angles of the $\tau$ pair decay products can be used to significantly suppress the background when one $\tau$ decays to $\mu\gamma$ and the other one to $\pi\nu$, which is the decay channel most sensitive to the polarization of the $\tau$ lepton. Similar background suppression can also be obtained with the other decay modes, which vary in their sensitivity to the $\tau$ polarization.
In general, the polar angles in the center-of-mass frame times the charge of the $\tau$ decay  provide maximal background suppression.
The “irreducible background” from $\tau \to \mu \nu \bar{\nu} \gamma$
decays are studied in  Figure~\ref{fig:tau-lfv-exp}~\cite{Hitlin:2008gf}.
While the distributions of the backgrounds differ significantly with and without beam polarization, the signal distribution modelled by phase-space does not change. By varying the selected set of events based on such a distribution, the background can be reduced significantly, 
corresponding to a small loss in signal efficiency. 
An optimization study has shown that this would result in approximately a 10\% improvement in the sensitivity.
Similar analyses can be expected to yield comparable gain sensitivities for the $\tau\to e \gamma$ LFV decay mode, 
based on the published \babar\ analysis~\cite{BaBar:2009hkt}.
However, it should be noted that the phase space model of the signal 
is chosen due to the lack of current knowledge on the underlying theory behind LFV decays. By far, the most important aspect
of having the polarization is the possibility to determine the helicity
structure of the LFV coupling from the final state momenta
distributions, for example in $\tau\to\mu\mu\mu$ decays~\cite{Matsuzaki:2007hh, Dassinger:2007ru}.

%% file: QCD.tex
\newcommand{\id}{{\mathbb{I}}}
\newcommand{\vect}[1]{\vec{#1}} 



\begin{figure}[tb]
	\centering
	\includegraphics[width=0.35\linewidth]{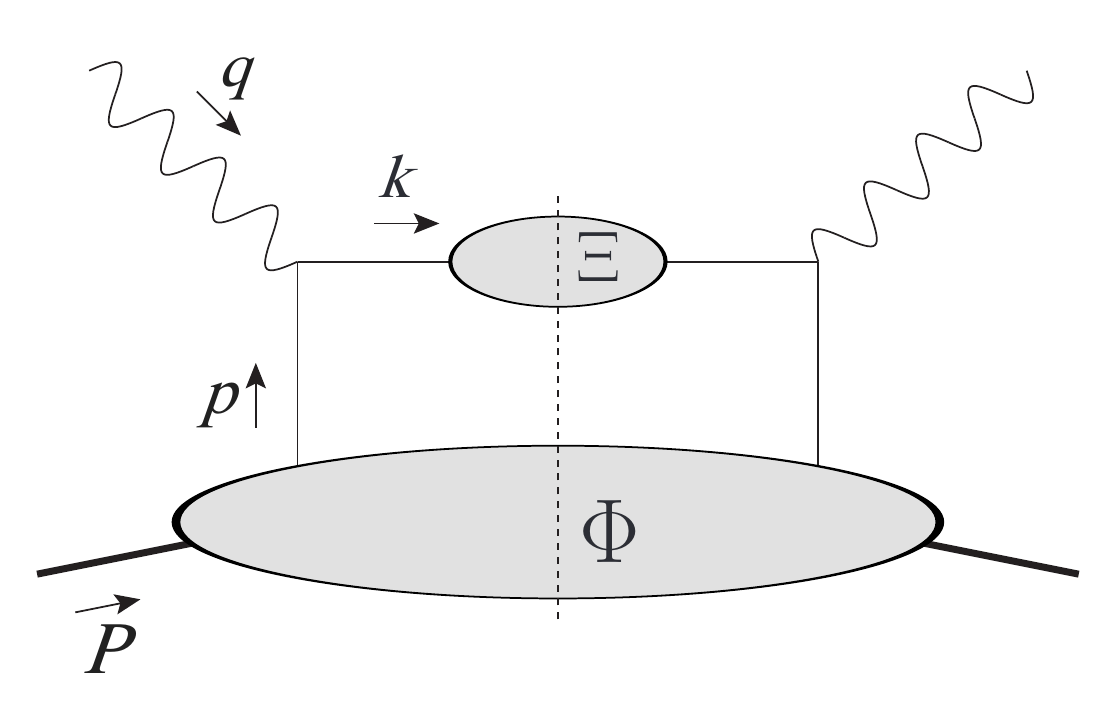}
	\quad
	\includegraphics[width=0.45\linewidth]{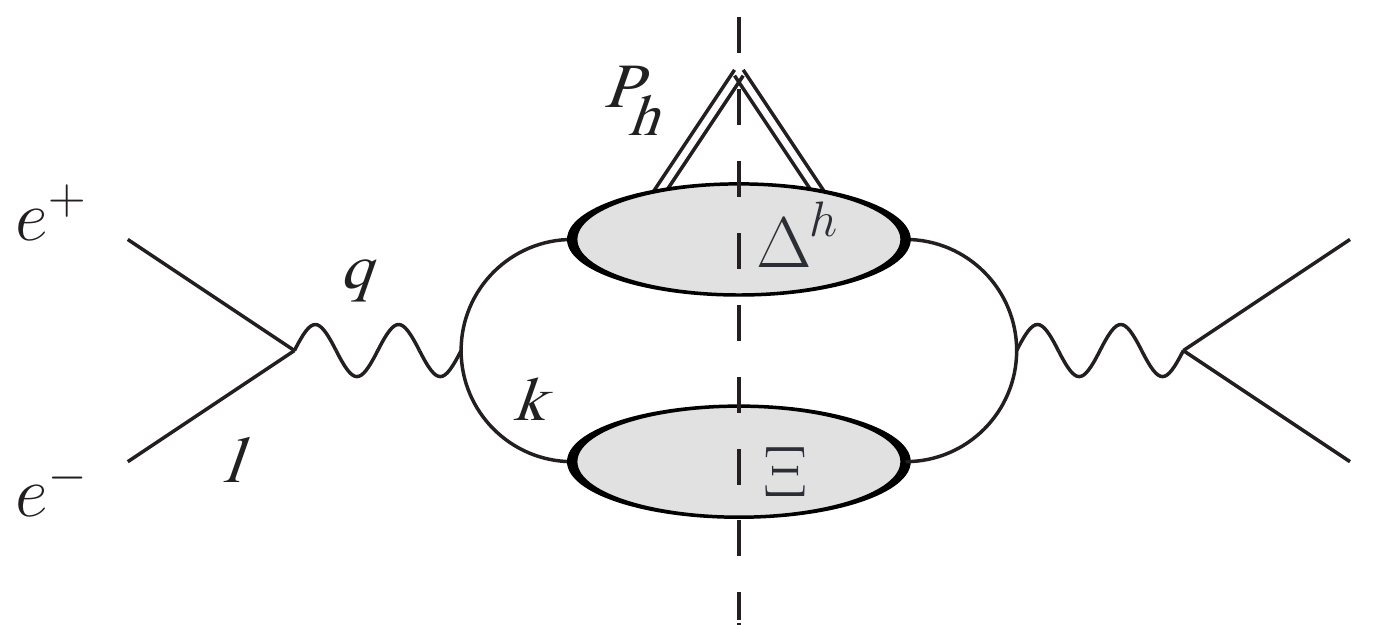}
	\caption{Inclusive DIS (left) and $\Lambda$ production from SIA (right) diagrams with an inclusive jet correlator $\Xi$ replacing unobserved  perturbative quarks in the final state. 
	$\Phi$ and $\Delta^h$ are the correlation functions that encode information on hadron structure and hadronization, respectively. 
	The jet correlator $J = \frac12 \int dk^+\,\Xi(k)$ is defined in Eq.~\eqref{eq:ijet_correlator} and further discussed in the main text. In this document, the detected hadron $h$ is a self-polarizing $\Lambda$ baryon. }
	\label{fig:jetmass_processes}
\end{figure}

Due to color confinement, the quarks created in a hard collision cannot appear as on shell particles in the final state, but rather decay into a jet of hadrons whose mass is dynamically generated, but the details of the  quark-to-hadron transition are still unknown. As proposed in \cite{Accardi:2019luo,Accardi:2020iqn} dynamical mass generation can be studied even without observing the produced hadrons, but instead studying the Dirac decomposition of the (color averaged) \textit{gauge-invariant} quark correlator 
\begin{align}
    J_{ij}(k^-,\vect{k}_T)
    & \equiv \frac{\text{Tr}_c}{2N_c} \int dk^+\, \text{Disc} \int \frac{d^4 \xi}{(2\pi)^4} e^{i k \cdot \xi} \,
    \langle\Omega| \psi_i(\xi) \bar \psi_j(0) W(0,\xi;n_+) |\Omega\rangle
    \nonumber \\
    & = \frac{\theta(k^-)}{4(2\pi)^3\, k^-} \, 
    \bigg\{ k^-\, \gamma^+ + \slashed{k}_T + M_j \id + \frac{K_j^2 + \vect{k}_T^2}{2k^-} \gamma^- \bigg\} \ .
\label{eq:ijet_correlator}
\end{align}  
where $|\Omega\rangle$ is the nonperturbative QCD vacuum, $\psi_i$ is the quark field, $W$ a Wilson line. This correlator describes the nonperturbative propagation and hadronization of a quark \cite{Accardi:2019luo,Accardi:2020iqn}, and generalizes the perturbative quark propagator contributing to particle production in lepton-nucleus deep inelastic scattering (DIS) at large Bjorken $x$ values \cite{Accardi:2008ne,Accardi:2017pmi} as well as in the semi-inclusive annihilation (SIA) of electrons and positrons, see Figure~\ref{fig:jetmass_processes}.

\begin{figure}[tbh]
	\centering
	\includegraphics[width=0.32\linewidth]{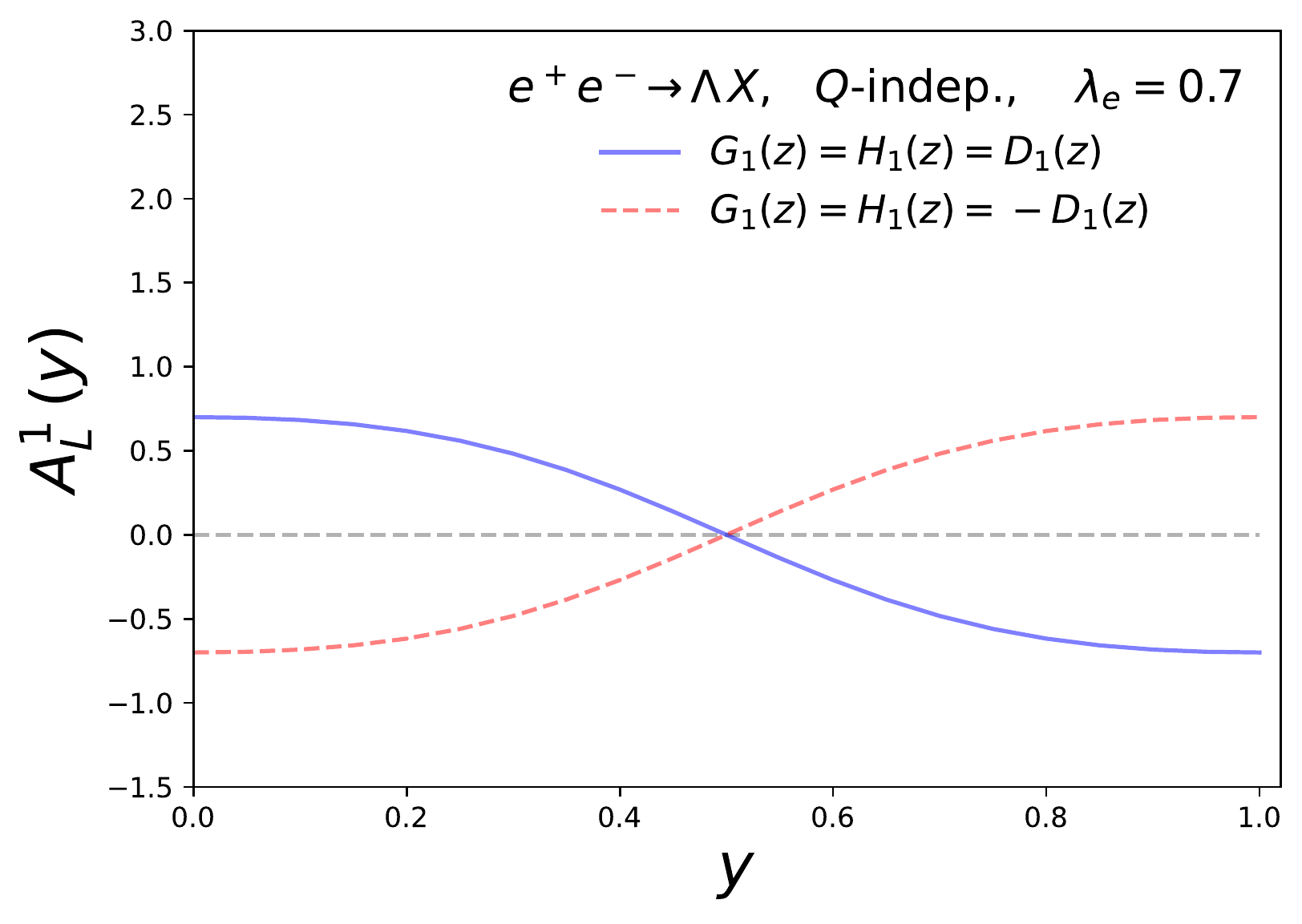}
	\includegraphics[width=0.327\linewidth]{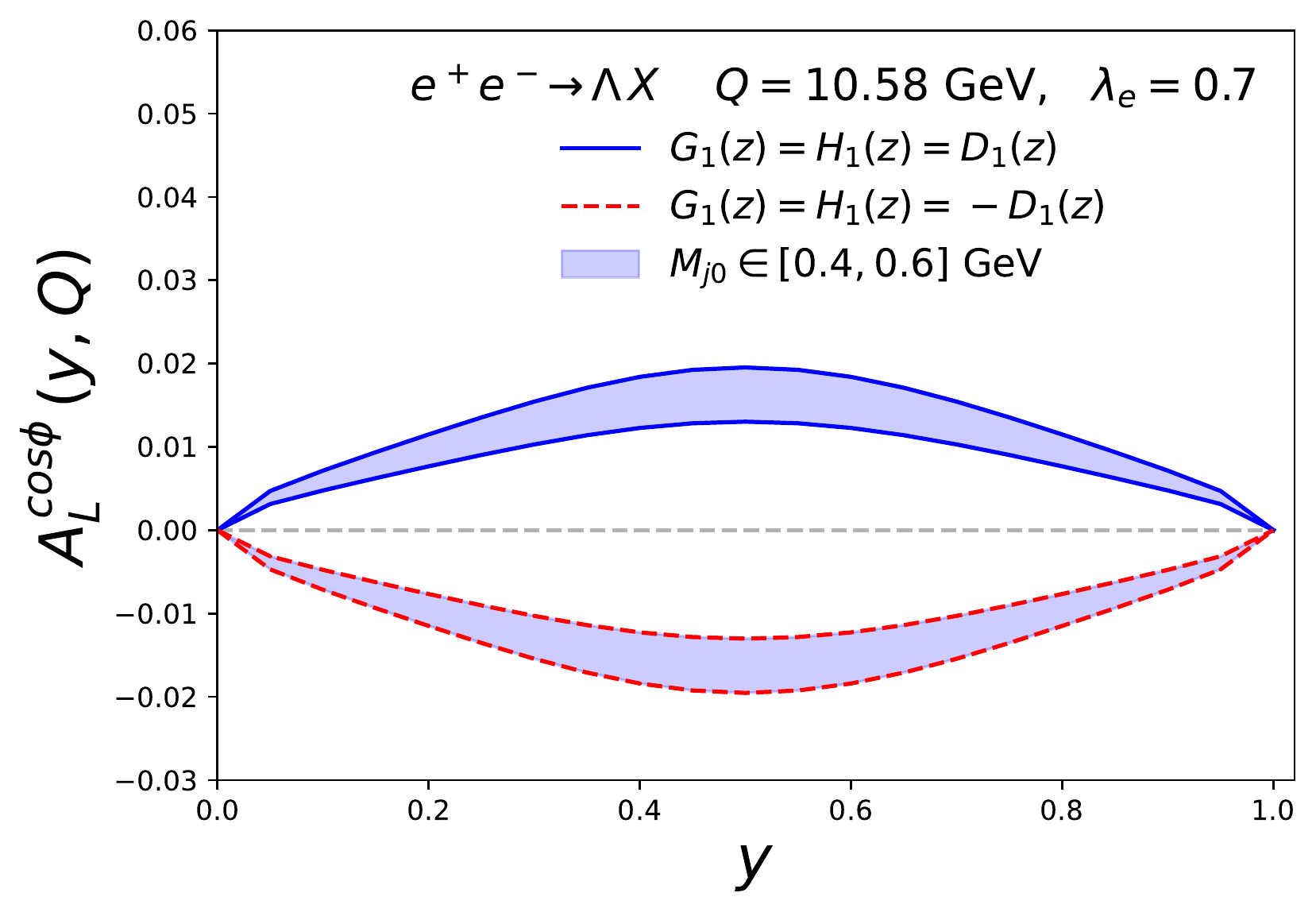}
	\includegraphics[width=0.32\linewidth]{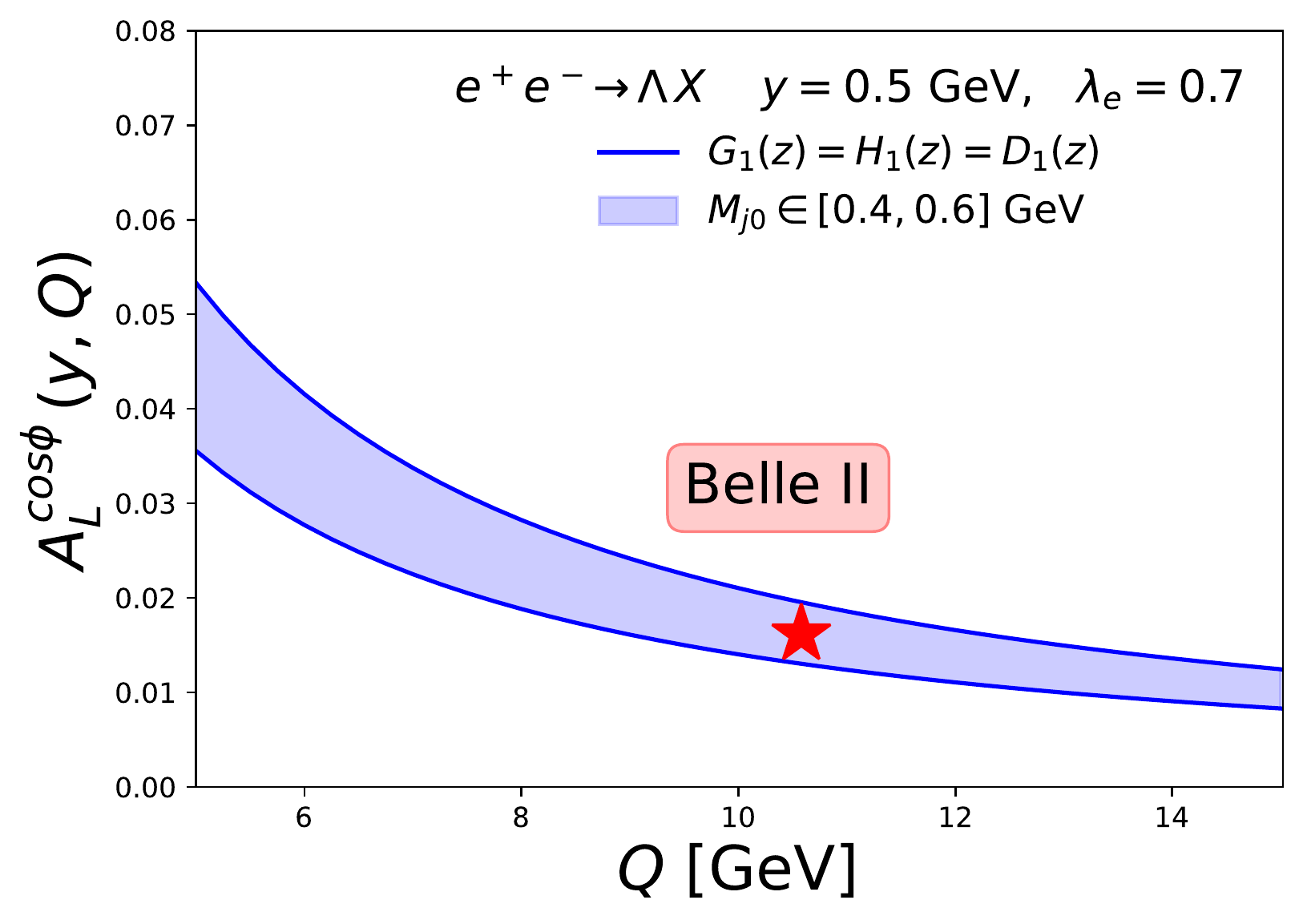}
	\caption{The Fourier components $A_L^1(y)$ and $A_L^{\cos\phi}(y,Q)$ of the longitudinal electron spin asymmetry as a function of $y$ at the SuperKEKB nominal energy $Q=10.58$ GeV. The band in the $\cos\phi$ modulation indicates the sensitivity of the measurement to $\pm20\%$ variation in the jet mass at the initial scale. The rightmost panel shows the $A_L^{\cos\phi}$ modulation as a function of $Q$ at fixed $y=0.5$, along with its $20\%$ sensitivity to $M_j$, which also slightly increases at lower energies due to QCD evolution.}
	\label{fig:AL_AT_Lambda}
\end{figure}

In particular, In SIA production of a self-polarizing $\Lambda$ hadron the chiral-odd jet mass $M_j$ couples to the $H_1^\Lambda$ transversity fragmentation function and contributes to the longitudinal beam spin asymmetry of the process, 
\begin{align}
\label{e:asymmetry}
    A_L & = \frac{d\sigma^{R_e} - d\sigma^{L_e}}{d\sigma^{R_e} + d\sigma^{L_e}} \ ,
\end{align}
where $R_e$ and $L_e$ refer to the handedness of the electron. 
The $\Lambda$'s longitudinal and transverse spin contributions to the asymmetry can be separated studying the $y = P_\Lambda \cdot l / P_\Lambda \cdot q$ dependence of the asymmetry, where $l$, $q$, and $P_\Lambda$ are the four-momenta of the incoming electron, the exchanged photon, and the $\Lambda$ baryon respectively~\cite{Boer:1997mf,Boer:2008fr}.
Assuming the saturation of the positivity bounds and Wandzura-Wilczek approximation for
the polarized leading twist fragmentation functions $G_1^{\Lambda}(z)$, $H_1^{\Lambda}(z)$, and the higher twist $G_T^{\Lambda}(z)$ fragmentation function~\cite{Mulders:1995dh} one obtains:
\begin{align}
\label{e:asymmetry_simpl}
    A_L(y,Q) 
    & = 
    \pm \underbrace{ \Big( \lambda_e\, \frac{C(y)}{2A(y)} \Big)}_{A_L^1(y)}  
    \lambda_\Lambda\, 
    \pm
    \underbrace{ \Big( 2\lambda_e\, \frac{M_j(Q)}{Q} \frac{D(y)}{A(y)}   \Big)}_{A_L^{\cos\phi}(y,Q)} |{S_T}_\Lambda| \cos(\phi) \ ,
\end{align}
where $\lambda_e$ and $\lambda_\Lambda$ are the electron's and $\Lambda$ hadron's helicities, respectively, ${S_T}_\Lambda$ is the transverse spin vector of the detected $\Lambda$ hadron (see the
``Opportunities for precision QCD physics in hadronization at Belle II'' Snowmass 2022 contribution for details~\cite{Accardi:2022oog}). 
The $\pm$ signs in Eq.~\eqref{e:asymmetry_simpl} refer to the case where the polarized leading twist fragmentation functions $G_1^{\Lambda}(z)$, $H_1^{\Lambda}(z)$ saturate the respective positivity bounds with a plus or a minus sign~\cite{Mulders:1995dh}. 
The configuration with the plus (minus) sign corresponds to the solid blue (dashed red) curves in Fig.~\ref{fig:AL_AT_Lambda}.

The jet mass $M_j$ can then be extracted from the Fourier coefficient $A_L^{\cos\phi}$.
With the expected 70\% beam polarization at the polarized SuperKEKB upgrade, this is found to be of ${\cal O}(1\%)$, reaching a maximum at $y=0.5$. 
At the same value of $y$ the constant modulation $A_L^1$ displays a node. This specific value allows one to separate the two modulations $A_L^1$ and $A_L^{\cos\phi}$, related to the longitudinal and transverse polarization of the detected $\Lambda$ hadron respectively. 
The blue band in Fig.~\ref{fig:AL_AT_Lambda} displays the sensitivity of this observable to a 20\% variation in the jet mass at the non-perturbative scale, $M_{j0} = 0.4-0.6$ GeV. 

In summary, the $A_L^{\cos\phi}$ modulation of the beam spin asymmetry $A_L$ in Eqs.~\eqref{e:asymmetry},~\eqref{e:asymmetry_simpl} for production of a $\Lambda$ hadron in polarized $e^+e^-$ annihilation provides access to the dynamical component of the jet mass, allowing one to experimentally measure the contribution of the non-perturbative QCD dynamics at play in the hadronization mechanism. 
If the positivity bounds for the polarized fragmentation functions turn out not to be saturated, the signal may drop below the ${\cal O}(1\%)$ estimated in Fig.~\ref{fig:AL_AT_Lambda}. 
However, being a twist three effect suppressed as $\sim 1/Q$, the signal and its sensitivity to $M_j$ can increase significantly at lower center of mass energies, as displayed in the right-most panel of Fig.~\ref{fig:AL_AT_Lambda}. Similar measurements with di-hadron production instead of a self-polarizing $\Lambda$ baryon are also under consideration.

%% file: PolarizedSource.tex

Development of the polarized source aims at creating and delivering a high-quality beam to the Main Ring via the injector linac.  A highly polarized beam is desired, with emission parameters at injection in line with the SuperKEKB's HER values  (see Table~\ref{table:beam_params}).

\begin{table}[hbt!]
\centering
\begin{tabular}{|c|c|}
\hline 
\multicolumn{2}{|c|}{SuperKEKB HER Beam Parameters} \\
\hline
Energy [GeV] & 7 \\ 
\hline 
Bunch charge [nC] & 4 \\ 
\hline 
Normalized Emittance [mm $\cdot$ mrad] & 40/20 (Hor./Vert.) \\ \hline 
\end{tabular} 
\caption{SuperKEKB Beam Parameters.}

\label{table:beam_params}
\end{table}

\subsection{Beam Generation}

We consider electron beam generation using a GaAs cathode activated with a circularly-polarized laser. GaAs has been shown since the 1970's to produce highly polarized electron beams\cite{Pierce_1975}.  Strain super-lattice GaAs crystals have been demonstrated to show $\ge$ 90\% polarization with a quantum efficiency of 1.6\%~\cite{Jin_2014}. However, the 1.43\,eV gap between the valence and conduction bands in GaAs presents an obstacle to efficient acceleration of electrons; application of a thin Negative Electron Affinity (NEA) surface may be used to permit electrons from the conduction band minimum to escape into the vacuum and thence to be accelerated through the beamline.

\begin{figure}[hbt!]
\begin{center}
\includegraphics[scale=0.5]{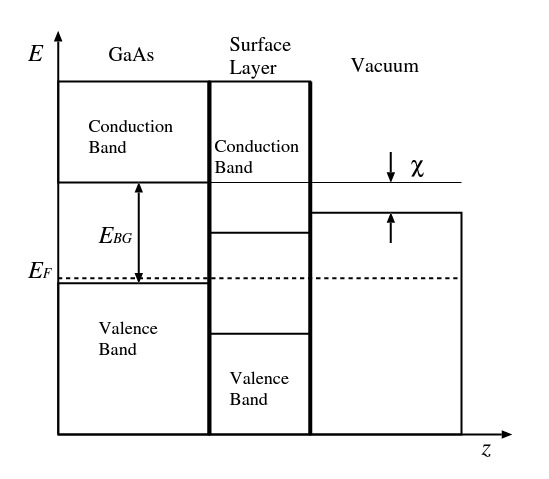}
\caption{Representation of a GaAs cathode with a thin-film NEA layer.  }
\end{center}
\end{figure}

NEA films are susceptible to several modes of degradation, including residual gas adsorption, ion back bombardment, and thermal desorption, and consequently have relatively short lifetimes. As a practical matter, however, this necessitates using a DC electron gun as opposed to the relatively harsh environment of an RF gun. Because the charge density of DC guns is relatively low, a buncher must also be considered to achieve the desired bunch charge. 

We are currently investigating increasing the lifetime of NEA cathodes by investigating novel materials in the thin film surface as well as improving the application procedure to produce more robust cathodes.  Results from the test bench at Hiroshima University have shown that lifetimes for CsKTe-activated cathodes, for example, have a lifetime approximately one order of magnitude greater than that of previous-generation CsO-activated cathodes, as shown in Table~\ref{table:cathode_lifetimes}.
\begin{table}[hbt!]
        \centering
\begin{tabular}{|c|c|}
\hline
    Cathode & Lifetime [10$^{-3}$\,Pa $\cdot$ s] \\ \hline
    CsO/GaAs & 0.29 $\pm$ 0.03 \\
    CsO/GaAs & 0.40 $\pm$ 0.02 \\ 
    CsKTe/GaAs & 6.50 $\pm$ 0.01 \\  \hline
        \end{tabular}
        \caption{Measured lifetimes of thin-film cathodes. Data taken from~\cite{Kuriki-JoP}.}
\label{table:cathode_lifetimes}
    \end{table}


\subsubsection{Cathode Production and Testing}

Cathode production is carried out at the Hiroshima University test bench. GaAs wafers are prepared by washing with an oxygenated ammonium bath (Semicoclean 23 solution) for five minutes, followed by rinsing with distilled water and ethanol and flushing any remaining impurities from the surface with He gas. The wafer is then mounted to a stage and inserted into the vacuum chamber.

The vacuum chamber is baked at 200\,$^\circ$C with a vacuum pump connected to flush outgassed hydrogen. At the level of $10^{-6}$\,Pa, an angle valve is closed and an ion pump is activated to increase the vacuum to approximately $10^{-9}$\,Pa and the cathode is heated again to remove any remaining surface contaminants.

Following baking, NEA surface materials are heated and dispersed evenly into the vacuum chamber and their deposition is monitored at the \r{A}ngstrom level. A 50\,\r{A} layer of Sb is deposited onto the face of the GaAs wafer directly, after which layers of other materials to be tested are deposted in 5\,\r{A} segments. 


\subsection{Linac Transport}

\begin{figure}[hbt!]
    \centering
    \includegraphics[width=.4\textwidth]{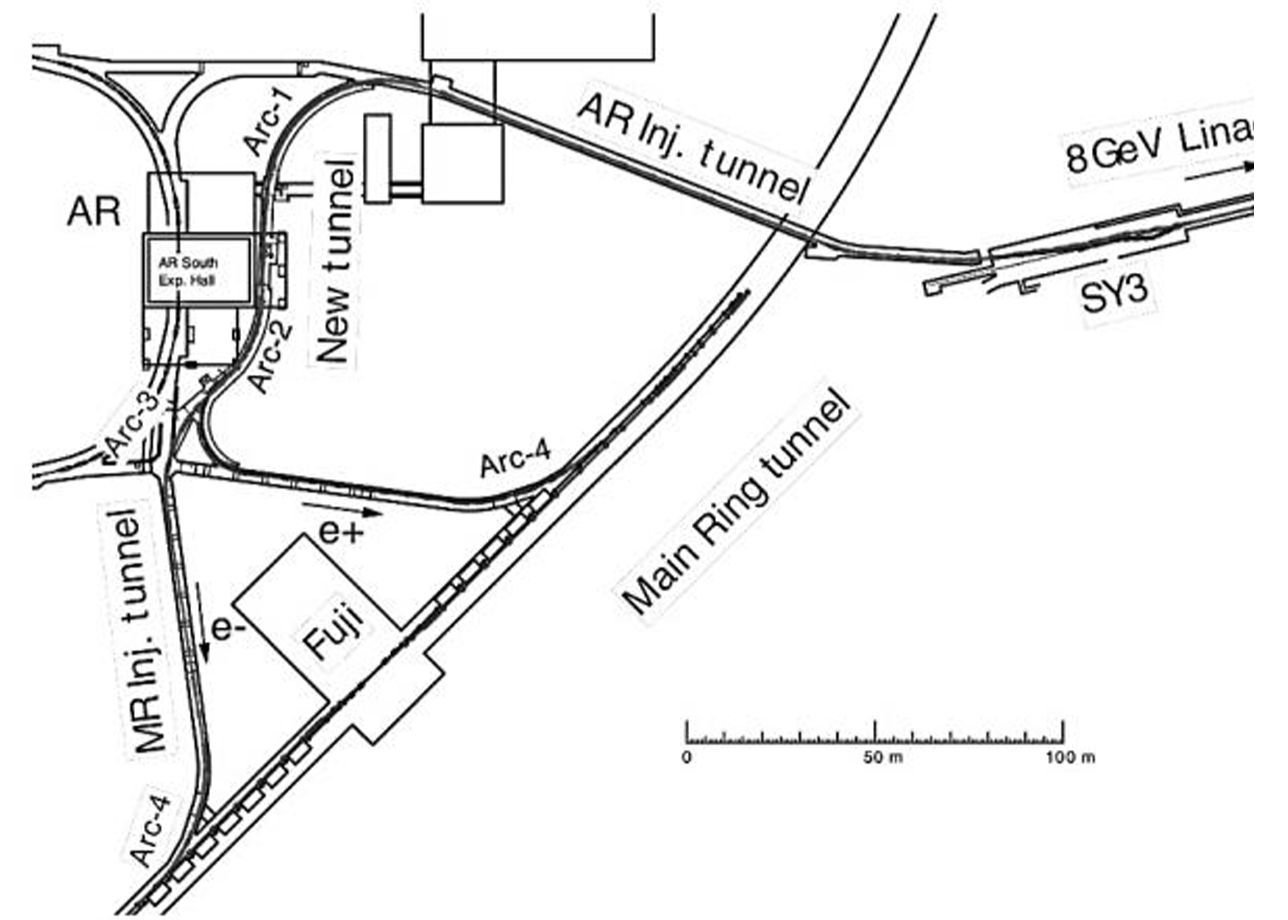}
    \caption{Overhead view of the beam transport lines, showing both electron and positron lines. Only the electron arc is to be polarized.}
    \label{fig:BT}
\end{figure}

After beam generation, the polarized electrons must be aligned to the vertical configuration and delivered through the injector linac into the HER while maintaining spin polarization. Spin-tracking simulations must thus be carried out to ensure that the polarization is maintained throughout, particularly in the 180$^\circ$ J-arc of the linac and in the beam transport line, which has a vertical displacement of approximately 10\,m as well as four horizontal arcs before injection into the HER. An overhead view of the beam transport line is shown in Figure~\ref{fig:BT}.

For purposes of aligning the spin orientation to the vertical, we consider the existing lattice when investigating installation of a spin rotator. Downstream of the beam source sit two focusing-defocusing quadrupole magnet cells and an X-Y steering magnet; before these beam elements, there is approximately 0.9\,m of open space which may be possible to use for beam alignment.

\begin{figure}[hbt!]
    \centering
    \includegraphics[width=.4\textwidth]{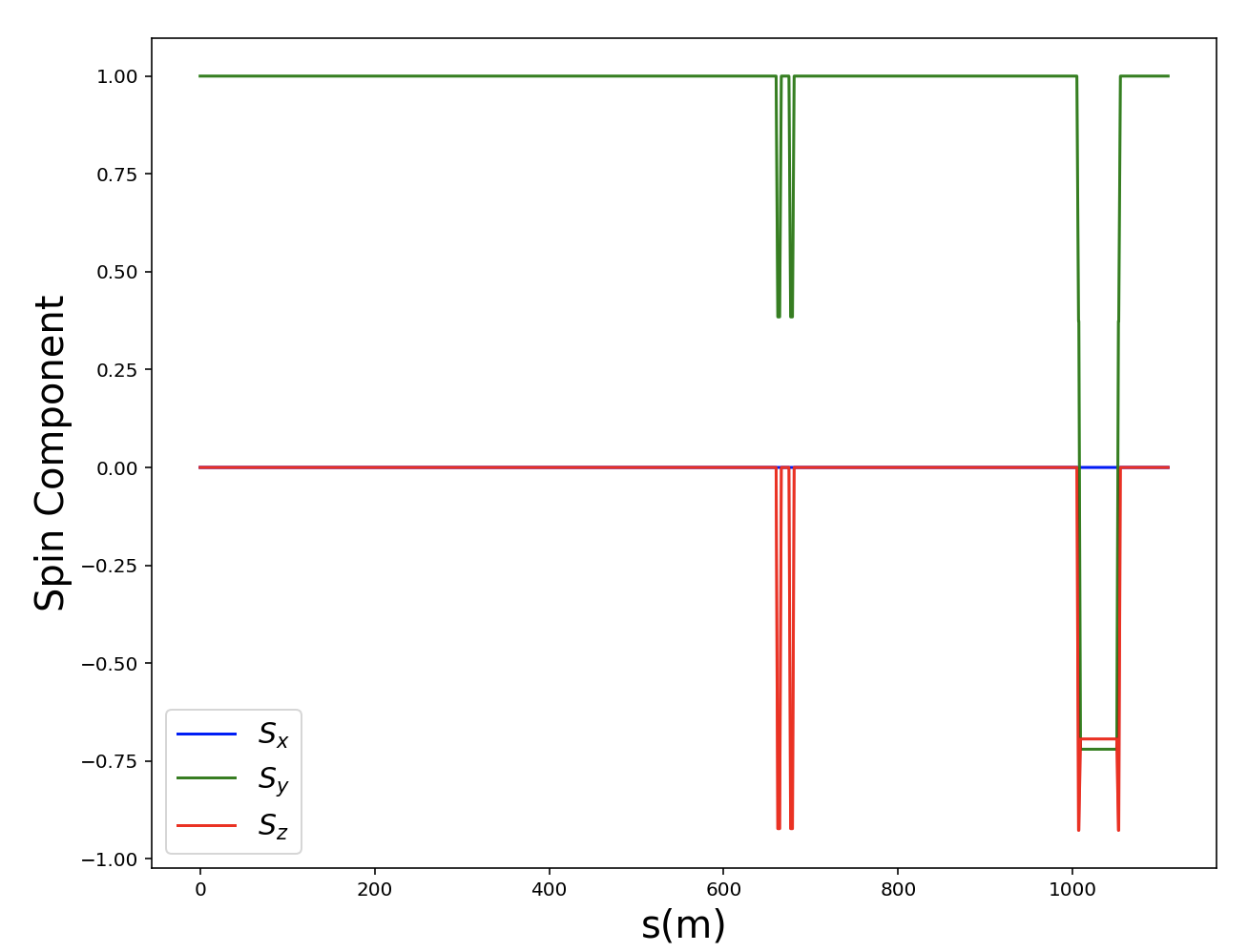}
    \caption{Spin motion of the electron in the Linac section with the beam vertically polarized at the early stage. s=0m corresponds to the approximate source position and s = 1100m corresponds to the position of injection into the HER. }
    \label{fig:Ls}
\end{figure}

Prior to entering the RF cavities traversing the J-arc of the linac, the beam spin vector should be aligned to the vertical to avoid depolarization due to the bend in the linac. To achieve this, we consider the addition of a Wien Filter in the early stages of the linac, immediately after source generation  and prior to entering the first RF cell. Although there exists four pairs of vertical bends before the pipe goes into the tunnel, the vertical polarization is re-established by the time the beam enters the HER, due to anti-symmetric structure of the bends, as shown in Figure~\ref{fig:Ls}.

Beam dynamics simulations are to be carried out for the entirety of the linac, including spin-tracking; however, while the beam lattice model exists in SAD, spin tracking has not been reliably demonstrated. An alternate solution to ensure consistency with tracking performed in the HER is to create the same model in BMAD, with simulations performed in the same manner. The goals of this simulation are to ensure that spin polarization can be maintained acceptably up to injection, and also to determine what, if any, changes must be made to the existing setup in order to ensure that such preservation is possible.

%% file: BeamBeamEffects.tex
The effect of beam-beam interactions on the polarization will have to be studied in simulations. To first-order, the beam-beam effect is a focusing force that affects spin-transparency. At HERA it was observed that the optimum polarization at strong beam-beam required slightly different optimization of the machine but was recoverable to a large extent~\cite{Boge:1995jc,Bieler:1999iz}. Beam-beam in SuperKEKB will be stronger, but only by a modest factor, not by an order of magnitude as the luminosity is increased by extremely small $\beta*$, not by an extremely large beam-beam parameter. We note that the beam-beam effects experienced by the electrons in HERA was not particularly small, due to the strong proton bunches, and was one of the factors limiting the luminosity~\cite{Shi:2003jx}. At SuperKEKB, with short beam lifetime and constant injection of freshly polarized electrons, a high equilibrium polarization is a realistic expectation.

%% file: SpinRotator-BINPDesign.tex

This section discusses how to implement a Spin Rotator within the SuperKEKB lattice. 
Due to the extremely small coupling coefficient of transverse oscillations in SuperKEKB~\cite{Ohnishi-IPAC2011,Ohnishi-ICHEP2016}
 it would be very challenging to use any spin rotator schemes using transverse dipole fields.
 Only the scheme with the use of a longitudinal magnetic field has no effect on the value of the vertical emittance, which is formed by quantum fluctuations of synchrotron radiation in the main dipole magnets of the ring and by parasitic coupling of transverse oscillations. 

\subsubsection{The concept of a scheme for obtaining longitudinal polarization}

\begin{figure*}[h]
	\centering
	\includegraphics[width=1\textwidth]{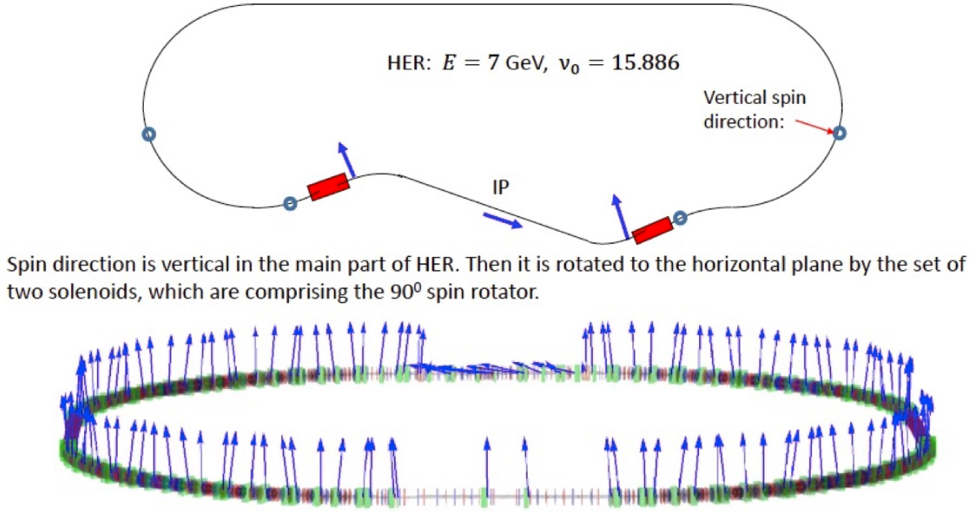}\hfill
	\caption{Scheme of spin rotations with restoration of their vertical direction in the main arcs of the ring. Each spin rotator consists of two solenoids and several skew-quadrupole lenses that compensate for the coupling of betatron oscillations introduced by these solenoids.}
	\label{fig:BINP-SR-Fig1}
	
\end{figure*}

Below, we discuss the simplest version of such a scheme of rotations of the direction of the electron spin vector, when the vertical direction of the spin in the main arcs is completely restored when the beam passes through a long experimental straight section of the HER ring, see Figure~\ref{fig:BINP-SR-Fig1}.

The left and right spin rotators have the opposite sign of the longitudinal magnetic field, and the whole rotation scheme is generally antisymmetric in the signs of magnetic fields at an equilibrium orbit. This antisymmetry ensures almost complete absence of the dependence of the spin orientation in the loops on the particle energy, which is very important for obtaining a long beam depolarization time. The presence of a nonzero dispersion function in spin rotators and magnets between them does not allow to completely suppress the chromatic energy dependence of the spin orbit in the loops. We will discuss this issue in greater details later.

To rotate the electron spin by 90 degrees, the field integral in the solenoids is proportional to the particle momentum:
\begin{equation*}
Bl = \frac{\pi}{2(1+a_e)} BR
\end{equation*}

Here   $BR=p~c/e$ is the magnetic rigidity and $a_e =1.16\times10^{-3}$ is the anomalous magnetic moment of an electron.
The subsequent rotation of the spin by 90$^o$ in the horizontal plane occurs in the section from the rotator to the point of intersection of the beams with the total angle of rotation of the velocity vector equal to: 
\begin{equation*}
\theta =  \frac{\pi}{2\nu_0}
\end{equation*}
where $\nu_o = \gamma a_e$  is the dimensionless spin frequency (or tune) proportional to the gamma factor of the particle.

To detune from resonances with betatron oscillation frequencies, which can quickly depolarize the beam due to beam-beam collision effects, we chose the optimal value of the electron energy equal to E = 7.15~GeV, which corresponds to the spin tune  $\nu_0 = 16.226$ which is sufficiently distant both from close to half-integer values of transverse oscillation frequencies, and from integer resonances with their synchrotron satellites. This value of the spin tune dictates to us the required total angle of all turns in the horizontal plane from the rotator to the interaction point equal to $\theta=0.0968$.

\begin{figure*}[h]
	\centering
	\includegraphics[width=1\textwidth]{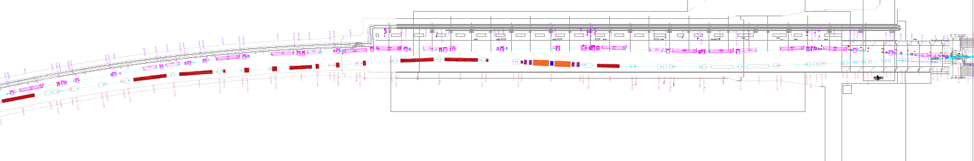}\hfill
	\caption{To the left from the IP half of experimental straight section. The modified magnetic elements of the HER ring are painted in dark brown, and the solenoids of the spin rotator are painted in dark yellow. The distance between the rings is great }
	\label{fig:BINP-SR-Fig2}
\end{figure*}

\begin{figure*}[h]
	\centering
	\includegraphics[width=1\textwidth]{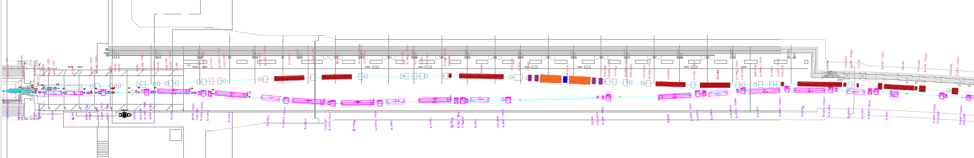}\hfill
	\caption{To the right from the IP half of straight section. At the entrance to the tunnel, the magnets of the rings are very close, but such technical problems can be solved.}
	\label{fig:BINP-SR-Fig3}
\end{figure*}

In the current geometry of the complicated wavy HER orbit, there is no suitable place for a spin rotator. Moreover, given that the length of the rotator is about 10 meters, several dipoles had to be moved from their places, simultaneously changing their angles of rotation. Such transformations of the long insertion connecting the left and right arches were calculated and optimized taking into account the preservation of the storage ring perimeter. The geometries of the separation of the trajectories of the LER and HER rings are slightly different on the left and right sides from the interaction point, see respectively Fig. \ref{fig:BINP-SR-Fig2} and \ref{fig:BINP-SR-Fig3}.

\begin{figure*}[htb!]
	\centering
	\includegraphics[width=.9\textwidth]{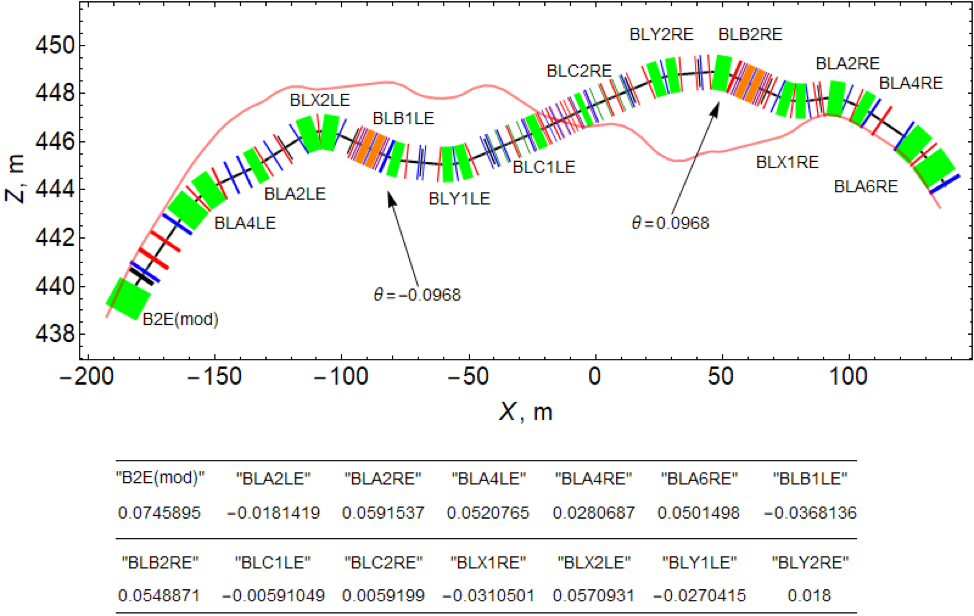}\hfill
	\caption{Optimized SuperKEKB ring intersection scheme for longitudinal polarization. Spin rotators are located in convenient places, away from the tunnel walls and the magnetic elements of the positron ring. The angles of rotation of all dipole magnets are given at the bottom of the diagram, and in Tables~\ref{tab:BINP-SR-Tab1} and \ref{tab:BINP-SR-Tab2}. The lengths and angles of only modified dipoles are given.
	}
	\label{fig:BINP-SR-Fig4}
\end{figure*}

In a condensed form, the scheme of intersection of the collider rings is shown in Fig.~\ref{fig:BINP-SR-Fig4}.

\begin{table*}[hbt!]
	\centering 
	\begin{tabular}{|l|l|l|l|l|l|} \hline
           &           & \multicolumn{2}{c|}{Original parameters of dipoles} &  \multicolumn{2}{c|} {New parameters} \\ \hline
      Name &  Quantity & Length, m & Angle, rad                            & Length, m     &  Angle, rad \\ \hline
      B2E.4&   1       &   5.90220 &  0.0557427                            &  5.90220      &     0.0745895 \\
      BLA4LE & 2       &   5.90220 &  0.0663658                            &  5.90220      &     0.0520765 \\
      BLA2LE & 1       &   5.90220 & 0.0206421                             & {\bf 3.96143} &    -0.0181419\\
      BLX2LE & 2       &   3.96143 & 0.0259281                             & {\bf 5.90220} &     0.0570931  \\
      BLB1LE & 1       &   3.96143 & -0.0229996                            & 3.96143       &    -0.0368136 \\ \hline
	\end{tabular}
	\caption{Lengths and rotation angles of the dipoles to the left of the intersection of the beams.}
	\label{tab:BINP-SR-Tab1}
\end{table*}
\begin{table*}[hbt!]
	\centering 
	\begin{tabular}{|l|l|l|l|l|l|}\hline
           &           & \multicolumn{2}{c|}{Original parameters of dipoles} &  \multicolumn{2}{c|} {New parameters} \\ \hline
      Name &  Quantity & Length, m & Angle, rad                            & Length, m     &  Angle, rad \\ \hline
     BLA6RE & 2        & 5.90220  & 0.0501497 &                         5.90220        & 0.0501498 \\
    BLA4RE  &  1    &   5.90220 &  0.0480687                            &  {\bf 3.96143}      &     0.0280687 \\
    BLA2RE  &  1    &   3.96143 &  0.0348280                           &  {\bf 5.90220}   &     0.0591537 \\
    BLX1RE  &  2    &   3.96143 & -0.0221788                             & 3.96143      &     -0.0310501 \\
    BLB2RE  &  1    &   3.96143 & 0.0234696                             & {\bf 5.90220}       &     0.0548871 \\
    BLY2RE   & 2    &   3.96143 & 0.0270000                             & 3.96143       &     0.0180000 \\ \hline
	\end{tabular}
	\caption{Lengths and rotation angles of the dipoles to the right of the intersection of the beams.}
	\label{tab:BINP-SR-Tab2}
\end{table*}

Note that due to a significant change in some of the rotation angles, the lengths of the corresponding dipoles also changed. In the new scheme, the number of long dipoles is increased by two units, while the number of short dipoles is decreased also by two units. We note that the right half of the long experimental region in the new geometry is lengthened by 14 mm, which is fully compensated by the corresponding shortening of its left half. Spin rotators are inserted into specially widened gaps about 10 m long, between the structural blocks of compensation for the local chromaticity of the triplets of the strong final quadrupole lenses. Each such SX or SY-block\cite{Ohnishi-IPAC2011} consists of a pair of identical dipole magnets and symmetrically spaced quadrupole lenses, which provide minus-unity of the diagonal elements of the transport matrix of the section between the centers of the same sextupole lenses $T_{11}=T_{22}=T_{33}=T_{44}=-1$. Also in this matrix the elements $T_{12}=T_{34}=0$   equal to zero.  Such optics ideally provide complete suppression of geometric aberrations for particles with equilibrium energy. This scheme of compensation for the local chromaticity of strong lenses of the final focus by pairs of non-interleaved sextupoles is currently generally accepted and we have kept this approach intact, changing only the dipole angles in both chromatic blocks.

\subsubsection{Spin rotators}
There are several options for compensating for the coupling of betatron oscillations introduced by the longitudinal magnetic field of the solenoids. The general idea is that in a system of reference, which rotates with a half Larmor frequency   around the longitudinal axis, motion in two transverse degrees of freedom becomes uncoupled if the angles of rotation of all skew-quadrupole lenses around their axis are chosen to be equal to the integral of such twist\cite{Zholents}:
\begin{equation*}
    \phi(s) = \int^s_{s_0} \kappa(s)ds
\end{equation*}

Here, the reference longitudinal coordinate $s_0$ is chosen in such a way that exactly half of the integral of the longitudinal field over the entire rotator is accumulated from the beginning of the rotator to this point. As a result, the angles of rotation of lenses at azimuths $s>s_0$ have a sign determined by the sign of the torsion $\kappa(s)$, while lenses at azimuths $s<s_0$ are rotated around their axis in the opposite direction.

In the thus introduced rotating Cartesian coordinate system (x,y,s), the equations of motion for transverse deflections x,y take on an extremely simple form:
\begin{equation*}
    \begin{matrix}
    x^"+(\kappa(s)^2+g(s))x = 0\\
    y^"+(\kappa(s)^2+g(s))y = 0
    \end{matrix}
\end{equation*}

where $g(s)=G(s)/BR$  is the transverse field gradient normalized to the magnetic rigidity. It is shown in reference~\cite{Zholents} that the full 4x4 transport matrix of a spin rotator has zero anti-diagonal 2x2 blocks if the matrices for x,y oscillations counted in the rotating coordinate system are equal to each other with the opposite sign: $T_x=-T_y=T$.  Under this condition, the full matrix M of the rotator has the same form both in the rotating basis and in the fixed one: 
\begin{equation*}
M = 
\begin{pmatrix}
T & 0\\
0 & -T
\end{pmatrix}
\end{equation*}

\begin{figure*}[htb!]
	\centering
	\includegraphics[width=.6\textwidth]{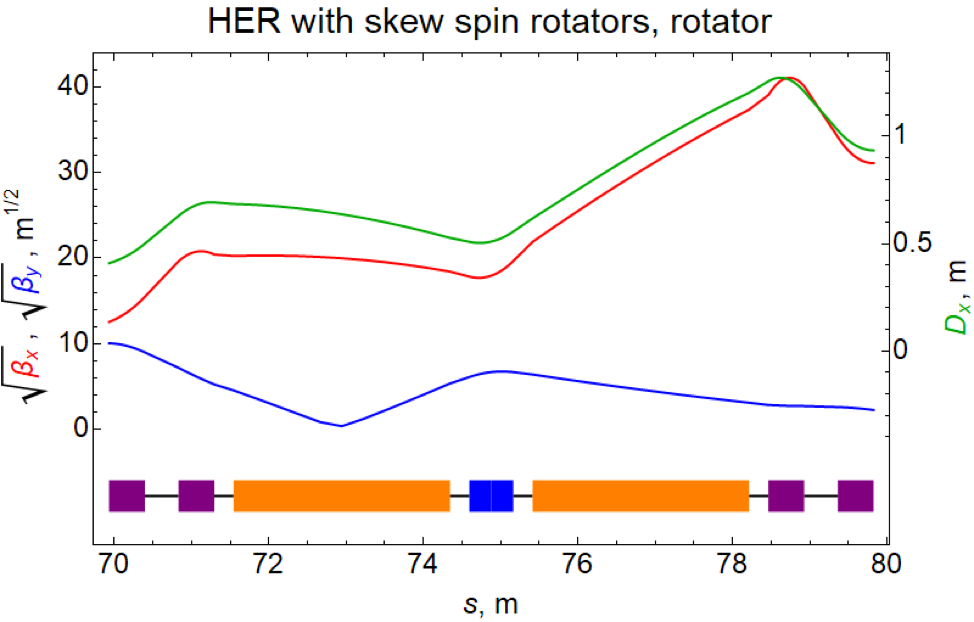}\hfill
	\caption{
Optical functions of the spin rotator for the left half of the long interaction region. 
	}
	\label{fig:BINP-SR-Fig5}
\end{figure*}

Note that there is great freedom in choosing the form of 2x2 matrices $T_x=-T_y=T$. 
In our choice, we settled on a mirror-symmetric arrangement of lenses and two sections of solenoids, shown in Fig.~\ref{fig:BINP-SR-Fig5}. Rotator optics based on such a scheme was successfully implemented in the 1990s on the AmPS storage ring at NIKHEF, Amsterdam\cite{Luijckx}.

The middle lens is not rotated around its axis, while the lenses of the two outermost doublets are rotated at an angle $\phi = \pm \pi/8(1+a_e)=\pm22.474^{o}$

The $X$-box 2x2 matrix is:
\begin{equation*}
T = 
\begin{pmatrix}
0.4134   & 7.13 \\
-0.14025 & 0.4134
\end{pmatrix}
\end{equation*}

The total length of the rotator is 9.89112 meters. Table~\ref{tab:BINP-SR-Tab3} shows the main parameters of the solenoids and lenses of this spin rotator. 

\begin{table*}[h]
	\centering 
	\begin{tabular}{|l|l|l|} \hline
Element &
Length, m &
Field or Gradient:  T,  T/m \\ \hline

Quadrupole \#1, \#5 &
0.46227 &
-29.4792 ($\phi1=-\phi5=-22.474^{o}$) \\ \hline

Drift 1 &
~~~~~~~~0.436   & \\ \hline

Quadrupole \#2, \#4 &
0.46227 &
28.5569  ($\phi2=-\phi4=-22.474^{o}$)  \\ \hline

Drift 2 &
~~~~~~~~0.25  & \\ \hline

Solenoid & 
2.8 & 
6.54197  \\ \hline

Drift 3  &
~~~~~~~~0.25  & \\ \hline

Quadrupole \#3 &
0.57004    & 
-25.3736   ($\phi3=0$)) \\ \hline
	\end{tabular}
	\caption{
	Basic parameters of lenses and solenoids for BR = 23.3495 T·m (E = 7.15 GeV).
	}
	\label{tab:BINP-SR-Tab3}
\end{table*}

Alternatively, we consider the option of combining the solenoidal and quadrupole fields. This approach, but with a more radical proposal, to combine three types of fields (solenoidal, quadrupole, and dipole) is also considered and  described in the following section. But, as mentioned above, the presence of a dipole field can lead to the excitation of too large a vertical emittance and we consider here the solenoidal-quadrupole combined version. Table \ref{tab:BINP-SR-Tab4} shows the parameters of quadrupole lenses of such a solenoidal-quadrupole combined version, in which the solenoidal field is continuous and occupies the entire length of the rotator of 9.2~m. 

The windings of the quadrupole lenses are superimposed on a cylindrical mandrel over the solenoid winding. Moreover, there are two sets of windings: straight-oriented windings and weaker second windings rotated at an angle of 45$^o$ to them, creating a skew-quadrupole field. The total number of lenses in the rotator is increased to 7 in order to have the freedom to reproduce the same optics in the versions with the longitudinal field on and off. 
\begin{table*}[hbt!]
	\centering 
	\begin{tabular}{|l|l|l|} \hline
	
	Element &
Length, m &
Field or Gradient:  T,  T/m \\ \hline

Drift \#1, \#17 &
0.34556 & 
$B_s=G=0$ \\ \hline

Pure Solenoid \#2, \#16 &
0.15  &
$B_s=4.067373$ \\ \hline

Quadrupole plus Solenoid: \#3, \#15  &
0.7  & 
-20.067768 ($\phi3=-\phi15=-19.822^o$) \\ \hline

Solenoid \#4, \#14  &
0.4  &
$B_s=4.067373$ \\ \hline

Quadrupole plus Solenoid: \#5, \#13  &
0.7  &
23.232294  ($\phi5=-\phi13=-14.5297^o$)  \\ \hline

Solenoid \#6, \#12 &
0.8 &
$B_s=4.067373$ \\ \hline

Quadrupole plus Solenoid: \#7, \#11  &
0.7  &
-5.385630   ($\phi7=-\phi11=-7.3598^o$)  \\ \hline

Solenoid \#8, \#10 &
0.8 & \\ \hline

Quadrupole plus Solenoid: \#9 & 
0.7 &
-22.806964  ($\phi9=0^o$) \\ \hline

		\end{tabular}
	\caption{
	Basic parameters of lenses and solenoids of a spin rotator with a superposition of solenoidal and quadrupole fields for BR = 23.3495 T·m (E = 7.15 GeV). The sequence of numbering of structure elements: 1, 2,..., 17.
	}
	\label{tab:BINP-SR-Tab4}
\end{table*}

The option with combined fields has several attractive points. Its main advantage over the variant with separated functions is the acceptably small value of the longitudinal field: $B_s=4.067373$~T versus the
alternative  $B_s=6.54197$~T. Also, the use of a common cylindrical frame supporting both the solenoidal and quadrupole windings may be more technologically advanced as compared to their separate fixation in space. In addition, the need to create two types of lens windings - straight-oriented and skew-rotated - makes the cosine-theta technology the preferred choice.

When calculating the transport matrix of a section of the structure with superposition of the quadrupole and solenoidal fields constant along s, we used numerical methods for calculating the exponent of the matrix of a linear system of equations of motion:

\begin{equation*}
\frac{dX}{ds}= A \cdot X
~~~~A=
\begin{pmatrix}
0 & 1 & \kappa & 0 \\
-\kappa^2-g & 0 & 0 & \kappa \\
-\kappa & 0 & 0 & 1 \\
0 & -\kappa & -\kappa^2+g & 0
\end{pmatrix}
\end{equation*}

The transformation transport matrix of vector $X = (x,p_x,y,p_y)^T$ is:
\begin{equation*}
    T(s,g,\kappa) = R(-s\kappa) \text{exp}(sA)R(s\kappa) 
\end{equation*}
where  $R(s\kappa)$ is the rotation matrix, and the arguments of the matrix $A$ are as defined above. Note that  $p_x$ and $p_y$  are canonical momenta, defined as: 
\begin{equation*}
\begin{matrix}
p_x &=& x' - \kappa~y \\
p_y &=& y' + \kappa~x
\end{matrix}
\end{equation*}

Canonical momenta do not experience any jumps at the boundaries of the solenoidal field, in contrast to kinetic impulses $x'$,$y'$ which subjected a discontinuity at solenoid edges. 

The angles of rotation of the lenses were selected in such a way as to make all elements of antidiagonal 2x2 coupling boxes of the complete transport matrix of the rotator vanish. To fulfill this condition, taking into account the mirror symmetry of the placement of all seven lenses and with the antisymmetric rotation of six of them around the axis, we used a procedure for numerical optimization of three rotation angles and gradients of four lens families (the central lens is not rotated!). The X-block matrix was selected the same as in the rotator version with separated longitudinal and transverse fields:
\begin{equation*}
T=
\begin{pmatrix}
0.4134 & 7.24 \\
-0.14025 & 0.4134
\end{pmatrix}
\end{equation*}
The matrix of the Y-block is equal to it with the opposite sign. 

At the edges of the rotator, short drift gaps with a length of 0.34556 m are left free from fields, connecting the cold superconducting magnetic system with room temperature of the adjacent areas. The total length of the entire rotator is 9.89112 m, same as in the first version.

\subsubsection{Beam depolarization time}
When a photon of synchrotron radiation is emitted, there is an abrupt change in the equilibrium direction of the spin and the magnitude of the projection of the spin on the new equilibrium direction. The spin relaxation time of the beam due to this process and the equilibrium degree of polarization are determined by the well-known Derbenev-Kondratenko formulas~\cite{Derbenev}:
\begin{equation*}
\begin{matrix}
\tau_{rad}^{-1} & = & \frac{5 \sqrt{3}}{8} \lambda_e r_e c \gamma^5
\left< |r|^{-3} \left( 1 - \frac{2}{9}(\vec{n} \vec{v}) )^2 + \frac{11}{18} (\vec{d})^2 \right)  \right>  \\
\xi_{rad} & = &
-\frac{8}{ 5 \sqrt{3}} \left< |r|^{-3} \vec{b}  (\vec{n}-\vec{d})  \right> /
\left< |r|^{-3} \left( 1- \frac{2}{9}(\vec{n} \vec{v} )^2 + \frac{11}{18} ( \vec{d} )^2 \right)  \right>

\end{matrix}
\end{equation*}
where $\lambda_e$, $r_e$, $c$  and $\gamma$ are the Compton wavelength of an electron, its classical radius, speed of light and gamma factor, respectively. Other parameters stands for:  $r$ – radius of curvature of the orbit at the point of emission of the photon, $\vec{b}$ is the  unit vector indicating the direction of the magnetic field,  $\vec{n}$ the  unit vector along the equilibrium spin direction at a given azimuth, and $\vec{d} = \gamma (d\vec{n}/d\gamma)$ the spin-orbit coupling vector, showing the direction of the jump of vector  $\vec{n}$ when a photon is emitted, and the magnitude of this jump.

\begin{figure*}[htb!]
	\centering
	\includegraphics[width=.6\textwidth]{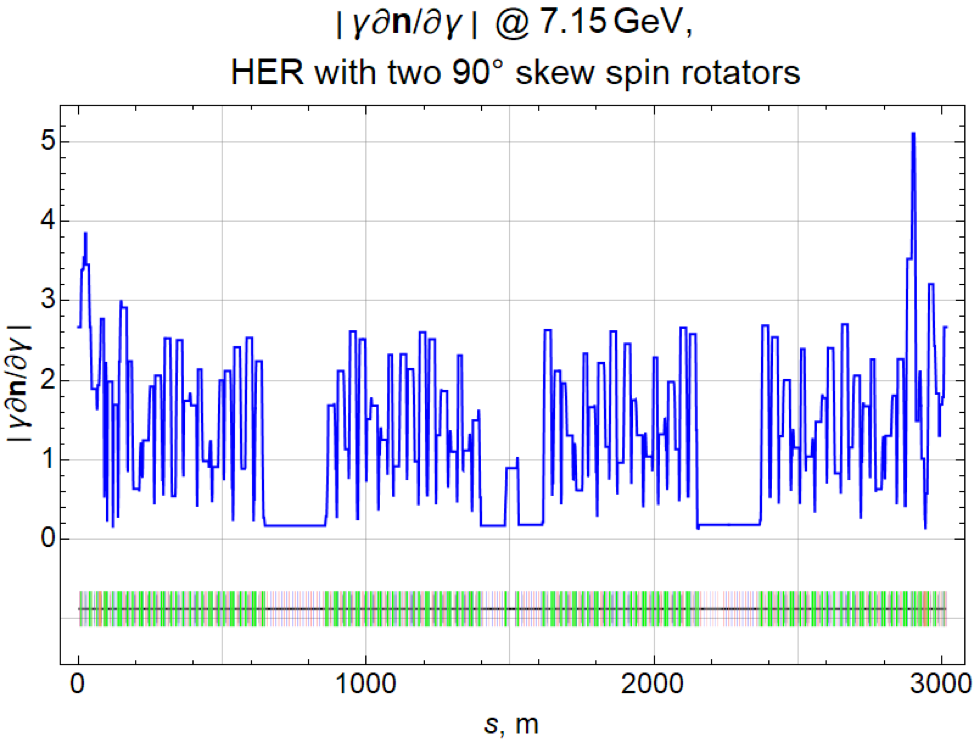}\hfill
	\caption{
	Graph of the modulus of the spin-orbit coupling vector in the HER ring with a modified rotation geometry in the experimental section. Rotator parameters 
	from Table~\ref{tab:BINP-SR-Tab3} and Fig.~\ref{fig:BINP-SR-Fig5}.
	}
	\label{fig:BINP-SR-Fig6}
\end{figure*}

Modulus distribution of vector $\vec{d}$ over the azimuth of the storage ring was calculated by the ASPIRRIN program, created in the 1990s\cite{Ptitsyn1,Ptitsyn2}. Figure~\ref{fig:BINP-SR-Fig6} shows the graph of $|\vec{d}(s)|$ calculated for the rotator optics option with the parameters in Table~\ref{tab:BINP-SR-Tab3}.

\begin{figure*}[htb!]
	\centering
	\includegraphics[width=.6\textwidth]{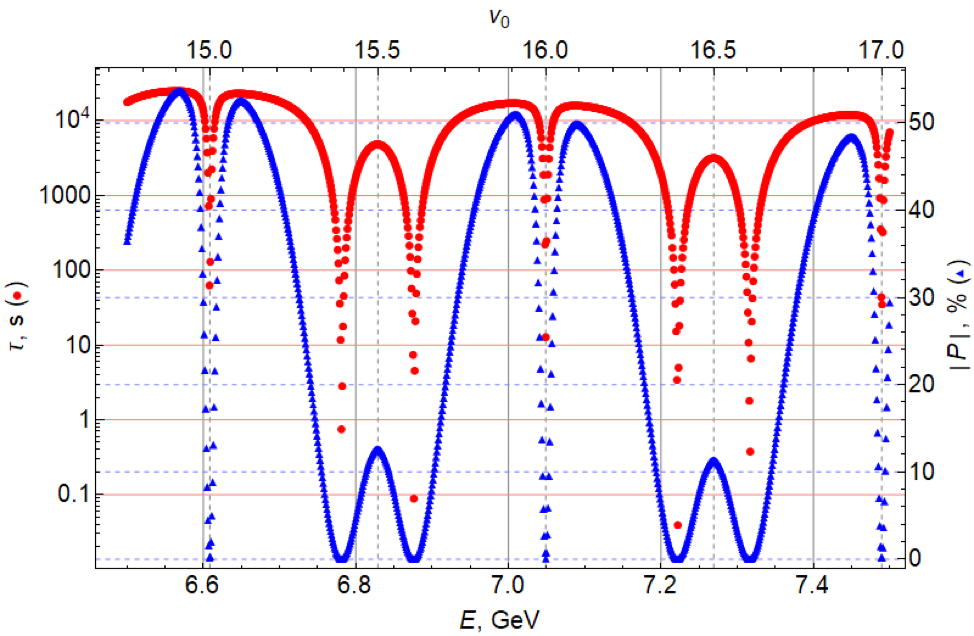}\hfill
	\caption{
Dependence of the radiation spin relaxation time on energy with the rotator version 
from Table~\ref{tab:BINP-SR-Tab3} 
with rotated extreme doublets of the lenses, see Fig.~\ref{fig:BINP-SR-Fig5}.
	}
	\label{fig:BINP-SR-Fig7}
\end{figure*}

It is essential that for a given optimized value of the beam energy E = 7.15 GeV the modulus  $|\vec{d}(s)|$ almost everywhere does not exceed the factor 2. Thereby the spin relaxation time falls from the initial values of the Sokolov-Ternov polarization time 
$\tau_{ST}=32000$s  to a quite acceptable $\tau_{rad} = 10000$s. This time remains very long in comparison with the time for refreshing the beam with new polarized electrons, which is less than  $\tau_{beam}<1000$s. The energy dependence of the radiation spin relaxation time is shown in Fig.~\ref{fig:BINP-SR-Fig7}. Resonances with integer values 
of the spin frequency occurred at the energy E = 6.61, 7.05, 7.49 GeV, and the so-called “intrinsic” spin resonances with the betatron vibration frequencies are located at the spin tunes $\{\nu_0\}$=0.4 and 0.6. You have to choose the beam energy and operate somewhere in between these two types of resonances, for example at E=7.15 GeV, or in other words at spin tune ${\nu_0}=16.226$.

In principle, resonances with betatron frequencies can be made narrower. For this, it is necessary to fulfill the condition of the spin transparency of both: the rotator itself and the entire experimental straight section as a whole 
- from the rotator to the rotator~\cite{Koop2}. But, as our study has shown, the fulfillment of the conditions for spin transparency requires several times stronger lenses in comparison with the variants of rotator optics considered above. Therefore, we have come to a compromise variant of using relatively weak lenses providing, nevertheless, a sufficiently long spin relaxation time. Note that electrons with the opposite sign of polarization with respect to the equilibrium one dictated by the Sokolov-Ternov self-polarization mechanism will depolarize much faster, tending exponentially to their natural state with a positive degree of polarization shown by the blue curve in Fig.~\ref{fig:BINP-SR-Fig7}. Apparently, bunches of electrons with a negative sign of polarization need to be updated more often so that their degree of polarization averaged over time is as high as for its positive sign.

\begin{figure*}[htb!]
	\centering
	\includegraphics[width=.6\textwidth]{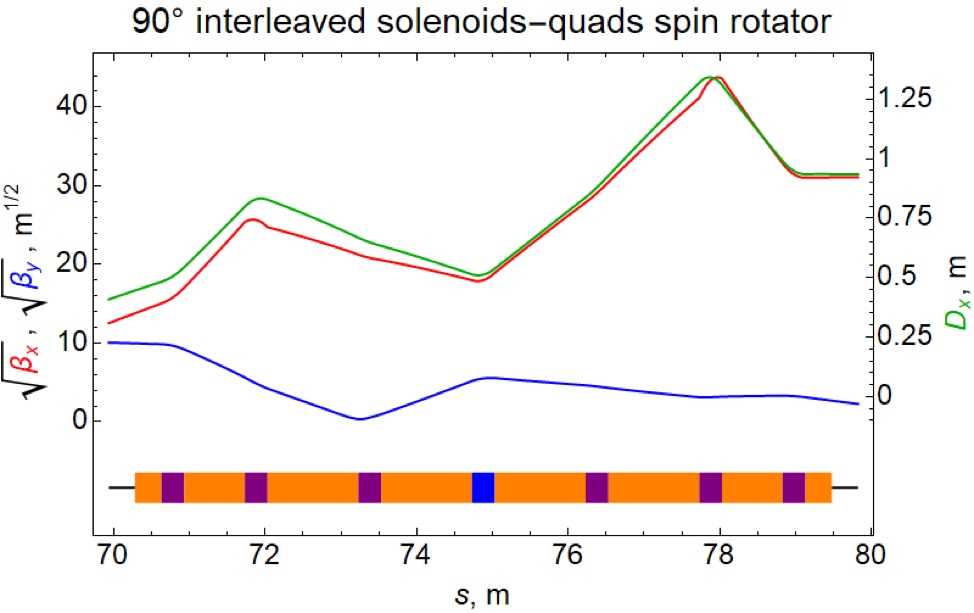}\hfill
	\caption{
Amplitude functions in a rotator with alternating sections of solenoids and quadrupole lenses rotated around the axis. 
	}
	\label{fig:BINP-SR-Fig8}
\end{figure*}

Unfortunately, all programs of accelerator optics calculations available to us, such as MADx or RING, as well as the program for calculating the spin response functions ASPIRRIN, do not support the calculation of matrices of optical elements with combined longitudinal and quadrupole fields. Therefore, to calculate the time and degree of self-polarization in the variant of optics with combined longitudinal and quadrupole fields, we replaced the structure of rotators with a continuous longitudinal field (Table~\ref{tab:BINP-SR-Tab4}) with a structure similar to it presented in Table~\ref{tab:BINP-SR-Tab5}. In this structure, longitudinal field discontinuities are made, in which somewhat shortened lenses are located at the same distances from each other between their centers, as in the variant of Table~\ref{tab:BINP-SR-Tab4}, see Fig.~\ref{fig:BINP-SR-Fig8}. The presence of four families of quadrupole lenses and complete freedom in choosing the angles of their rotation around the axis make it possible to obtain any betatron phase advances and periodic beta functions of the rotator, both with the longitudinal field turned on and off. We reproduced the X-box matrix the same as the rotator shown in Fig.~\ref{fig:BINP-SR-Fig5}.

\begin{table*}[hbt!]
	\centering 
	\begin{tabular}{|l|l|l|} \hline
	
	Element &
Length, m &
Field or Gradient:  T,  T/m \\ \hline

Drift \#1, \#17 &
0.34556 & 
$B_s=G=0$ \\ \hline

Solenoid: \#2, \#16 &
0.35  &
$B_s=5.15983$ \\ \hline

Quadrupole: \#3, \#15  &
0.3  & 
-41.5055 ($\phi3=-\phi15=-20.25818^o$) \\ \hline

Solenoid \#4, \#14  &
0.8  &
$B_s=5.15983$ \\ \hline

Quadrupole: \#5, \#13  &
0.3  &
45.5005  ($\phi5=-\phi13=-15.19364^o$)  \\ \hline

Solenoid \#6, \#12 &
1.2 &
$B_s=5.15983$ \\ \hline

Quadrupole: \#7, \#11  &
0.3  &
-7.56501      ($\phi7=-\phi11=7.59682^o$)  \\ \hline

Solenoid \#8, \#10 &
1.2 & 5.15983 \\ \hline

Quadrupole: \#9 & 
0.3 &
-53.2734     ($\phi9=0^o$) \\ \hline

		\end{tabular}
	\caption{
	Parameters of lenses and solenoids of a spin rotator with alternating solenoidal and quadrupole fields for BR = 23.3495 T·m (E = 7.15 GeV).
	}
	\label{tab:BINP-SR-Tab5}
\end{table*}

\begin{figure*}[htb!]
	\centering
	\includegraphics[width=.6\textwidth]{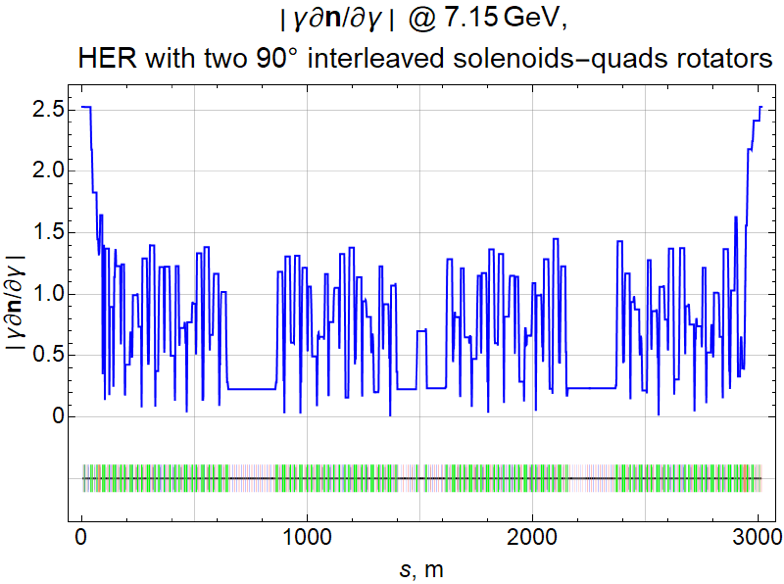}\hfill
	\caption{
	Modulus of the spin-orbit coupling vector in the HER ring with the rotator parameters from Table~\ref{tab:BINP-SR-Tab5}.
	}
	\label{fig:BINP-SR-Fig9}
\end{figure*}

\begin{figure*}[htb!]
	\centering
	\includegraphics[width=.6\textwidth]{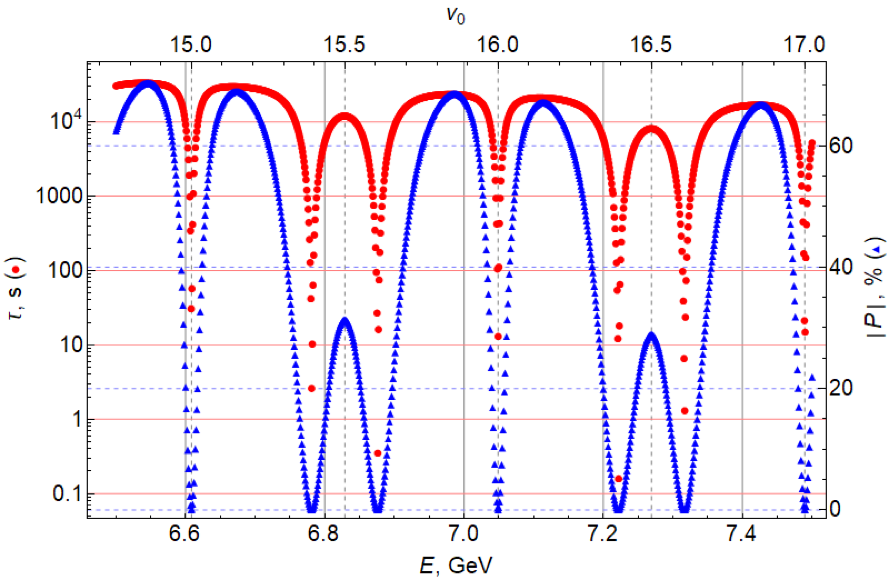}\hfill
	\caption{
	Radiation spin relaxation time and equilibrium degree of polarization versus energy for the rotator optics option from Table~\ref{tab:BINP-SR-Tab5}.
	}
	\label{fig:BINP-SR-Fig10}
\end{figure*}

The magnitude of the modulus of the spin-orbit coupling vector calculated by the ASPIRRIN program is shown in Fig.~\ref{fig:BINP-SR-Fig9}. Its average value in the main part of the ring is less than one and it can be seen that this version of the rotator optics is the best of all those presented earlier. This  was reflected in a significant increase in the spin relaxation time: $\tau_{rad}$=18800~s versus 10000~s for the rotator version from Table~\ref{tab:BINP-SR-Tab3}. Both these tau-values are given for beam energy E = 7.15 GeV. The degree of equilibrium polarization also increased from P = 40\% to P = 60\%. The graphs of the dependence of the relaxation time and the equilibrium degree of polarization on energy are shown in Fig.~\ref{fig:BINP-SR-Fig10}. We believe that the above estimates of radiative self-polarization refer not only to the version from Table~\ref{tab:BINP-SR-Tab5}, but also to the version of the parameters taken from Table~\ref{tab:BINP-SR-Tab4} with a continuous longitudinal field.

\begin{figure*}[htb!]
	\centering
	\includegraphics[width=.55\textwidth]{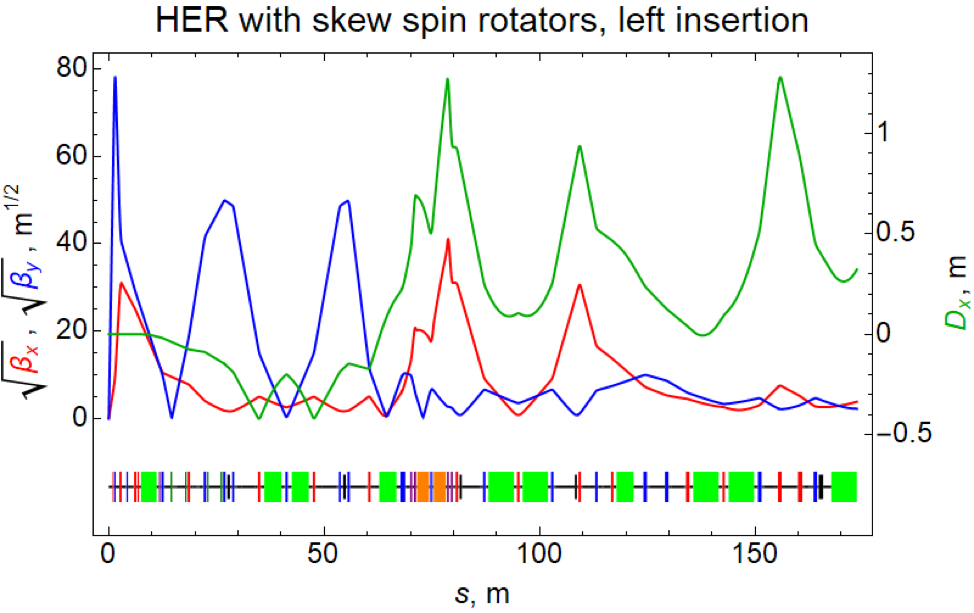}\hfill
	\caption{
	Optical functions of the left half of the long experimental section.
	}
	\label{fig:BINP-SR-Fig11}
\end{figure*}

\begin{figure*}[htb!]
	\centering
	\includegraphics[width=.55\textwidth]{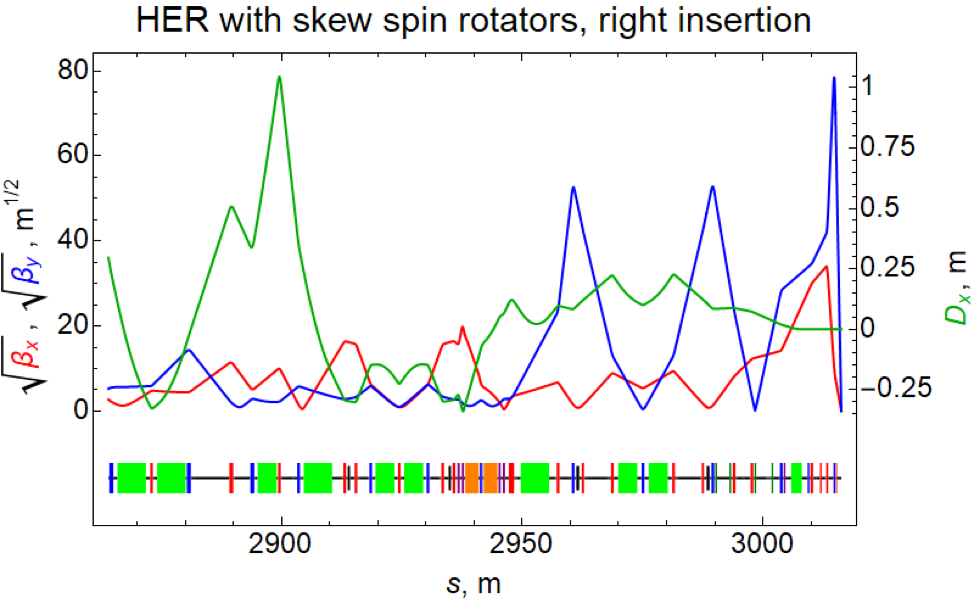}\hfill
	\caption{
	Optical functions of the right half of the long experimental section.
	}
	\label{fig:BINP-SR-Fig12}
\end{figure*}

Finally, Figures~\ref{fig:BINP-SR-Fig11}-\ref{fig:BINP-SR-Fig12} show the optical functions of the left and right halves of the long experimental straight section. They show spin rotators with an optical structure from Table~\ref{tab:BINP-SR-Tab3}. Outside the rotators, the beta functions and the dispersion function are the same for all the versions considered above, since the transport matrices of all these versions are made the same.

\subsubsection{Conclusions}
We found, in our opinion, an acceptable variant of converting the geometry of electron beam bends in the experimental section of the HER storage ring, which provides drift gaps with a length of about 10 meters for installing spin rotators in them. Two rotators serve to rotate electron spins by 90$^o$ from vertical to horizontal and then, after passing the crossing point, back to vertical.  
Various lattice options for spin rotators are considered. In the variant with separate longitudinal and quadrupole transverse fields, the optimal transport matrix of the rotator was found, which made it possible to minimize the disturbances in the optics of the storage ring by the changes introduced. The disadvantage of this option is the relatively large value of the longitudinal magnetic field: over 6.54 T. This disadvantage is eliminated in the variant with the superposition of the rotated quadrupole fields on the continuous longitudinal field of the solenoid, which now occupies the entire length of the gap allocated for the rotator. The value of the longitudinal field in this variant dropped to 4.067 T. 
Calculations of the relaxation time of polarization from the initial +90\% or -90\% to its equilibrium value near +60\% showed its sufficiently large value, about 19000 sec, which will make it possible to have a very high level of average polarization during the lifetime of the beam with continuous feeding with new polarized electrons from injector.
The urgent task at the next stage is to check the boundaries of the dynamic aperture by tracking the particles in order to get an answer to the question of whether it has worsened in comparison with the initial version of the ring optics.

%% file: SpinRotator-ANLDesign.tex
\begin{figure}[ht]
  \includegraphics[width = 6 in]{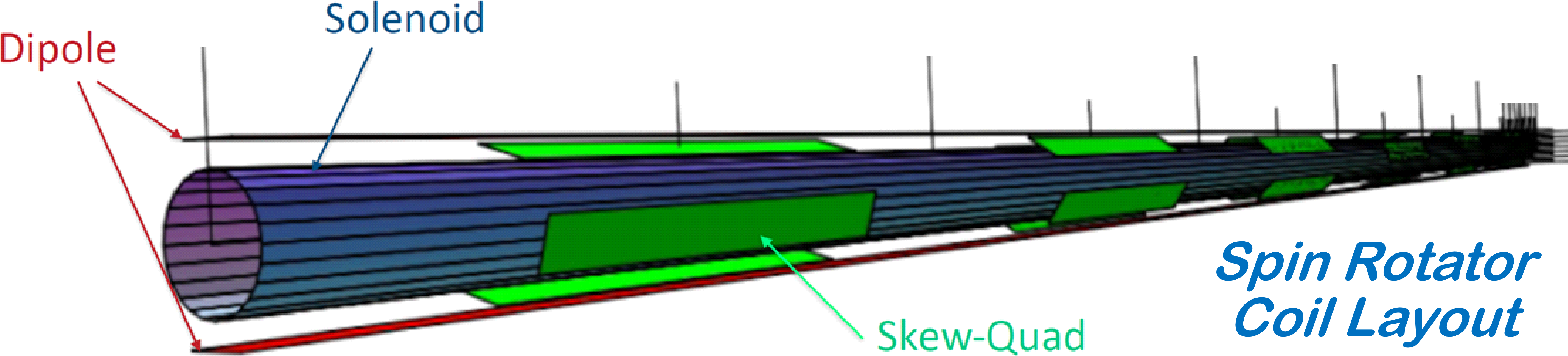}
  \caption{ Uli Wienands’ (ANL) concept for a compact combined function spin rotator unit with overlaid dipole, solenoid and skew-quadrupole superconducting coil fields. }
  \label{fig:spin_rotator_fig}
\end{figure}
\begin{figure}[!ht]
\centering
\includegraphics[width = 6 in]{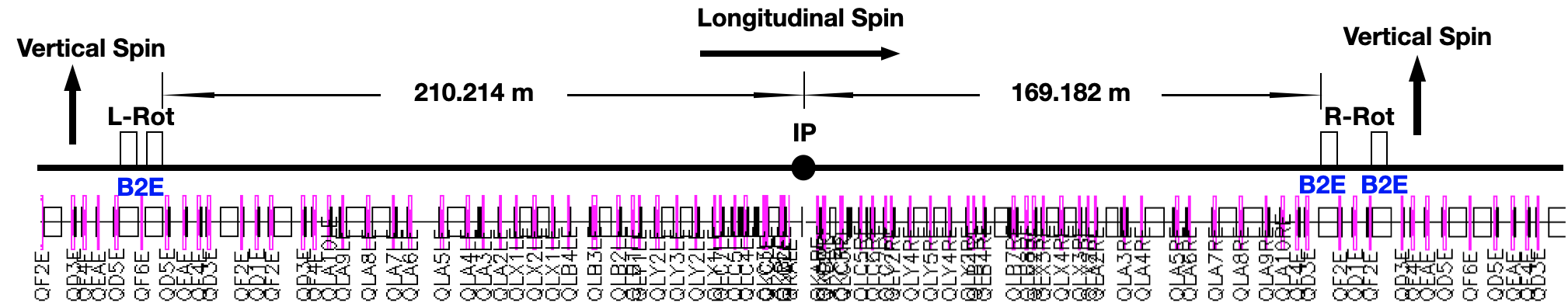}
\caption{Overview of the spin rotator structure: the four B2E dipoles indicated would be replaced by the combine function spin rotator magnets.}
\label{4B2E}
\end{figure}

This section describes the conceptual design of the spin rotator with combined function magnets proposed by Uli Wienands (Argonne National Laboratory). The basic idea is to replace a small number of existing HER dipoles on either side of the interaction point (IP) with the rotator magnets in a manner that renders the change transparent to the rest of the lattice and HER operation. This design is intended to introduce the spin rotators with minimal changes to the HER. In this design, the spin rotator magnet consists of dipole-solenoid combined function magnets with six solenoid-quadrupole magnets installed on the top to compensate for the x-y plane coupling caused by the solenoid \cite{decouple}, as shown in Figure~\ref{fig:spin_rotator_fig}. The strength of the dipole is maintained as the original to preserve the machine geometry. Also, this design allows the original machine to be recovered by minimizing any disruption to the existing ring, such as by turning off the solenoid-quadrupole field in the rotator magnet. As Figure~\ref{4B2E} shows, the spin rotator has two components: the left rotator (L-Rot) located at $\sim$ 210 m upstream of the IP, the right rotator (R-Rot) located at $\sim$ 169 m downstream of the IP. The L-Rot is to rotate the vertical spin of the incoming beam to some direction in the horizontal plane, and dipoles located between L-Rot and the IP continue to rotate spin until it reaches the longitudinal direction. Then, the R-Rot rotates the horizontal spin back to the vertical. The choice of the rotator's installation position has to consider the following constraints: 1) minimizing the impact on the machine dynamics caused by installing the spin rotator; 2) the rotator magnet strength not exceeding the technical limit. The installation position must avoid the region $\pm 100$ m near the IP and keep the area between L-Rot and R-Rot as narrow as possible because the vertical polarization is most stable in the ring due to the vertically induced dipole field. Also, it is beneficial to minimize the number of dipoles that will be replaced. The technical limit is imposed based on considering the possible technology applied to manufacture the combined functions magnets, such as the direct wind technology \cite{direct-wind}; the technical limit is 5 T for the solenoid and 30 T/m for the skew-quadrupole. Considering all the constraints listed above, the four B2E dipoles (field: 0.22075 T, length: 5.9 m) in Figure~\ref{4B2E} are determined to be the optimal positions to install and the detail is given in the Reference \cite{Yuhao}. 


The installation of the rotator is for the e$^-$ polarization purpose only; the original machine dynamics must be preserved as much as possible, which is called ``transparency". Procedures performed to achieve transparency include the decoupling, optical rematching, and restoring of ring parameters. The x-y plane is decoupled at the exit of the rotator by fitting the skew-quads. The optical functions, such as the beta, alpha, and dispersion, need to be matched to the original at the exit of the rotator region by tuning nearby existing quadrupoles to restore the beam dynamics. Also, the overall ring parameters such as the Tunes and the chromaticities need to be the same as in the original lattice to make the rotator fully transparent to the ring. 

\subsubsection{Rotator Modelling with BMAD}
\begin{figure}[ht]
\centering
\includegraphics[width = 4 in]{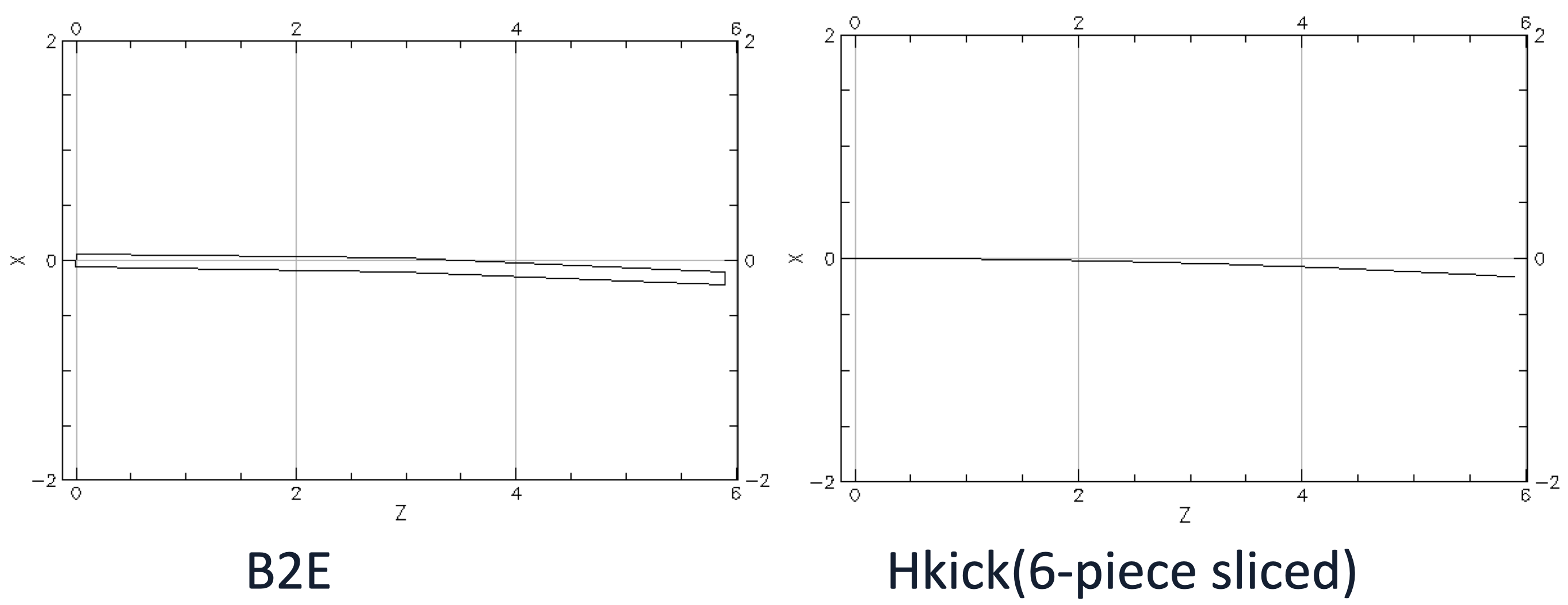}
\caption{Comparison of the geometry of B2E and 6-piece sliced hkick with patches to shift the reference orbit.}
\label{geo}
\end{figure}

The lattice simulation tool applied to run optimizations for this project is BMAD \cite{BMAD}, an open-source, subroutine library created/maintained by David Sagan at Cornell University. The BMAD lattice file of the current SuperKEKB High Energy Ring (HER) with crab waist and a $\beta_y^*=$1~mm was translated from SAD \cite{SAD}. Note that the positive y-direction of the HER lattice points downward in the KEK frame and the $e^+$ default simulation particle is used. The rotator magnet modelling requires a combination of the dipole-solenoid-quadrupole element. BMAD has the solenoid-quadrupole element (Sol\_Quad), but it does not have the attribute of a dipole because Sol\_Quad is a straight element, and a dipole (Sbend) is a curved element. Following the advice of David Sagan, we use the `hkick' (horizontal kick) feature to simulate the dipole. The Sol\_Quad has an hkick attribute since it is a straight element. To simulate the curved element (dipole) with the straight element(hkick), the hkick is initially sliced into six pieces (stand-alone model) to match the number of skew-quad, and the geometry of the dipole is recovered by applying patch elements to shift the reference orbit at the exit of each piece, as shown in Figure~\ref{geo}.

Also, the strength of the hkick pieces is adjusted to match the horizontal orbit. To validate the dipole simulation with the hkick, the four B2E dipoles mentioned in the previous section are replaced with the hkicks in the ring and we compare the optical functions, spin, and ring parameters such as the Tunes and chromaticities. The validation result is shown in the Reference \cite{Yuhao}. However, using hkick and patches causes non-physical orbit excursion due to the shift of the reference orbit. It might affect the performance of the rotator since the particle would experience the solenoid-quad field if it is not following the ideal orbit. This problem is addressed by applying the ``slice model", which is to further slice each piece of the hkick into smaller pieces. With this method, the non-physical orbit excursion is significantly reduced but makes the optimization more computationally expensive. The optimal slicing number of the ``slice model" is determined to be 16, 96 slices in total. 
Also, the `vkick' (vertical kick) is introduced to match the vertical orbit since solenoids cause the x-y plane coupling.

\subsubsection{Open-geometry Optimization}
\begin{figure}[!ht]
\centering
\includegraphics[width = 4 in]{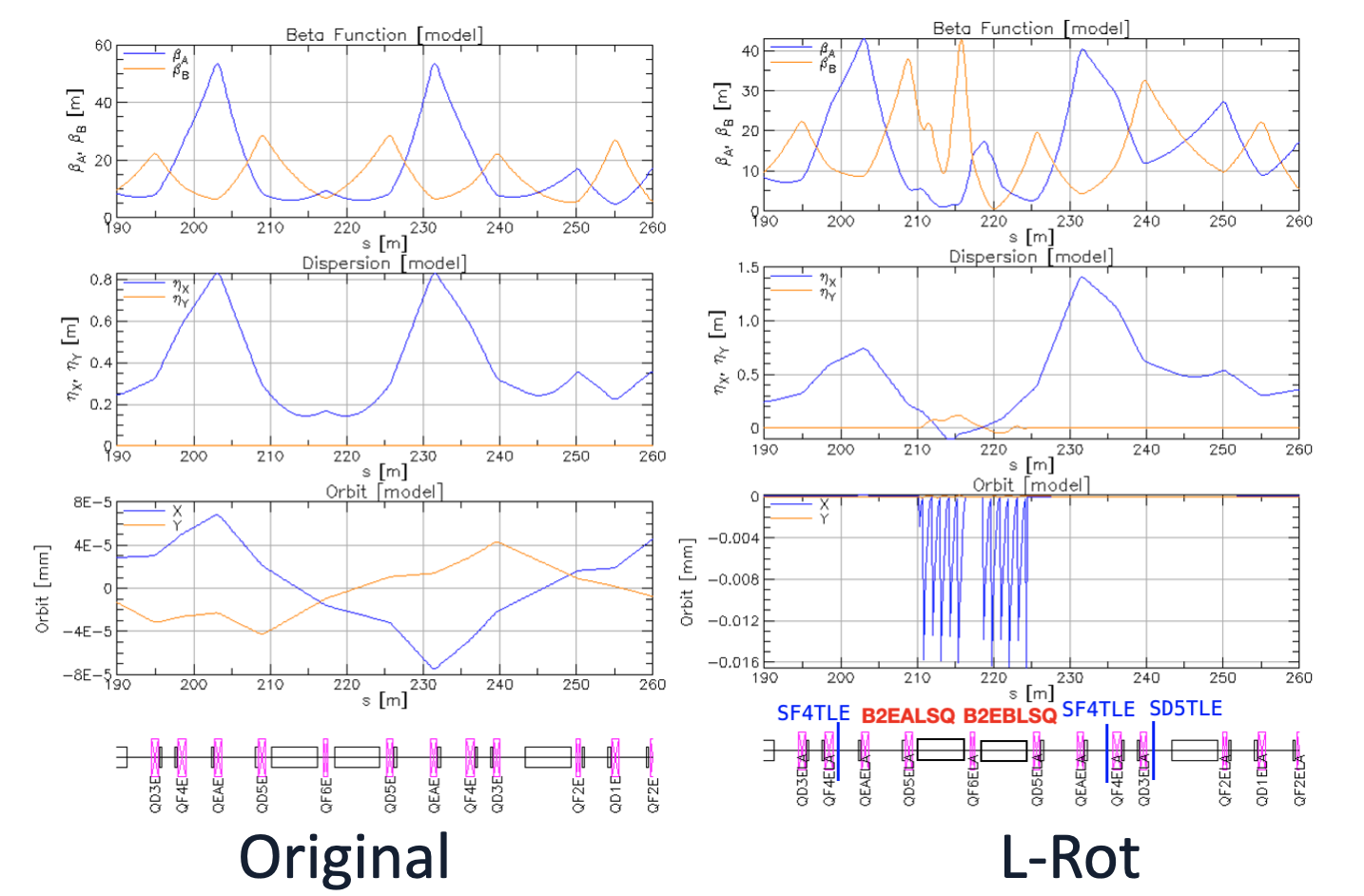}
\caption{Comparison of the L-Rot tuning region: optical functions are matched to the original at the exit. Note that the orbit excursions in the bottom right plot in the region of the L-Rot are non-physical artifacts of the slice model used to describe the dipole component of the spin rotator, as discussed in the text.}
\label{Lrot}
\end{figure}

\begin{figure}[!ht]
\centering
\includegraphics[width = 4 in]{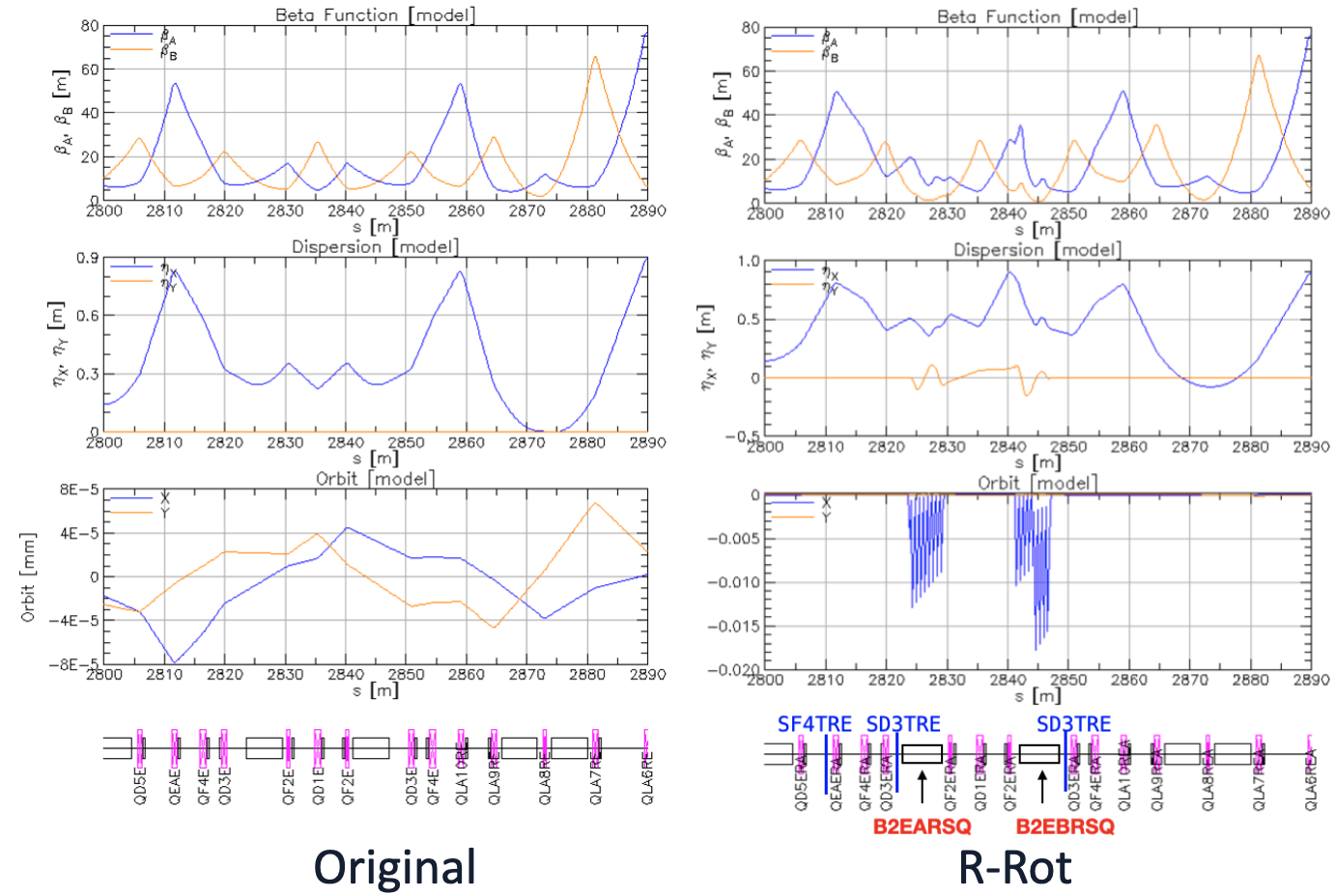}
\caption{Comparison of the R-Rot tuning region: optical functions are matched to the original at the exit. Note that the orbit excursions in the bottom right plot in the region of the R-Rot are non-physical artifacts of the slice model used to describe the dipole component of the spin rotator, as discussed in the text.}
\label{Rrot}
\end{figure}

Four B2E dipoles, indicated in Figure~\ref{4B2E}, are replaced with the rotator elements, and the lattice segments which contain the L-Rot/R-Rot are taken out from the full lattice. The open-geometry optimizations are performed to the lattice segments. The first goal is to fit the value of the solenoid fields to achieve the polarization purpose. The strengths of the two solenoids of the L-Rot are determined to be $k_s = - 0.20720$ $m^{-1}$, $-0.11027$ $m^{-1}$, or $B_s = -4.8431$ T, $-2.5774$ T. For the R-Rot, solenoids are determined to be $k_s$ = -0.15438 $m^{-1}$, -0.16865 $m^{-1}$, or $B_s = -3.6084$ T, $ -3.9420$ T. All solenoids are below the technical limit of 5 T. In the next step,  the skew-quads are turned on to perform the decoupling, with  the goal of making every element in the $\textbf{C}$ matrix \cite{Cmat} zero at the exit of each rotator magnet. The skew-quad strength is correlated to the solenoid. The strongest skew-quad in the L-Rot is about 20 T/m and approximately 14 T/m for the R-Rot; all solenoids are below the limit of 30 T/m. The final step of the open-geometry optimization is to recover the original optical functions to achieve transparency. The lattice segments containing the rotators are extended to involve nearby quadrupoles. Existing quadrupoles and skew-quads are adjusted to match the optical functions to the original at exit of the lattice segments. However, it is challenging to match the vertical dispersion because it lacks the horizontal bending effect in the ring. The way to address this problem is to abandon the decoupling at the exit of the first rotator magnet in the L-Rot/R-Rot, giving extra freedom. In addition, the vertical emittance must be kept low to guarantee the high luminosity. The finalized optimization result is shown in Table~\ref{Lq}, Table~\ref{Lsq} for the L-Rot, and Table~\ref{Rq}, Table~\ref{Rsq} for the R-Rot. The maximum strength of new quadrupoles is about 14 T/m, which is achievable. The lattice segments are put back into the ring, with  Figure~\ref{Lrot} and Figure~\ref{Rrot} showing the comparison of the L-Rot/R-Rot tuning region with the original. It can be seen that the optical functions are well matched to the original at the exit of the tuning region. Note that the orbit excursions in the L-Rot/R-Rot regions are non-physical; they are artificial effects associated with the use of a series of hkicks  to model the dipole component of the spin rotator. 

\subsubsection{Closed-geometry Optimization}

\begin{figure}[!ht]
\centering
\includegraphics[width = 4 in]{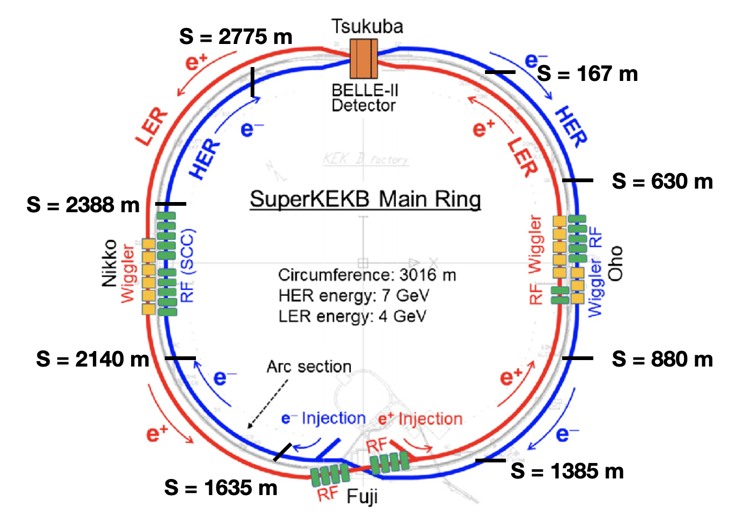}
\caption{Quadrupoles in the ``Nikko" section of the HER are adjusted to match the Tunes; sextupoles in four arc sections between  s =167 to 630 m, s = 880 to 1385 m, s = 1635 to 2140m, and s = 2338 to 2775 m are adjusted to match the chromaticities }
\label{ring}
\end{figure}

The optical functions are matched to the original lattice outside the rotator regions but are not restored inside those regions, which causes Tunes and chromaticities to be shifted from the original lattice. To make the spin rotator fully transparent to the ring, they must be re-matched. The match of the Tunes is achieved by adjusting existing quadrupoles in the straight section region having zero dispersion, allowing local quadrupoles to be adjusted without changing the dispersion. The straight area satisfying the zero-dispersion requirement is the ``Nikko" section shown in Figure~\ref{ring} \cite{superkekb}, and the original chromaticities are restored by tuning sextupoles in four arc sections. The adjustment of the sextupoles does not change the beta function. Thus, Tunes must be matched before chromaticities. Also, Tunes and chromaticities are overall ring parameters that can only be calculated in closed-geometry. 

\subsubsection{Match the Tune}
\begin{figure}[htb!]
\centering
\includegraphics[width = 6 in]{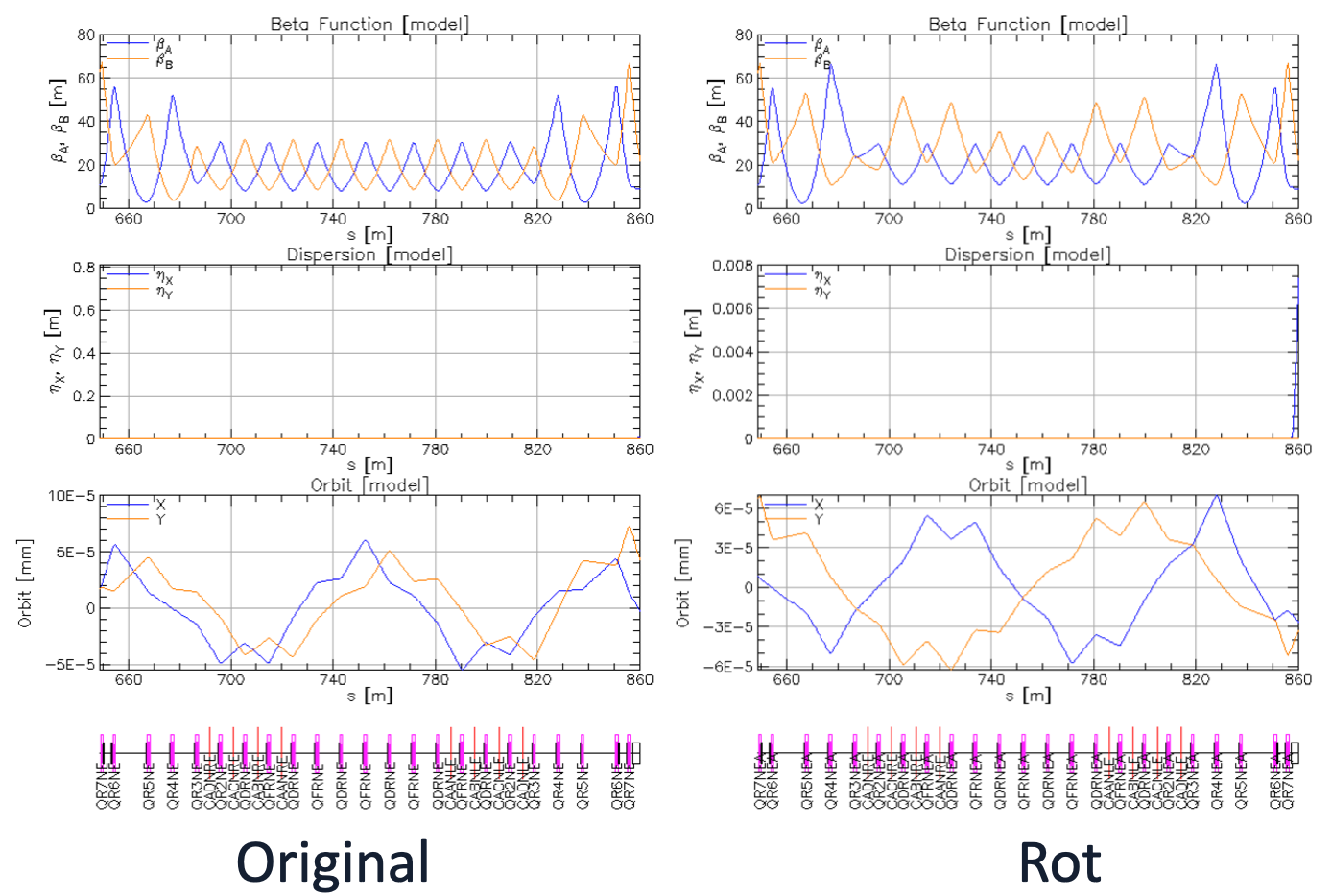}
\caption{Comparison of the ``Nikko" section after the Tune match}
\label{nikko}
\end{figure}

In the ``Nikko" section, the beta function has a symmetric structure because of the symmetric distribution of the quadrupole pairs. The optimization goal is to recover the original Tunes $Q_x$, $Q_y$ and match the Twiss parameters ($\beta_x$, $\beta_y$, $\alpha_x$, $\alpha_y$) at the exit of the ``Nikko" section. In order to preserve the symmetric structure, quadrupoles with the same name must be kept identical. There are 8 variables available, and 6 parameters to match, which gives 2 degrees of freedom. To address the problem of simultaneously matching the Tune and the optical functions, we start with the following approximation. For a small quadrupole deviation $\Delta k$, the Tune shift is given by \cite{wille}: \\

\begin{equation}
    dQ = \frac{1}{4 \pi} \Delta k \beta ds,
\end{equation}\\

where $\beta$ is the local beta function. Notice that the local beta function remains changed. Similarly, when adjust a series of quadrupoles slightly, the total Tune shift $\Delta Q$ can be approximated by: \\

\begin{equation}
    \Delta Q \sim  \frac{1}{4 \pi} \sum_i \beta_i \Delta k_i L_i,
\end{equation}\\

where $L_i$ is the length of the quadrupole. This equation indicates that the Tune can be slightly shifted without significantly changing the local beta function if the variation of quadrupoles is sufficiently small. Thus, the ``Ladder method" is created to approach the original Tunes by taking small steps. This method allows the fitter to match the Tune and the beta function simultaneously since the variation of the quadrupole is small. As Table~\ref{LD} shows, it takes 15 steps to approach the original Tunes. At each step, the Tunes and the beta function at the exit of the ``Nikko" section are matched to the original with an extra constraint to add an upper limit to the local beta function. With this method, the original Tunes are successfully restored. Also, the symmetric structure of the beta function in the ``Nikko'' section is maintained, and the local beta function stays in a reasonable range, as shown in Figure~\ref{nikko}. Table~\ref{quad} shows the optimization result and there is no significant adjustment. \\
\\
\\

\begin{table}[htb!]
	\label{}
    \begin{center}
    \begin{tabular}{ |c||c|c|c|c|c|c| } 
    \hline
    Step	&	0th	&	1st	&	2nd	&	3rd	& ... &	15th	\\
    \hline
    $Q_x$	&	45.777566	&	45.761128	& 45.744690 & 45.728252  &...	& 45.530994	\\
    \hline
    $Q_y$	&	44.446774	&	44.389036	& 44.331299 & 44.273561  &...	& 43.580709	\\
    \hline
    \end{tabular}
    \caption{Ladder Method: taking steps to approach the original Tunes }
    \label{LD}
    \end{center}
\end{table}

\begin{table}[htb!]
	\label{}
    \begin{center}
    \begin{tabular}{ |c||c|c|c| } 
    \hline
    Quadrupole 	&	Length (m)	&	$k_1$ ($m^{-2}$) original	& $k_1
    $ ($m^{-2}$) Rot 	\\
    \hline
    QFRNE	&	1.080	&	0.122	&	0.099	\\
    QDRNE	&	1.080	&	-0.118	&	-0.085	\\
    QR7NE  	&	0.826	&	-0.252	&	-0.249	\\
    QR6NE  	&	1.015	&	0.196	&	0.202	\\
    QR5NE   &	1.080	&	-0.110	&	-0.091	\\
    QR4NE	&	1.080	&	0.144	&	0.127	\\
    QR3NE	&	1.080	&	-0.145	&	-0.071	\\
    QR2NE   &	1.080	&	0.110	&	0.067	\\
    \hline 
    \end{tabular}
    \caption{Quadrupoles adjusted to match the Tunes for the rotator ring. The quadrupole magnet labels are those used in the SAD lattice file.}
    \label{quad}
    \end{center}
\end{table}

\subsubsection{Match the Chromaticity}
Matching the original first-order chromaticity is achieved by adjusting sextupole strengths of existing HER sextupoles in a manner that introduces the required additional quadrupole strength from the quadrupole field component of those sextupoles. The total effective chromaticity $\xi$ is given by \cite{wille}:\\

\begin{equation}
    \xi_{tot} = \frac{1}{4\pi}\oint \bigg [k(s) + m(s)\eta(s) \bigg] \beta(s) ds,
\end{equation}
\\
where $m$ is sextupole strength, $\eta$ is the dispersion, and $k$ is the quadrupole strength. Based on this equation, it can be seen that there exists a linear relation between the chromaticity and the sextupole strength since varying the sextupole strength does not change the beta function. The variation of chromaticity can be expressed as a linear combination of the variation of sextupole strength:

\begin{equation}
    \begin{cases}
        \begin{aligned}
            \Delta \xi_x = \sum_{i}   p_i \Delta x_i  \\
            \Delta \xi_y = \sum_{j}   q_j \Delta x_j 
        \end{aligned}
    \end{cases}
\end{equation}

with 
\begin{equation}
    \begin{cases}
        p_i = \frac{1}{4 \pi} \beta_{x,i} \eta_{x,i}  \\
        q_j = - \frac{1}{4 \pi} \beta_{y,j} \eta_{x,j}
    \end{cases},
\end{equation}
\\
where $\Delta \xi_x$, $\Delta \xi_y$ is the variation that needs to be achieved in order to match the original horizontal and vertical chromaticity, respectively; $\Delta x$ is the variation of integrated strength of the sextupole; $\eta_{x}$, $\eta_{y}$ is the horizontal and the vertical component of the dispersion, respectively. Notice that $b_2 = \frac{k_2 L}{2}$ represents the integrated strength of sextupole in the HER lattice. With this relation, an algorithm is developed to achieve a better staring point for running the BMAD optimization. Assume N (N $>$ 2) sextupoles are used to match the the chromaticity. Let $\vec{\text{P}} =(p_1, \; p_2,\; ..., \; p_\text{N})$, $\vec{\text{Q}} =(q_1, \; q_2, \; ...,\; q_\text{N} )$, $\vec{x} = (\Delta x_1, \; \Delta x_2, \; ... \;, \Delta x_N)$, and define following functions:

\begin{equation}
\begin{cases} 
    f_p (\vec{x})&= \vec{P}\cdot \vec{x} - \Delta \xi_x  \\
    f_q (\vec{x})&= \vec{Q}\cdot \vec{x} - \Delta \xi_y 
\end{cases},
\end{equation}
\\
Construct following square function:\\
\begin{equation}
    f(\vec{x})= {f_p^2(\vec{x})}+{f_q^2(\vec{x})}+\lambda \vec{x}^{2}.
\end{equation}
\\
The first two terms represent discrepancies to the original chromaticities. The last term prevents variables from having large values, and the choice of $\lambda$ must not have significant impact on discrepancies. When $f(\vec{x})$ reaches the global minimum (or close to it), the desired chromaticities are reached. A gradient descent algorithm is applied to approach the global minimum. We give random starting point to initiate the algorithm and at the $j$th iteration, compute following functions:\\

\begin{equation}
    f^{j}_p = \vec{P}\cdot \vec{x}^{j} - \Delta \xi_x,
\end{equation}

\begin{equation}
    f_q^{j} = \vec{Q}\cdot \vec{x}^{j} - \Delta \xi_y,
\end{equation}

\begin{equation}
    \nabla f^{j} = 2f^{j}_p \vec{P}+2f^{j}_q\vec{Q}+2\lambda\vec{x}^j,
\end{equation}
\\

and determine variables for the next iteration:

\begin{equation}
    \vec{x}^{j+1} = \vec{x}^{j} - \varepsilon \nabla f^{j},
\end{equation}
where $\varepsilon$ is the learning rate. The choice of $\varepsilon$ has vital impact on the performance of the algorithm, since large values may not converge to the optima, and small values require longer iteration process.

Sextupoles located at the rotator area and their identical pair are not used for matching the chromaticity. The phase advance between each sextupole pair needs to be fixed at $\pi$ to cancel out the leading order of their non-linear effect. However, the adjustment of the quadrupoles near the rotator region changes the local beta function thus changing the local phase, as shown in Table~\ref{phase}. Sextupole pairs SF4TLE, SD5TLE, and SD3TRE are turned off due to the large phase shift from $\pi$. SF4TRE pair is not adjusted since the phase advance is slightly shifted from $\pi$. 

\begin{table}[htb!]
	\label{}
    \begin{center}
    \begin{tabular}{ |c||c|c|c|c|c| } 
    \hline 
    sextupole    	& s-position (m)  	&		     $\Psi_x$ 	&	     $\Psi_y$	&	$\Delta \Psi_x$	&	$\Delta \Psi_y$	\\
    \hline \hline
    SF4TLE 	& 199.675	&		22.604	&	24.978	&		    &		   \\
    SF4TLE 	& 235.237	&		29.183	&	30.468	&	6.579	&	5.490 \\
	        &		    &		    &		    &		    &		   \\
    SD5TLE 	& 241.458	&		29.552	&	30.790	&		    &		   \\
    SD5TLE 	& 269.820	&		31.817	&	33.344	&	2.265	&	2.554 \\
	        &		    &		    &		    &		    &		   \\
    SF4TRE 	& 2779.934	&		263.620	&	262.070	&		&	   \\
    SF4TRE 	& 2815.495	&		266.770	&	265.220	&	3.150	&	2.950	\\
	        &		    &		        &		        &		&		\\
    SD3TRE 	& 2821.717	&		265.560	&	248.930	&		&		\\
    SD3TRE	& 2850.078	&		268.330	&	256.290	&	2.770	&	7.360	\\
    \hline 
    \end{tabular}
    \caption{Sextupoles located at the rotator tuning area. $\Psi_x$, $\Psi_y$ are the horizontal and vertical component of the phase, and $\Delta \Psi_x$, $\Delta \Psi_y$ are the horizontal and vertical component of advance between the sextupole pair.  
    }
    \label{phase}
    \end{center}
\end{table}

The gradient descent algorithm is implemented in Python. 45 pairs of sextupoles in 4 arc sections are adjusted. It was found that there is no significant adjustment required to match the chromaticities, thus taking $\lambda = 0$. 
The square function $f(\vec{x})$ drops exponentially and reaches the global minimum at $\sim$ 630th iteration. All the adjusted sextupoles stay below the technical limit, and their sign does not change. The output from Python is used as the starting point to run BMAD optimization, and the optimization result is shown in Table~\ref{sext}.


\subsubsection{HER With Spin Rotators: Rotator Ring}
\begin{figure}[htb!]
\centering
\includegraphics[width = 6 in]{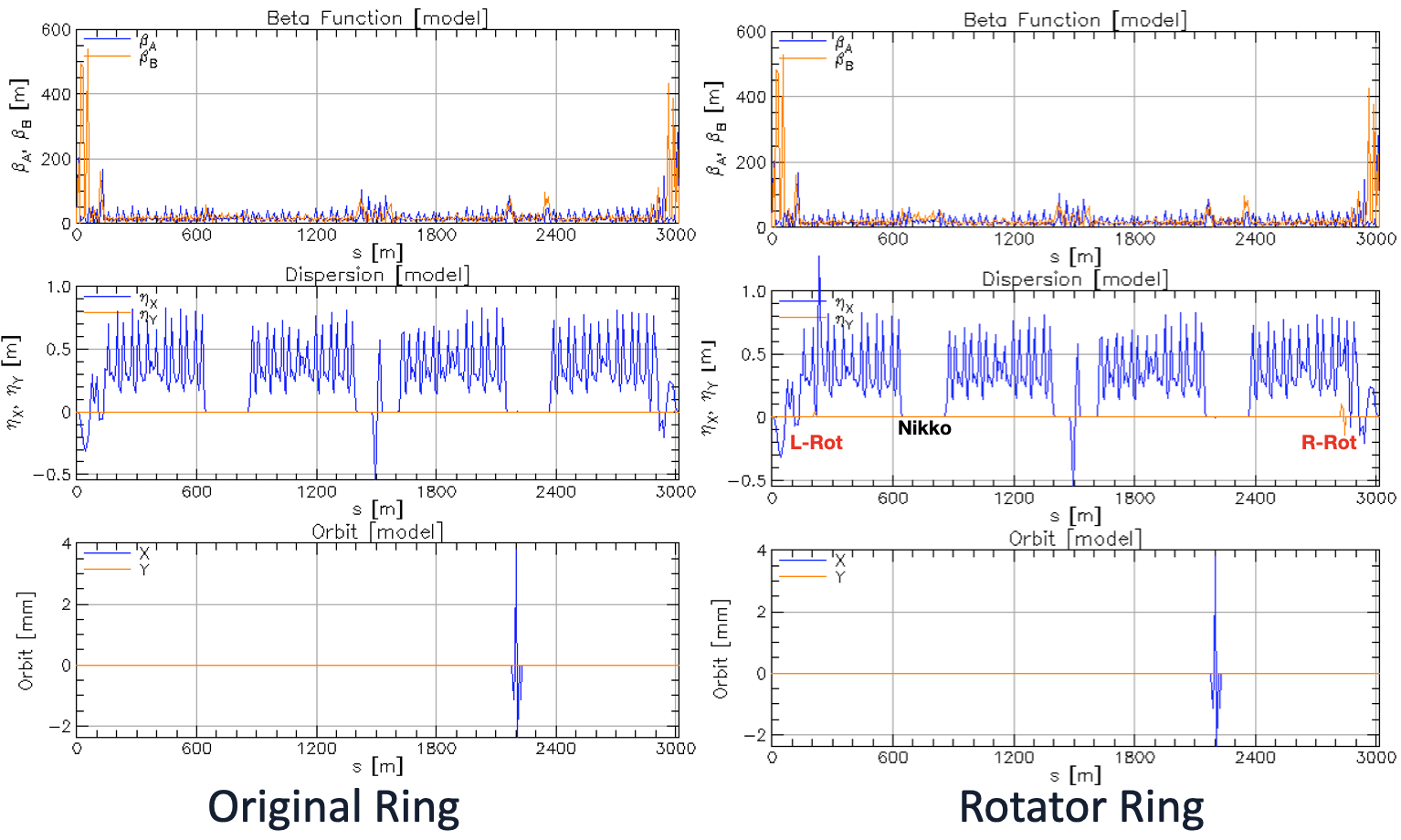}
\caption{Comparison of the HER with Rotator installed: the squiggle in the orbit plot at s $\sim$2200 m is the effect of wigglers }
\label{Rotring}
\end{figure}

\begin{figure}[htb!]
\centering
\includegraphics[width = 7 in]{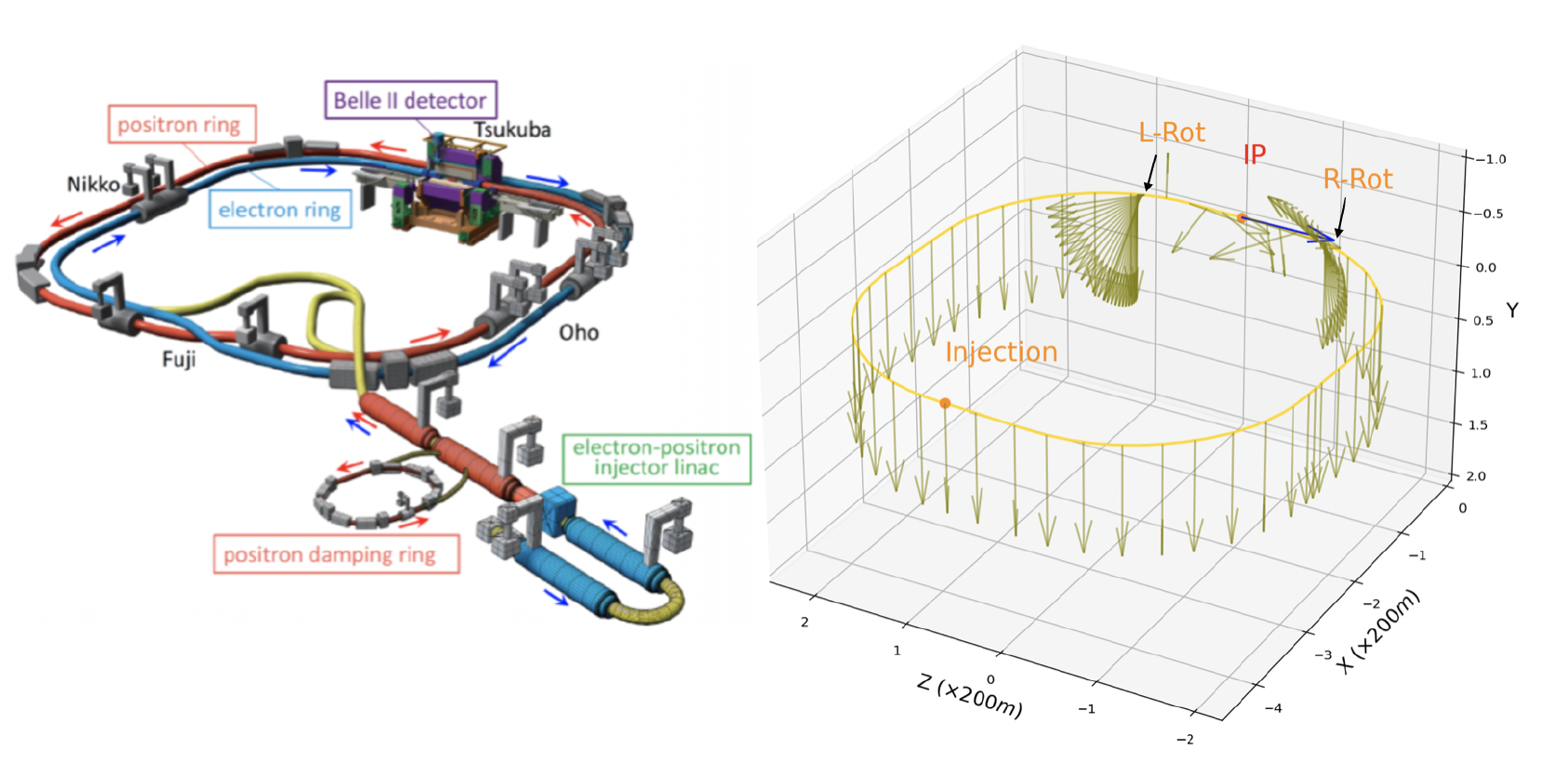}
\caption{The spin motion of the electron (KEK frame) in the HER with the spin rotator installed}
\label{KEK}
\end{figure}

With the rematching completed, we now turn to examining the full HER with the spin rotators - the `Rotator Ring'.
Figure~\ref{Rotring} shows the comparison of the entire lattice of the HER, and the optical functions are well preserved outside the rotator region. 
For example, the $\beta_y^*$ values for both the original and rotator ring are the same, 1~mm. The only noticeable change of the ring parameters in Figure~\ref{rp} is the vertical emittance, which is 
higher than the original but still smaller than the current design 12.9 pm \cite{kek}. These comparisons indicate that transparency is successfully achieved; the impact of installing the spin rotator is minimized to an acceptable level. The single-particle tracking result shows that the longitudinal polarization at the IP reaches a significantly high level, greater than 99.99\%, as shown in Table~\ref{sp}, and vertical polarization is achieved at the entrance of the L-Rot and exit of the R-Rot, which satisfies the polarization purpose. Figure~\ref{KEK} shows the spin motion of the electron in the KEK frame with the spin rotator installed in the HER. The IP polarization depends on the injection point as the injected beam is polarized aligned with the dipole field, leading to right-handed longitudinal polarization at the IP  and anti-aligned with the dipole field, leading to left-handed longitudinal polarization. 

\begin{table}[htb!]
	\label{}
    \begin{center}
    \begin{tabular}{ |c||c|c|} 
    \hline
    Machine Parameter & Original Ring& Rot Installed \\
    \hline 
    Tune $Q_x$ & 45.530994 & 45.530994   \\
    Tune $Q_y$ & 43.580709 & 43.580709 \\
    Chromaticity $\xi_x$ & 1.593508 &1.593508  \\
    Chromaticity $\xi_y$ & 1.622865 &1.622865 \\
    Damping partition $J_x$ & 1.000064  & 0.984216\\
    Damping partition $J_y$ & 1.000002 & 1.005266\\
    Emittance $\varepsilon_x$ (m) & $4.44061\times 10^{-9}$ & $4.89628\times 10^{-9}$\\
    Emittance $\varepsilon_y$ (m)& $5.65367 \times 10^{-13}$ & $3.96631 \times 10^{-12}$ \\
    \hline 
    \end{tabular}
    \caption{Comparison of the ring parameters after matching the Tunes and Chromaticities }
    \label{rp}
    \end{center}
\end{table}

\begin{table}[htb!]
	\label{}
    \begin{center}
    \begin{tabular}{ |c||c|c|c| } 
    \hline
    Spin Component & Entrance of the L-Rot & IP & Exit of the R-Rot\\
    \hline     
    X & -0.0000450734 &0.0000066698 &  0.0000538792 \\
    Y &  0.9999999959 &0.0000926945 &  0.9999999959 \\
    Z & -0.0000788085 &0.9999999957 & -0.0000728110 \\
    \hline
    \end{tabular}
    \caption{Single particle spin tracking result of the electron in the rotator ring }
    \label{sp}
    \end{center}
\end{table}

\subsubsection{Tracking Studies and Next Steps}
Having achieved a first-order BMAD spin rotator solution using this approach of replacing four existing dipoles with spin rotators, the next step is to consider the non-linear effects using long-term tracking studies.  
Two goals of these studies:  1)investigate the dynamic aperture and re-tune sextupoles to achieve the maximum dynamic aperture; 2)determine the beam lifetime and polarization lifetime. The beam tracking will be implemented by the Long Term Tracking \cite{LTT} program built upon BMAD with the Polymorphic Tracking Code (PTC) interface. 
The R\&D effort on this combined function spin rotator conceptual design has shown it to be promising and will be continue to be pursued with these long-term tracking studies, along with R\&D on how to implement the magnet realization using direct wind technologies, such as those available at BNL.

\begin{table}[htb!]
	\label{}
    \begin{center}
    \begin{tabular}{ |c||c|c|c|c|c| } 
    \hline
    Quads & L(m) & $k_1$L (Original) & $k_1$L (L-Rot) & $\text{B}_1$(Original) T/m& $\text{B}_1$(L-Rot) T/m \\
    \hline \hline
    QD3E &	0.826 &	-0.175 & -0.177 & -4.948 & -5.012 \\
    QF4E &	1.015 &	0.035 &	0.071 &	0.805 &	1.633 \\
    QEAE &	0.826 &	0.183 &	0.175 &	5.178 & 4.961 \\
    QD5E &	0.826 &	-0.179 & -0.286 & -5.074 & -8.079 \\
    QF6E &	0.557 &	0.163 &	0.342 & 6.855 & 14.366 \\
    QF2E &	0.557 &	0.192 &	0.145 & 8.050 & 6.067 \\
    QD1E &	1.015 &	-0.255 & -0.203 & -5.868 & -4.682 \\
    \hline 
    \end{tabular}
    \caption{Ring quadrupoles used for matching the optics in the L-Rot tuning region. The quadrupole magnet labels are those used in the SAD lattice file.}
    \label{Lq}
    \end{center}
\end{table}

\begin{table}[htb!]
	\label{}
    \begin{center}
    \begin{tabular}{ |c||c|c|c|c| } 
    \hline
    Skew-Quad & L(m)  & $k_1$ L & $\text{B}_1$ (T/m) & Tilt (rad)\\
    \hline \hline
    B2EALSQ1 	&	0.9837	&	0.511	&	12.133	&	-0.426	\\
    B2EALSQ2 	&	0.9837	&	0.510	&	12.130	&	1.053	\\
    B2EALSQ3 	&	0.9837	&	-0.314	&	-7.457	&	-0.988	\\
    B2EALSQ4 	&	0.9837	&	0.855	&	20.315	&	0.030	\\
    B2EALSQ5 	&	0.9837	&	0.688	&	16.350	&	-0.630	\\
    B2EALSQ6 	&	0.9837	&	0.814	&	19.340	&	1.383	\\
    \hline
    B2EBLSQ1 	&	0.9837	&	0.558	&	13.266	&	0.651	\\
    B2EBLSQ2 	&	0.9837	&	-0.482	&	-11.444	&	0.992	\\
    B2EBLSQ3 	&	0.9837	&	0.426	&	10.119	&	-1.494	\\
    B2EBLSQ4 	&	0.9837	&	0.338	&	8.024	&	-0.931	\\
    B2EBLSQ5 	&	0.9837	&	0.562	&	13.359	&	0.735	\\
    B2EBLSQ6 	&	0.9837	&	-0.185	&	-4.404	&	0.868	\\
    \hline 
    \end{tabular}
    \caption{Skew-qaudrupoles in the L-Rot.}
    \label{Lsq}
    \end{center}
\end{table}

\begin{table}[htb!]
	\label{}
    \begin{center}
    \begin{tabular}{ |c||c|c|c|c|c| } 
    \hline
    Quads & L(m) & $k_1$L (Original) & $k_1$L (R-Rot) & $\text{B}_1$(Original) T/m& $\text{B}_1$(R-Rot) T/m \\
    \hline \hline
    QD5E 	&	0.826	&	-0.179	&	-0.165	&	-5.074	&	-4.667	\\
    QEAE	&	0.826	&	0.183	&	0.154	&	5.178	&	4.362	\\
    QF4E	&	1.015	&	0.035	&	0.067	&	0.805	&	1.538	\\
    QD3E	&	0.826	&	-0.175	&	-0.251	&	-4.948	&	-7.088	\\
    QF2E	&	0.557	&	0.192	&	0.183	&	8.050	&	7.659	\\
    QD1E 	&	1.015	&	-0.255	&	-0.274	&	-5.868	&	-6.311	\\
    QLA10RE	&	0.826	&	0.202	&	0.185	&	5.718	&	5.234	\\
    QLA9RE	&	0.826	&	-0.237	&	-0.226	&	-6.703	&	-6.385	\\
    QLA8RE  &	0.557	&	0.203	&	0.169	&	8.527	&	7.106	\\
    QLA7RE 	&	0.826	&	-0.192	&	-0.195	&	-5.438	&	-5.522	\\
    QLA6RE	&	0.826	&	0.202	&	0.205	&	5.716	&	5.808	\\
    \hline
    \end{tabular}
    \caption{Ring quadrupoles used for matching the optics in the R-Rot tuning region. The quadrupole magnet labels are those used in the SAD lattice file.}
    \label{Rq}
    \end{center}
\end{table}

\begin{table}[htb!]
	\label{}
    \begin{center}
    \begin{tabular}{ |c||c|c|c|c| } 
    \hline
    Skew-Quad & L(m)  & $k_1$L & $\text{B}_1$ (T/m) & Tilt (rad)\\
    \hline \hline
    B2EARSQ1 	&	0.9837	&	0.435	&	10.341	&	-2.610	\\
    B2EARSQ2 	&	0.9837	&	0.600	&	14.258	&	2.290	\\
    B2EARSQ3 	&	0.9837	&	0.043	&	1.032	&	2.328	\\
    B2EARSQ4 	&	0.9837	&	-0.566	&	-13.451	&	-0.180	\\
    B2EARSQ5 	&	0.9837	&	0.600	&	14.258	&	-2.545	\\
    B2EARSQ6 	&	0.9837	&	-0.591	&	-14.038	&	0.618	\\
    \hline
    B2EBRSQ1 	&	0.9837	&	0.495	&	11.769	&	-2.480	\\
    B2EBRSQ2 	&	0.9837	&	0.532	&	12.648	&	2.238	\\
    B2EBRSQ3 	&	0.9837	&	0.280	&	6.663	&	-0.960	\\
    B2EBRSQ4 	&	0.9837	&	-0.565	&	-13.429	&	-0.197	\\
    B2EBRSQ5 	&	0.9837	&	0.600	&	14.258	&	-2.846	\\
    B2EBRSQ6 	&	0.9837	&	-0.383	&	-9.098	&	0.475	\\
    \hline 
    \end{tabular}
    \caption{Skew-quadrupoles of the R-Rot.}
    \label{Rsq}
    \end{center}
\end{table}

\begin{table}[htb!]
	\label{}
    \begin{center}
    \begin{tabular}{ |c||c|c|c| } 
    \hline
Name 	&	L (m)	&	b2 (original)	&	b2 (Rot)\\
    \hline
SD3TLE	&	1.030	&	-3.577	&	-3.789	\\
SF6TLE	&	0.334	&	0.818	&	0.869	\\
SD7TLE	&	1.030	&	-3.607	&	-3.819	\\
SF8TNE	&	0.334	&	1.751	&	1.554	\\
SD7NRE	&	1.030	&	-4.582	&	-4.788	\\
SF6NRE	&	0.334	&	1.467	&	1.539	\\
SD5NRE	&	1.030	&	-1.389	&	-1.573	\\
SF4NRE	&	0.334	&	2.092	&	2.175	\\
SD3NRE	&	1.030	&	-1.443	&	-1.628	\\
SF2NRE	&	0.334	&	0.371	&	0.403	\\
SF2NLE	&	0.334	&	0.077	&	0.109	\\
SD3NLE	&	1.030	&	-3.070	&	-3.281	\\
SF4NLE	&	0.334	&	0.497	&	0.535	\\
SD5NLE	&	1.030	&	-1.527	&	-1.714	\\
SF6NLE	&	0.334	&	0.660	&	0.705	\\
SD7NLE	&	1.030	&	-1.537	&	-1.724	\\
SD7FRE	&	0.334	&	-5.461	&	-5.652	\\
SF6FRE	&	0.334	&	2.296	&	2.384	\\
SD5FRE	&	1.030	&	-6.803	&	-6.954	\\
SF4FRE	&	0.334	&	0.691	&	0.737	\\
SD3FRE	&	1.030	&	-1.903	&	-2.099	\\
SF2FRE	&	0.334	&	1.226	&	1.289	\\
SF2FLE	&	0.334	&	0.856	&	0.897	\\
SD3FLE	&	1.030	&	-1.359	&	-1.542	\\
SF4FLE	&	0.334	&	0.541	&	0.581	\\
SD5FLE	&	1.030	&	-2.926	&	-3.136	\\
SF6FLE	&	0.334	&	2.260	&	2.353	\\
SD7FLE	&	1.030	&	-6.909	&	-7.055	\\
SF8FOE	&	0.334	&	1.871	&	1.770	\\
SD7ORE	&	1.030	&	-7.242	&	-7.375	\\
SF6ORE	&	0.334	&	0.217	&	0.245	\\
SD5ORE	&	1.030	&	-2.833	&	-3.043	\\
SF4ORE	&	0.334	&	1.686	&	1.761	\\
SD3ORE	&	1.030	&	-3.123	&	-3.335	\\
SF2ORE	&	0.334	&	0.362	&	0.397	\\
SF2OLE	&	0.334	&	2.296	&	2.384	\\
SD3OLE	&	1.030	&	-0.706	&	-0.868	\\
SF4OLE	&	0.334	&	0.585	&	0.628	\\
SD5OLE	&	1.030	&	-2.483	&	-2.689	\\
SF6OLE	&	0.334	&	0.415	&	0.435	\\
SD7OLE	&	1.030	&	-3.385	&	-3.598	\\
SF8OTE	&	0.334	&	0.353	&	0.216	\\
SD7TRE	&	1.030	&	-1.730	&	-1.921	\\
SF6TRE	&	0.334	&	0.829	&	0.876	\\
SD5TRE	&	1.030	&	-1.695	&	-1.885	\\
    \hline
    \end{tabular}
    \caption{Sextupoles pairs adjusted to match the chromaticity. The sextupole magnet labels are those used in the SAD lattice file.}
    \label{sext}
    \end{center}
\end{table}

\clearpage
\begin{figure}[htb!]
\centering
\includegraphics[width = 3 in]{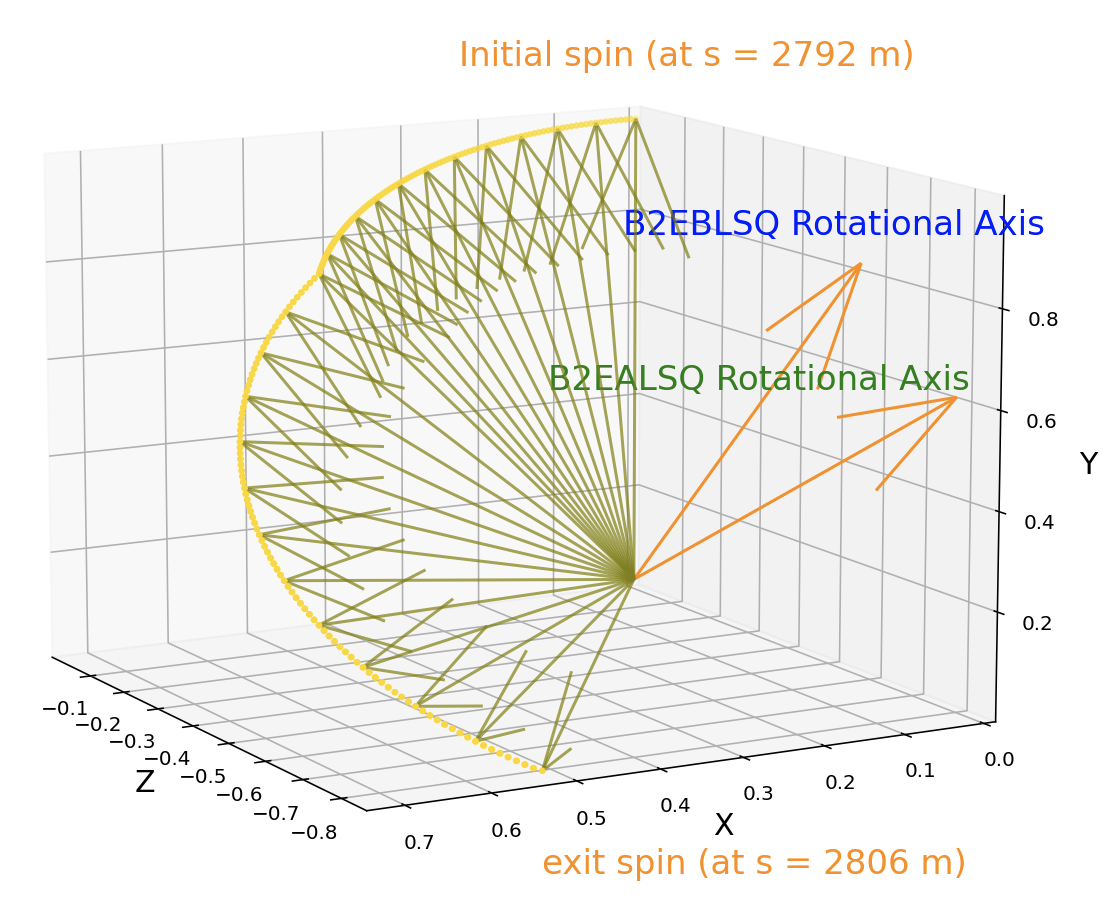}
\includegraphics[width = 3 in]{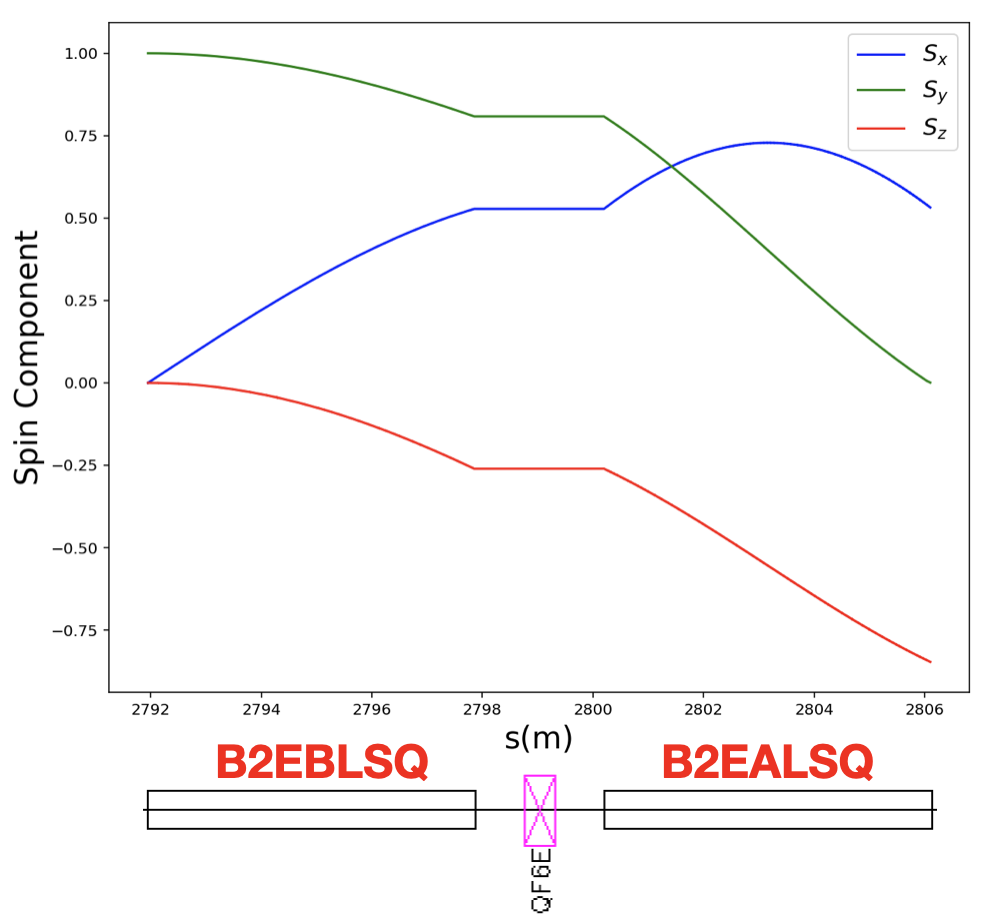}
\caption{Left: Spin motion of the electron in the L-Rot (rest frame).
Right:Spin motion of the electron in the L-Rot (co-moving frame). }
\end{figure}

\begin{figure}[htb!]
\centering
\includegraphics[width = 3 in]{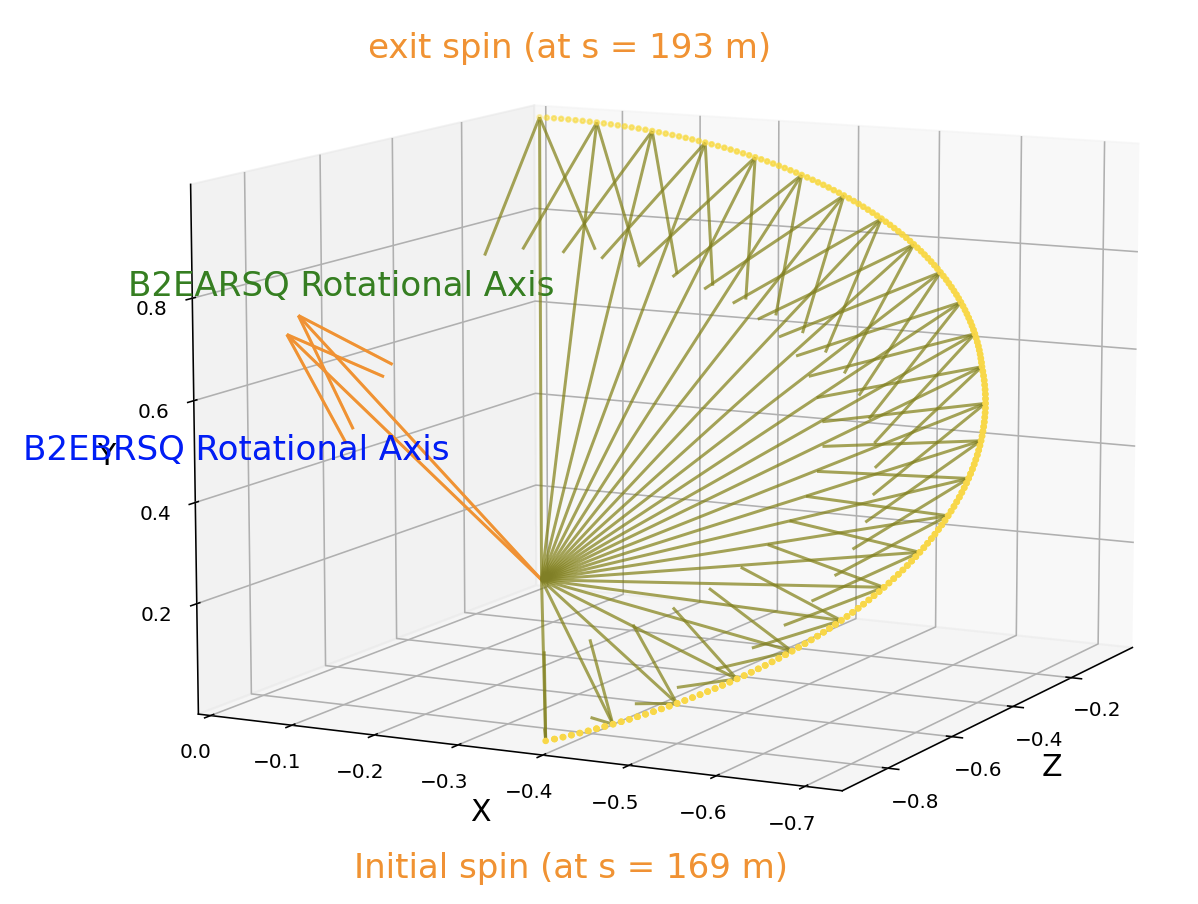}
\includegraphics[width = 3 in]{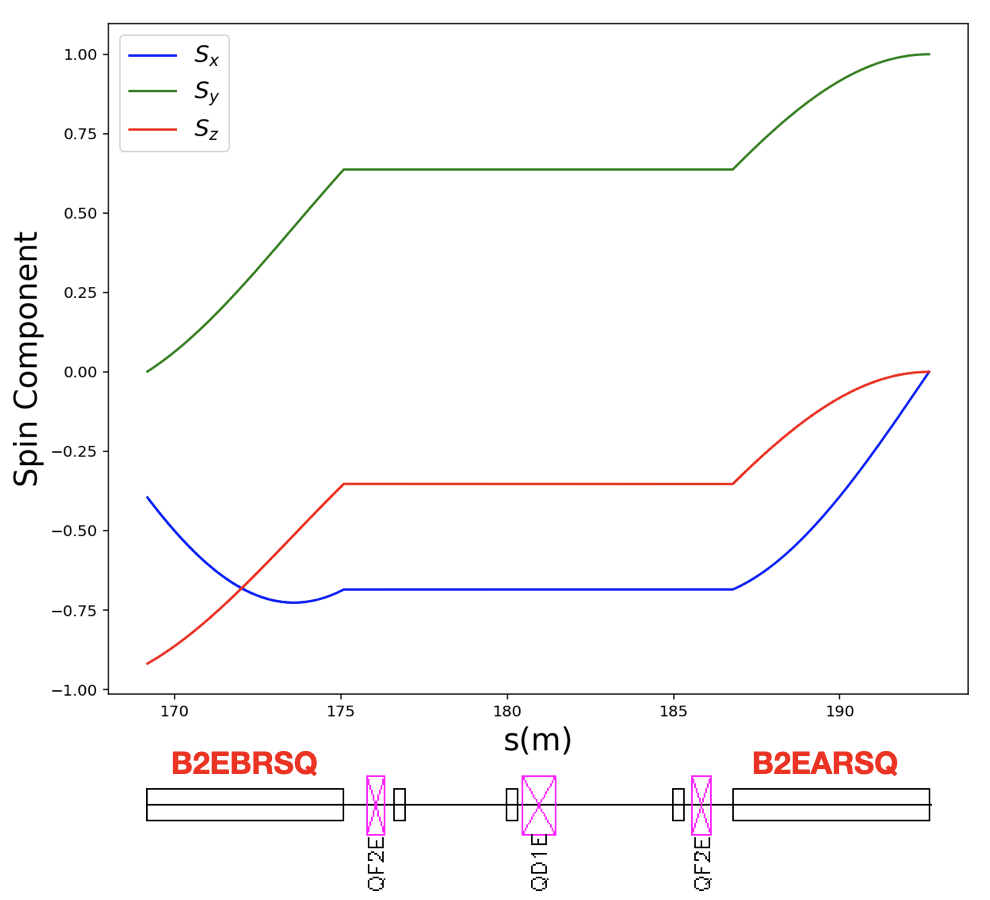}
\caption{Left: Spin motion of the electron in the R-Rot (rest frame). Right: Spin motion of the electron in the R-Rot (co-moving frame). }
\end{figure}

\begin{figure}[htb!]
\centering
\includegraphics[width = 3 in]{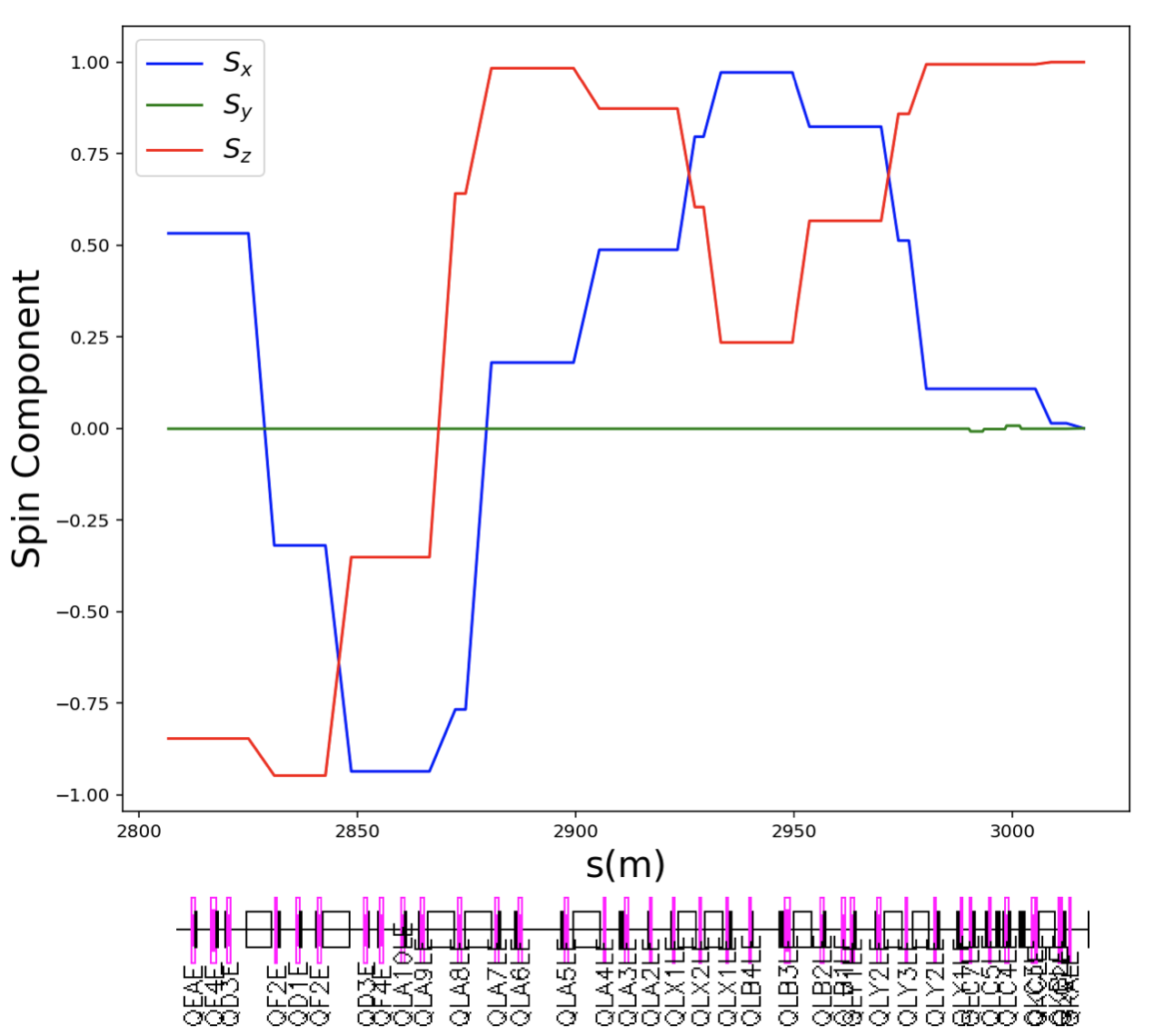}
\includegraphics[width = 3 in]{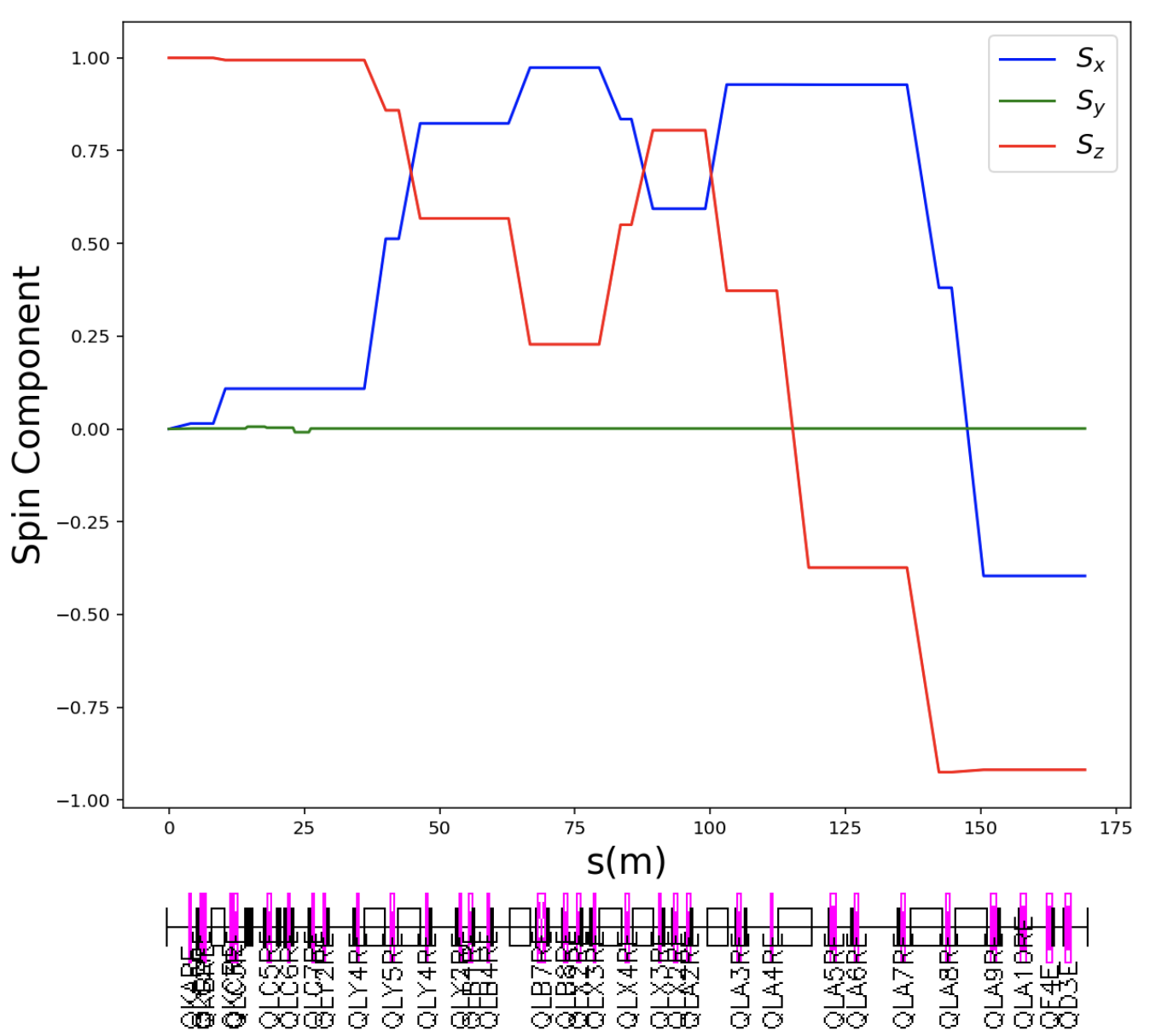}
\caption{Left: Spin motion of the electron (co-moving frame) in the area between the L-Rot and the IP at s $\sim$ 3016~m. Right: Spin motion of the electron in the area between the IP (at s = 0 m)  and the R-Rot (co-moving frame).  },
\end{figure}

\begin{figure}[htb!]
\centering
\includegraphics[width = 6 in]{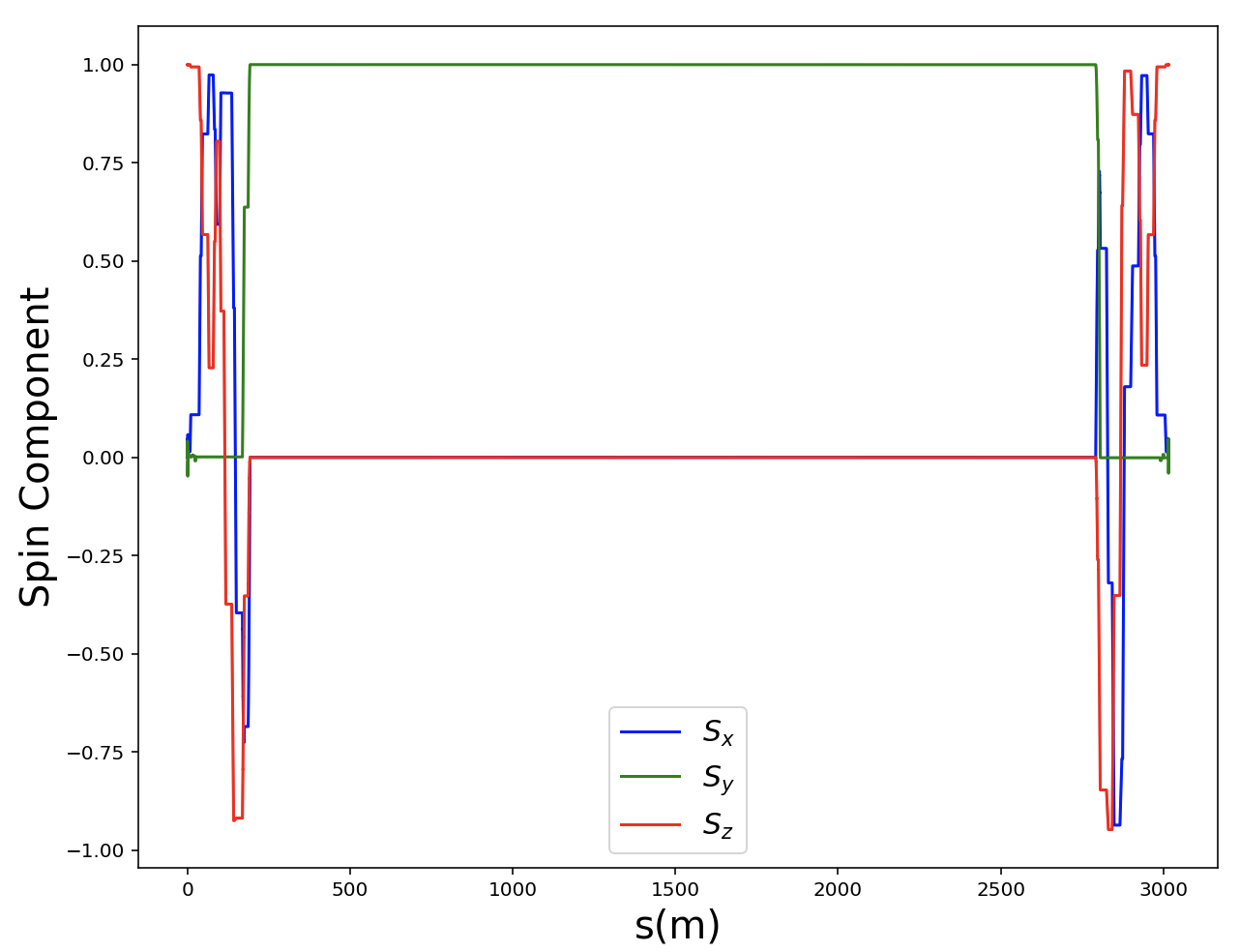}
\caption{Spin motion of the electron in the Rotator Ring (co-moving frame): Ip at s = 0 and 3016 m}
\end{figure}

%% file: DirectWindSpinRotator.tex
\par
 
The seminal concept for the compact spin rotator module invented by U. Wienands' is illustrated in Figure~\ref{fig:spin_rotator_fig}. The main idea is to start with a superconducting dipole coil having a field profile that matches that of an existing SuperKEKB High Energy Ring (HER) arc dipole. Then by overlaying the field from a 4.5~T solenoid, the electron spin direction starts to rotate away from its stable, dipole field anti-parallel, orientation in the rest of the HER. By using pairs of spin rotator units located immediately upstream and then downstream of the Belle~II IP in SuperKEKB, we can bring the spin to longitudinal and subsequently return it back to the anti-parallel direction. The spin rotator unit shown in Figure~\ref{fig:spin_rotator_fig} includes additional skew-quadrupole coils to compensate locally for undesirable coupling effects introduced by the solenoidal field. The gradients of the HER lattice quadrupoles around the spin rotator units are adjusted so as to re-match the present HER lattice optics. In this manner the replacement of HER warm dipoles by the spin rotator units is ``invisible" to the rest of the HER optics. \par
 
This spin rotator concept is especially attractive for use in an operating accelerator complex such as SuperKEKB because we minimize the change to the existing ring layout in the tunnel and disruption of SuperKEKB accelerator running. A standard spin rotator solution typically uses isolated solenoids separated by dipole bends designed to provide ``magic angles" between the solenoids for proper spin rotation. But such a standard solution eats up precious lattice space for the solenoids and at a minimum requires shifting accelerator components in the tunnel over extended sections. Our proposed multi-function spin rotator solution minimizes the number of components that have to be physically changed or moved and maintains the present HER geometry, so it should be possible to make these changes in a standard shutdown and with lowest possible project cost.  \par

In order to accomplish these goals our spin rotator modules must
\begin{itemize}
  \item satisfy accelerator physics and beam optics requirements,
  \item have reasonable superconducting coil magnetic designs,
  \item be housed in an appropriate cryostat with a suitable mechanical structure,
  \item and have low-enough heat leak that the units can be operated each with a small number of local cryocoolers within the existing SuperKEKB tunnel.
\end{itemize} \par
 
\begin{figure}[htb!]
\centering
  \includegraphics{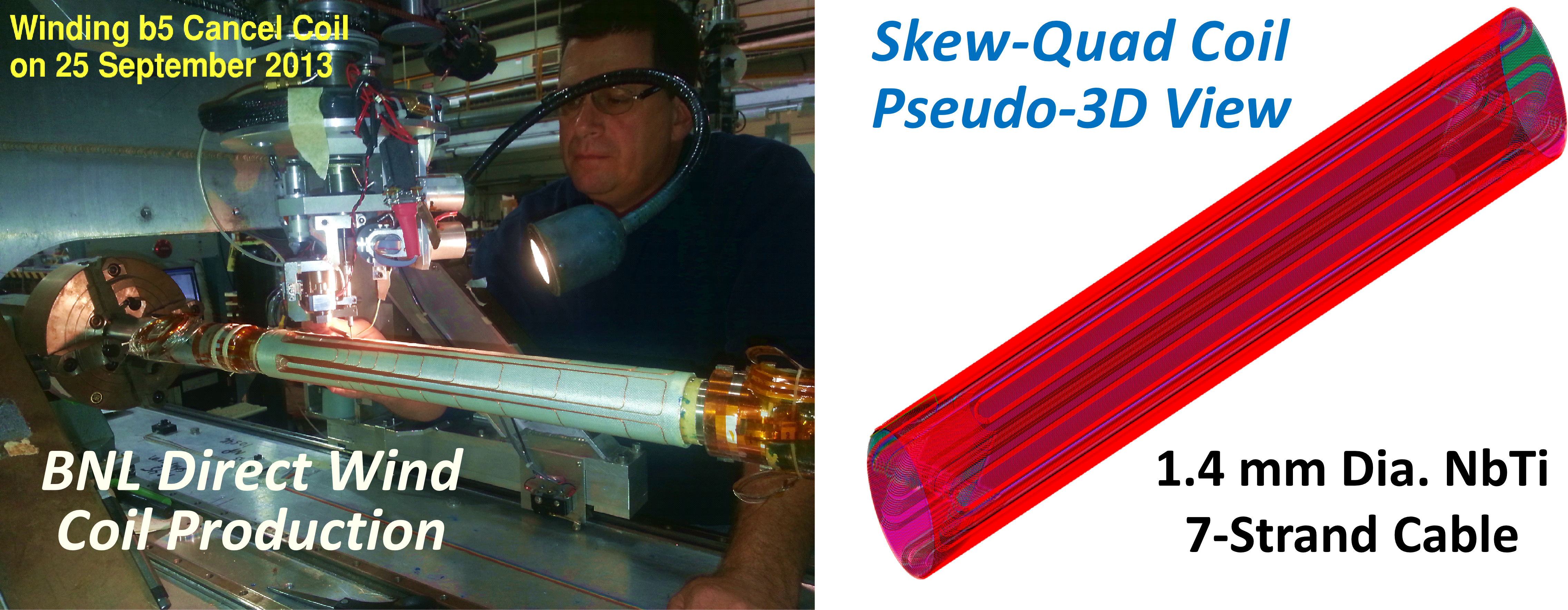}
  \caption{Direct Wind Coil Production Example. Photo on the left shows one layer of the SuperKEKB external field cancel coil during production on the BNL computer controlled winding machine. For comparison a conceptual coil winding pattern for a spin rotator skew-quadrupole is shown on the right.}
  \label{fig:direct_wind_fig}
\end{figure}

The compact spin rotator modules needed for IP longitudinal polarization will take advantage of BNL Direct Wind production technology to make overlapping dipole, solenoidal and skew-quadrupole coils according to HER optics requirements. Direct Wind coil production involves temporarily bonding superconducting wire or round cable to a support tube as shown in Figure~\ref{fig:direct_wind_fig} with a picture of one layer of the SuperKEKB b5 external field cancel coil during winding next to a rendition of the skew-quadrupole coil pattern needed for our spin rotator modules. Note for the Direct Wind process, once the desired coil pattern is laid down under computer control any spaces or gaps in the winding pattern are filled in with a combination of fiberglass and epoxy and then the coil structure is wrapped with tensioned fiberglass roving to provide coil prestress against the local electromagnetic Lorentz force when the coils are cooled down and energized. \par

\begin{figure}[htb!]
\centering
  \includegraphics[width = 3 in]{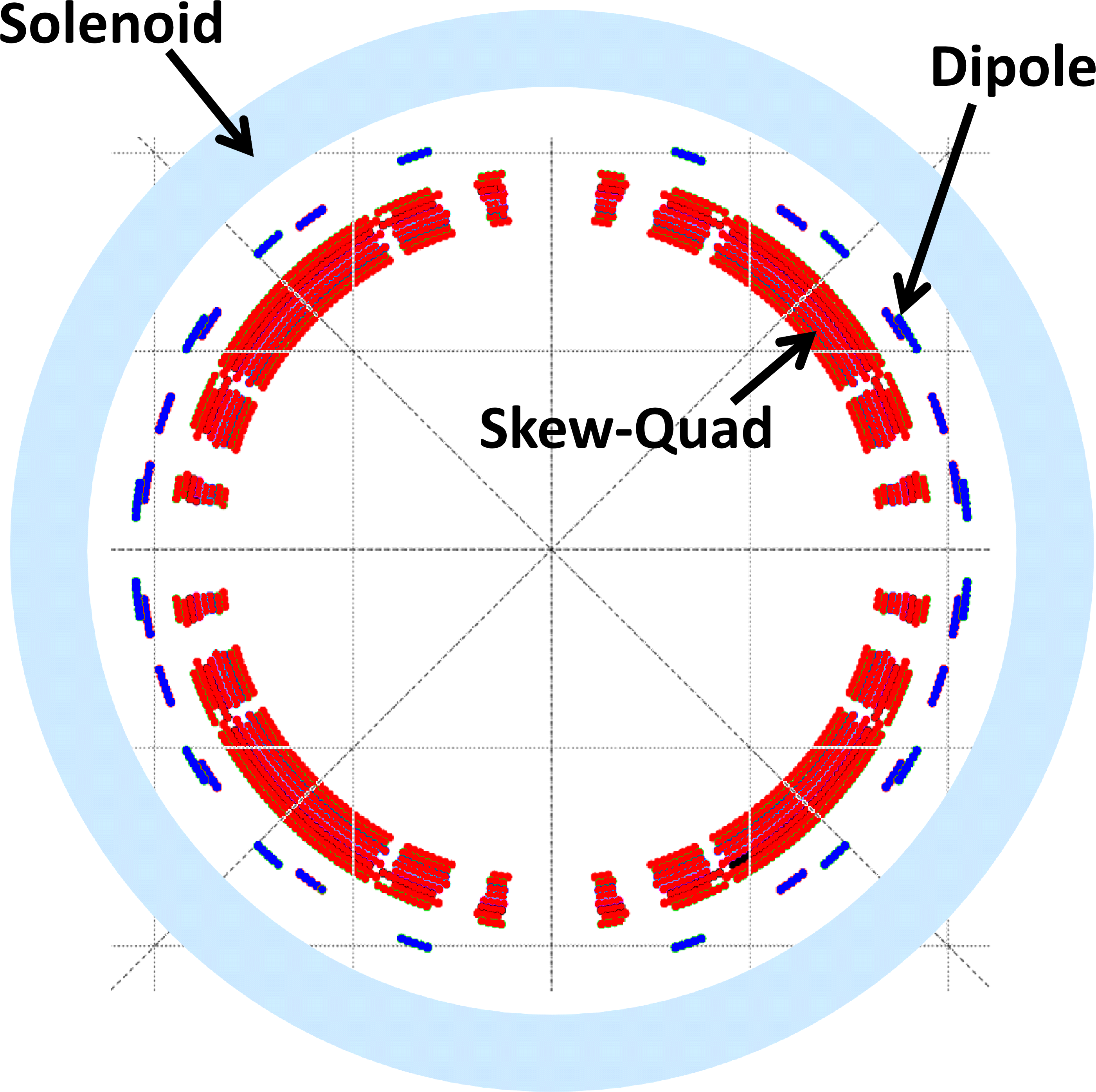}
  \caption{Conceptual Cross Section for the Proposed Multi-function Spin Rotator Coil Structure Utilizing BNL Direct Wind Technology. In this design the skew-quadrupole coil radius was minimized in order to reduce the coil magnetic peak field at the superconductor.}
  \label{fig:coil_cross_section_fig}
\end{figure}

In this manner we build up the desired multi-function superconductor coil structure shown schematically in Figure \ref{fig:coil_cross_section_fig} (for simplicity intermediate coil support tubes, cold mass containment and cryostat walls are not shown). Note that the four spin rotator modules are operated with different combinations of individual coil magnetic strengths but for the worst case, e.g. solenoid field 4.85~T, skew-quadrupole gradient 24~T/m and 0.2~T dipole field, the combined field at the skew-quadrupole conductor is 6.15 T which corresponds to 69\% of the predicted short sample quench value at a 729~A  operating current for the preliminary configuration shown and the assumed 1.4~mm diameter 7-strand cable (at 4.22~K). While these numbers are subject to later optimization once we have a chance to do a more careful and detailed system conceptual study, this first pass analysis indicates that the proposed magnetic design is reasonable and we do not find any obvious show stoppers. \par
 
\begin{figure}[htb!]
\centering
  \includegraphics[width = 3 in]{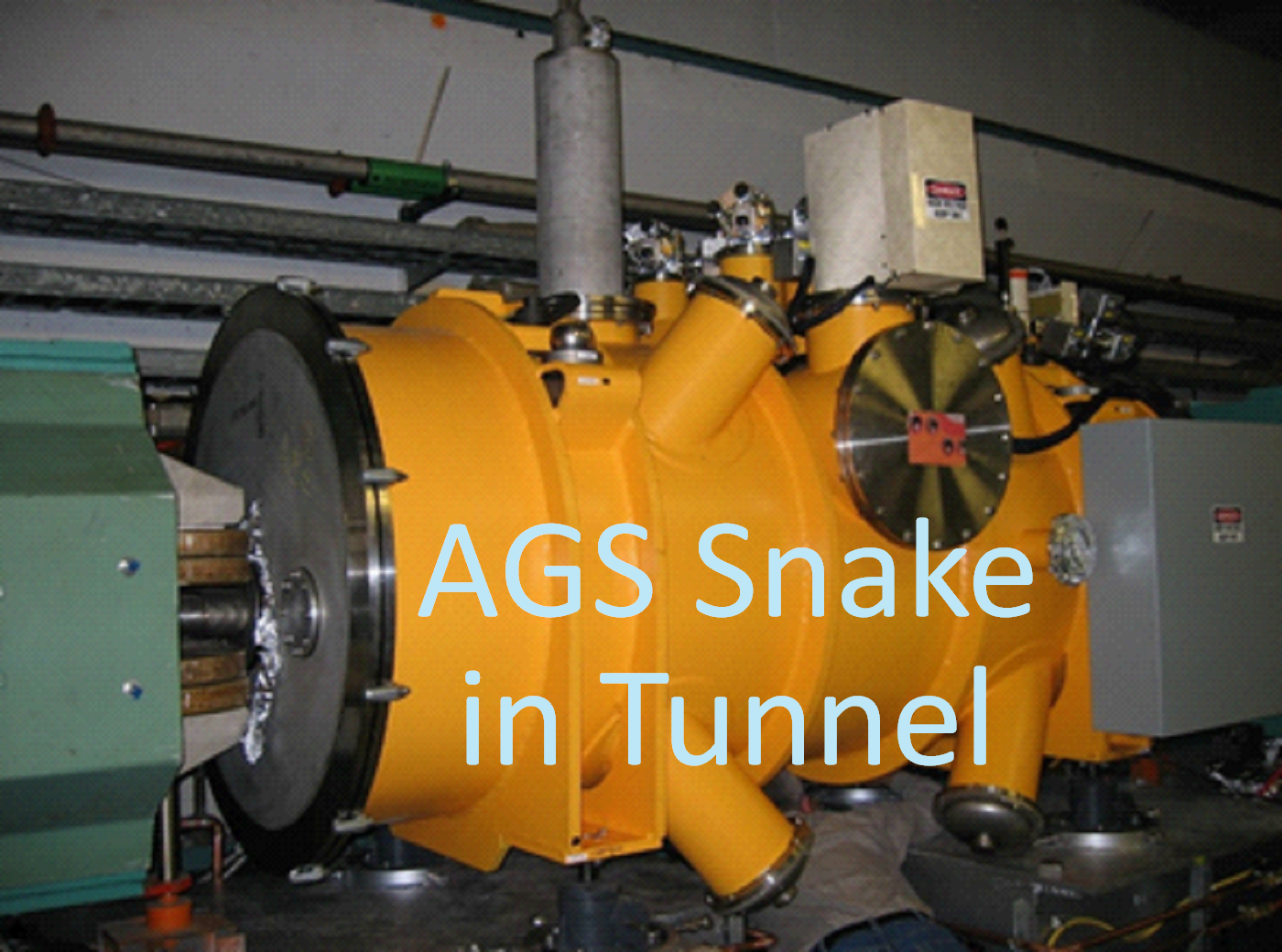}
  \caption{View of the BNL Manufactured Superconducting Snake Module Installed in the AGS Tunnel. The AGS Snake superconducting coil structure is housed in a standalone cryostat that is sandwiched in an otherwise normal conducting magnet accelerator lattice. This system uses local cryocooler based cooling along with a low heat leak support structure.}
  \label{fig:ags_snake_fig}
\end{figure}

\begin{figure}[htb!]
\centering
  \includegraphics[width = 3 in]{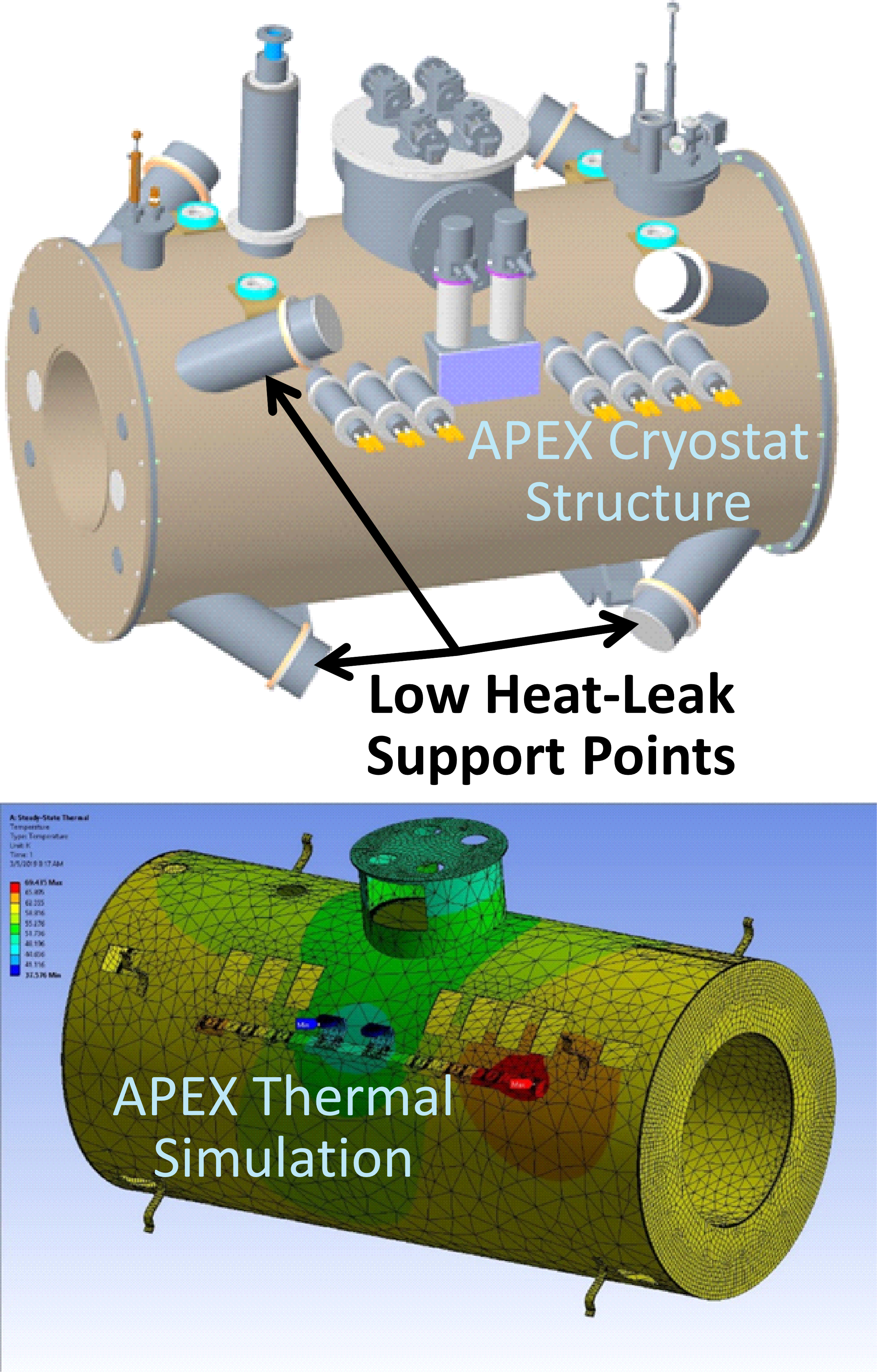}
  \caption{Cryostat Mechanical and Thermal Modeling Example. The SuperKEKB spin rotator modules will require optimization of the cryostat design with features similar to those shown here for the Oak Ridge APEX magnets. The required cryocooler capacity depends upon the thermal load from the current leads and the low heat leak support structure with intermediate temperature heat shields. The spin rotator cryostat needs a warm bore to accommodate inserting the HER beam pipe with features for synchrotron radiation heat load water cooling and beam vacuum pumping.}
  \label{fig:apex_fig}
\end{figure}

Note in order to set the coil radial buildup assumed in Figure \ref{fig:coil_cross_section_fig} we needed to know the radius of the inner cryostat wall and this in turn depends upon both the space required to accommodate a warm HER beam pipe with features to handle synchrotron radiation and vacuum as well as the space needed for an intermediate cryostat heat shield. Since the spin rotator locations are far from existing cryogenic supplies, the cryostat units must be self-contained with integrated current leads and cryocoolers similar to what is shown in Figure \ref{fig:ags_snake_fig} for the BNL AGS Snake Magnet. 
The specification of cryocooler capacity depends upon knowledge of the heat leak simulated using the cryostat inner cold mass support structure and the overall system design, as was done for the Oak Ridge APEX magnet system shown in Figure \ref{fig:apex_fig}. 
 Finally, we must show that the spin rotator module design satisfies KEK/Japanese safety and tunnel installation requirements (e.g. requirements that depend upon the net helium volume). \par
 
In summary the main spin rotator challenge that needs to be addressed is to develop a realistic, fully self-consistent conceptual design for our proposed system. To this end we submitted a proposal, ``R\&D for a New Belle-II Era of Polarization Physics at SuperKEKB'', for the required R\&D under the U.S.-Japan Science and Technology Cooperation Program in High Energy Physics. PIs in Japan and at BNL will coordinate setting spin rotator system requirements for the conceptual design to be developed. It is anticipated that the requested US/Japan support will act as seed funding and form an initial basis for requesting future international project funding. Assuming the research and development work demonstrates that a Direct Wind magnet spin rotator can be realized in SuperKEKB while maintaining high luminosity with a polarized source, and the polarization upgrade is approved by KEK, our Canadian collaborators, led by Michael Roney, intend to submit a request to the Canadian Foundation for Innovation (CFI) towards the construction of the spin rotator magnets at BNL. \par
 

 

%% file: ComptonPolarimetry.tex



\subsection{Introduction}
Compton backscattering is a process allowing one to measure accurately \cite{DENNER199958} the electron beam polarization at electron beam energies above 1~GeV. Indeed the cross-section of this process exhibits a polarization dependent behavior \cite{GINZBURG19845}. This technique has been successfully employed in the past at SLC \cite{199423}, at HERA \cite{BARBER199379,Baudrand_2010}, JLAB \cite{PhysRevX.6.011013} and was considered for SuperB \cite{SuperB:2013cxb}. In particular, it has enabled the demonstration and optimization of a high degree of polarization in the HERA ring \cite{BARBER1994166}. It is also considered in other projects as ILC \cite{Boogert_2009}, LHeC \cite{Abelleira_Fernandez_2012} and  EIC \cite{eicWS}. It is also considered for FCC-ee \cite{blondel2019polarization} in the context of energy calibration with resonant depolarization as it was done at LEP2 \cite{LEP2}. This project has large synergies  with HERA as well as with the EIC project, the latter being in the design phase.

The experimental setup consists of a circularly polarized laser beam which scatters off the electron beam. The photons are scattered within a narrow ($<$1mrad) cone around the tangent of the electron's trajectory at the interaction plane. A calorimeter can be implemented to measure the scattered gamma ray spectrum and/or a segmented electron counter placed after a dispersive element to measure the transverse electron distribution directly linked to their energy once the magnetic field is known. Both these distributions show sensitivity to the longitudinal electron beam polarization but with different sources of systematic uncertainties. It must be noted that vertical electron beam polarization may also be extracted but would require a vertically segmented detector since the sensitivity comes from spatial asymmetries in the measured energy spectra \cite{BARBER1994166}.

\subsection{Strategy}
The specificities of the SuperKEKB upgrade, and similarly of EIC, lie in the fact that no permanent regime is reached and that continuous top-up is realized to maintain a very high luminosity. According to these specificities,  and also following past experience \cite{Baudrand_2010} it will be necessary to measure the beam polarization for every bunch independently on a time scale similar to that of the top-up period in the SuperKEKB ring. It must be noted that this measurement cannot be made at the Belle~II interaction point itself, nor in a straight line to it due to too high backgrounds. This measurement of the polarization could be extrapolated to the Belle~II interaction point provided that the lattice and alignments are well known \cite{Beckmann_2014}. Two possibilities could be considered. The simpler one is to place the Compton interaction point relatively close to the interaction point but far enough from the polarization rotators such that the longitudinal projection electron beam polarization vector is sufficiently large. Alternatively, the polarimeter could be placed in a region of the ring where the polarization is nearly vertical. However in this case, the experience of HERA \cite{BARBER199379}, shows that alignments must be controlled within few tens of microns, in regions of the lattice where the vertical beam size, angular spread and dispersions are negligibly small. The longitudinal polarimeter is considered as a baseline solution at this stage of the project. 

This measurement may be complemented by measurements prior injection in the SuperKEKB ring since the beam lifetime is smaller than the polarization build-up time in the SuperKEKB ring and to avoid injection of badly polarized beams in the ring. In particular, it is sometimes valuable to implement Mott scattering technique at the injection for commissioning and troubleshooting purposes~\cite{doi:10.1063/1.1384229}. A dedicated short beam line right after the injector maybe built for this purpose. Additionally one may want to check that the high-energy beam polarization is correctly oriented with a non destructive manner, which could be easily implemented if a damping ring is used. This may be useful if it turns out that a bad polarization top-up injection cannot be extracted from real-time data of the main ring polarimeter. Detailed design of the latter is required to address this question.

\subsection{Main ring longitudinal polarimeter design}
\subsubsection{Location}
\begin{figure}[!ht]
\centering
\includegraphics[width=0.9\textwidth]{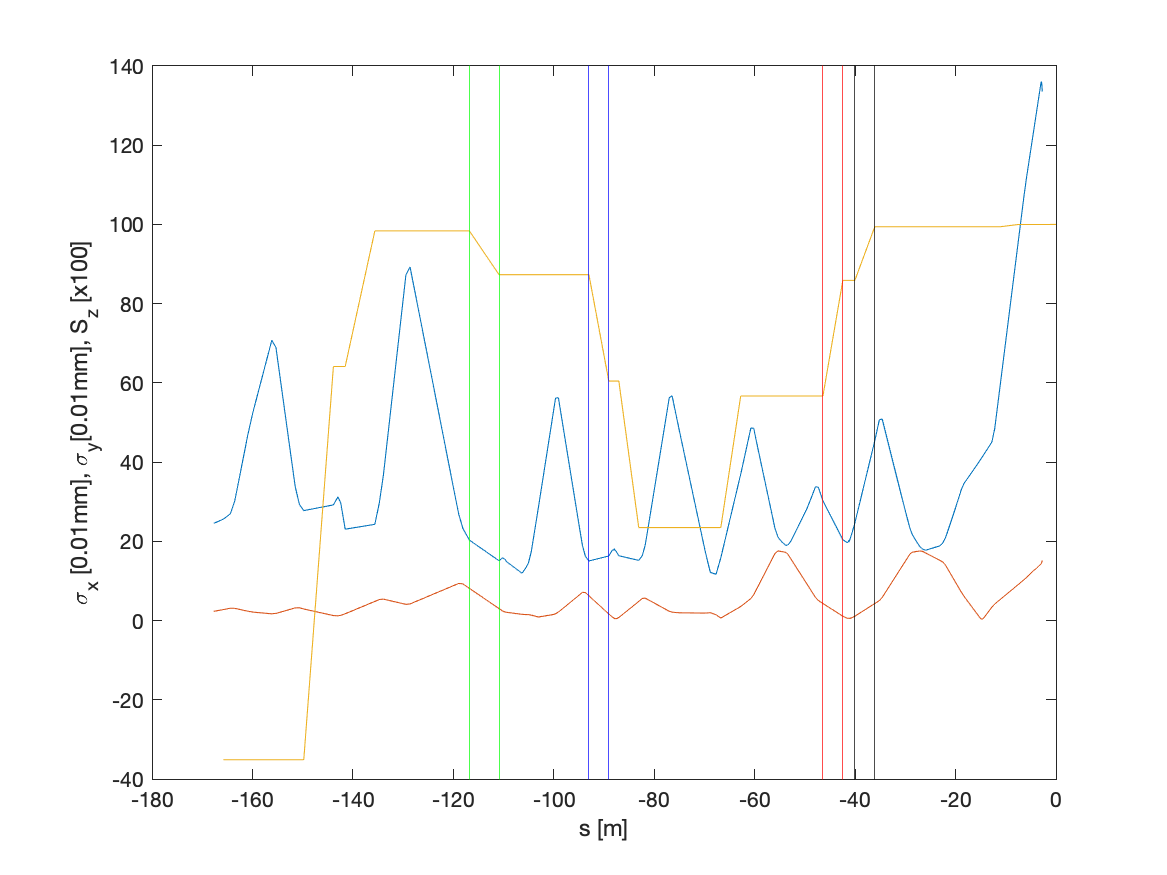}
\caption{Horizontal (blue) and vertical (red) beam size upstream the Belle II IP in the HER ring. The ratio of the longitudinal polarization to that at the Belle II IP is also given in yellow. Vertical bands locate the dipoles BLA2LE, BLX2LE.2, BLY2LE.2 and BLY2LE.1 in green, blue, red and black, respectively.}
\label{fig:beamsize}
\end{figure}

A location where the beam polarization is nearly parallel to that expected at the Belle II interaction point (IP) and as close as possible to it is looked for. The beam size and the ratio of the longitudinal polarization with respect to that expected at the IP is shown in Fig.~\ref{fig:beamsize} in this region. The several meters straight section before BLA2LE (represented by a green band in the figure) presents the advantage of providing a polarization vector aligned with that at the Belle II IP and that there is room for inserting the necessary elements for the Compton polarimeter. It is placed at about 130m upstream the IP in the HER ring.

\subsection{Laser system}
\begin{figure}[!ht]
\centering
\includegraphics[width=0.45\textwidth]{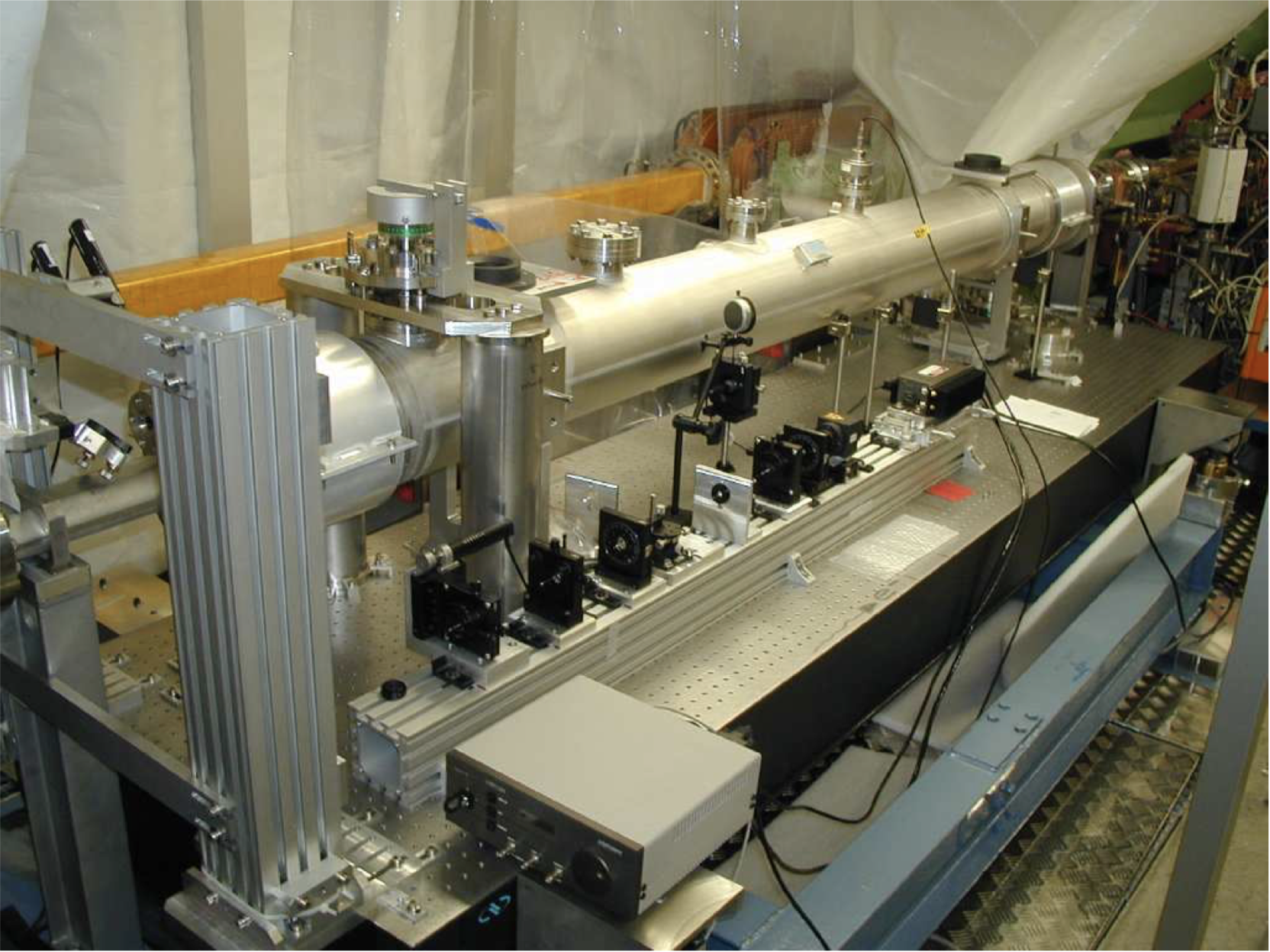}
\includegraphics[width=0.45\textwidth]{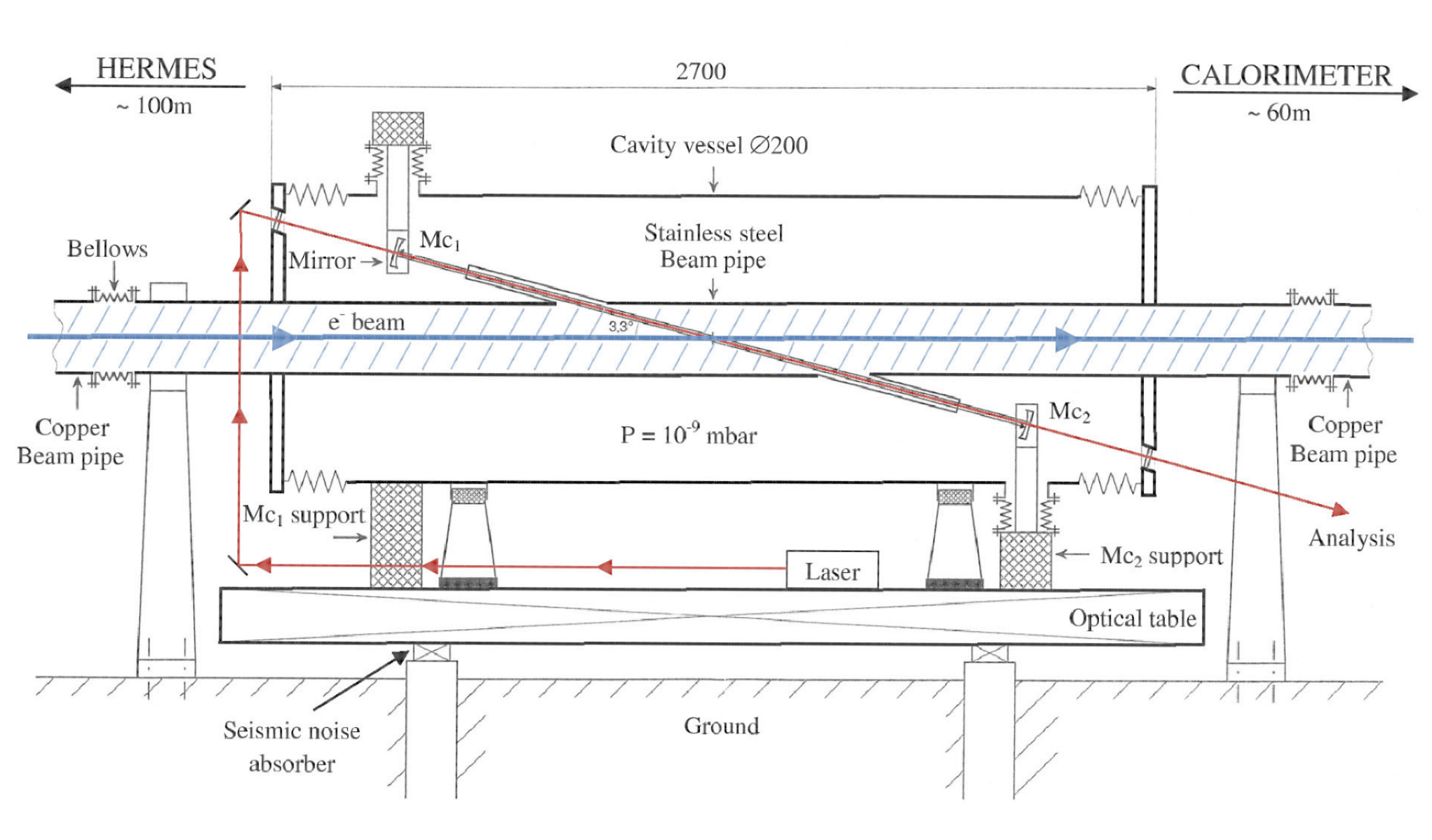}
\caption{(Left) picture of the LPOL2 cavity installed in HERA tunnel and (right) a schematic of its mechanical integration. No cavity mirrors are required for SuperKEKB but a similar mechanical system could be implemented. A particular was paid to the design of the vacuum pipe for beam impedance issues. Reproduced from Ref.~\cite{Baudrand_2010}.}
\label{fig:HERALPOL2cav}
\end{figure}

\begin{figure}[!ht]
\centering
\includegraphics[width=0.99\textwidth]{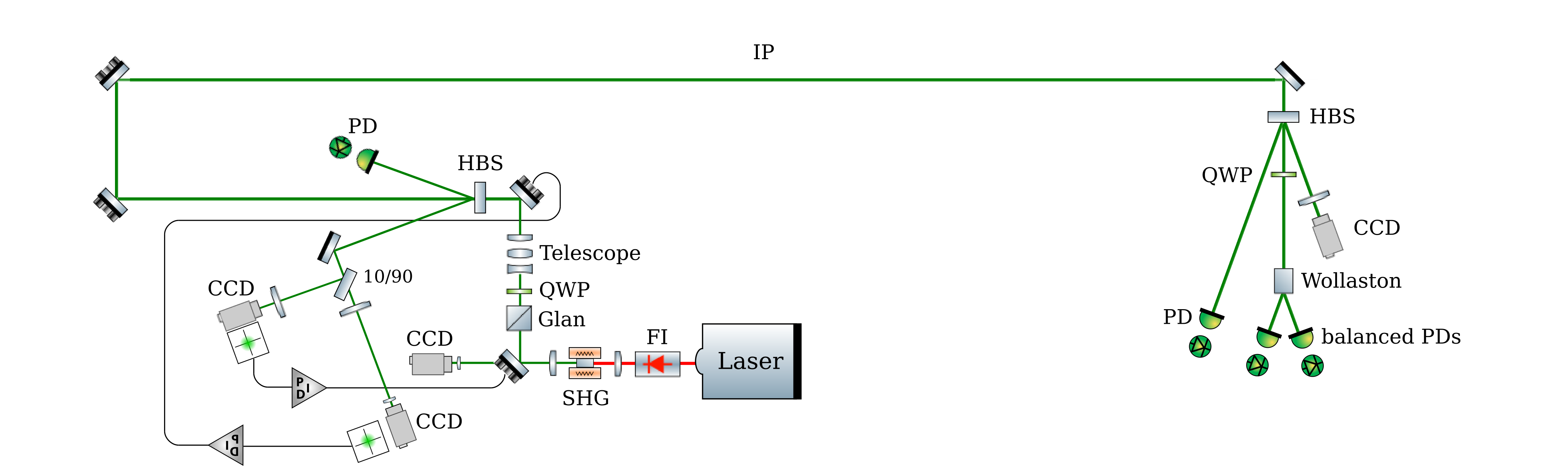}
\caption{Schematic describing the conceptual design of the laser system in case a \emph{green} laser is used. FI: Faraday isolator; SHG: second harmonic generation module; QWP: Quarter Wave plate; HBS: holographic Beam sampler allows to extract a small fraction of the beam power independently of the polarization; PD: photodiode. A position and pointing stabilization is integrated in the schematics.}
\label{fig:laserdesign}
\end{figure}

The laser system could be directly integrated in the accelerator in similar fashion as what was done for HERA LPOL2, see Fig.~\ref{fig:HERALPOL2cav}. The laser and and related diagnostics may be directly located below the beam pipe or close to it, assuming that proper shielding against radiations is implemented. The box in which this system will be embedded must be thermally controlled to limit any drifts that could affect the laser performances and particularly pointing and more importantly the laser polarization control that directly contributes as a systematic uncertainty. Industrial systems that can be locked onto the  accelerator reference clock delivering few Watts are commercially available. Such pulsed laser (frequency combs) systems are well suited since they naturally match the bunch pattern at 250MHz of SuperKEKB and may be sufficient to obtain one scattering in average per bunch crossing. These systems can be synchronized on an external reference clock.  This configuration is similar to that of HERA LPOL2 where real-time monitoring of the beam polarization was demonstrated \cite{Baudrand_2010}. The ability to operate the polarimeter in a regime of, in average,  one  scattering per bunch crossing needs to be confirmed with detailed estimates of background levels in the detectors. The choice of the laser wavelength, basically \emph{infra-red} or \emph{green} will mainly depend on the detailed design of the detectors. A green laser allows to better separate the scattered electrons from the main beam and the scattered photon maximum energy will be larger such that it is more immune to backgrounds. It however implies to implement frequency doubling of an infra-red laser in the accelerator environment which implies some manageable complications. Stable operation, compatible with a precise (per-mille level) control of the laser-beam circular polarization components, needs to be demonstrated in this environment to the best of our knowledge. A schematic describing the components needed for the laser system is shown on Fig.~\ref{fig:laserdesign}. The laser transport will be as short as possible by integrating the laser system in the vicinity of the Compton interaction point. It may be partly under high vacuum, depending on the length of the beam transport required. A careful vacuum chamber design compliant with the requirement to not significantly modify the impedance of the ring can be done, as it was in the past for instance for the HERA LPOL2 optical cavity, see Fig.~\ref{fig:HERALPOL2cav}.

\subsubsection{Detectors}
Two complementary detectors are being considered. They must comply with the need to measure each bunch independently of the others and potentially high radiation levels due to the signal itself.

\begin{figure}[!ht]
\centering
\includegraphics[width=0.45\textwidth]{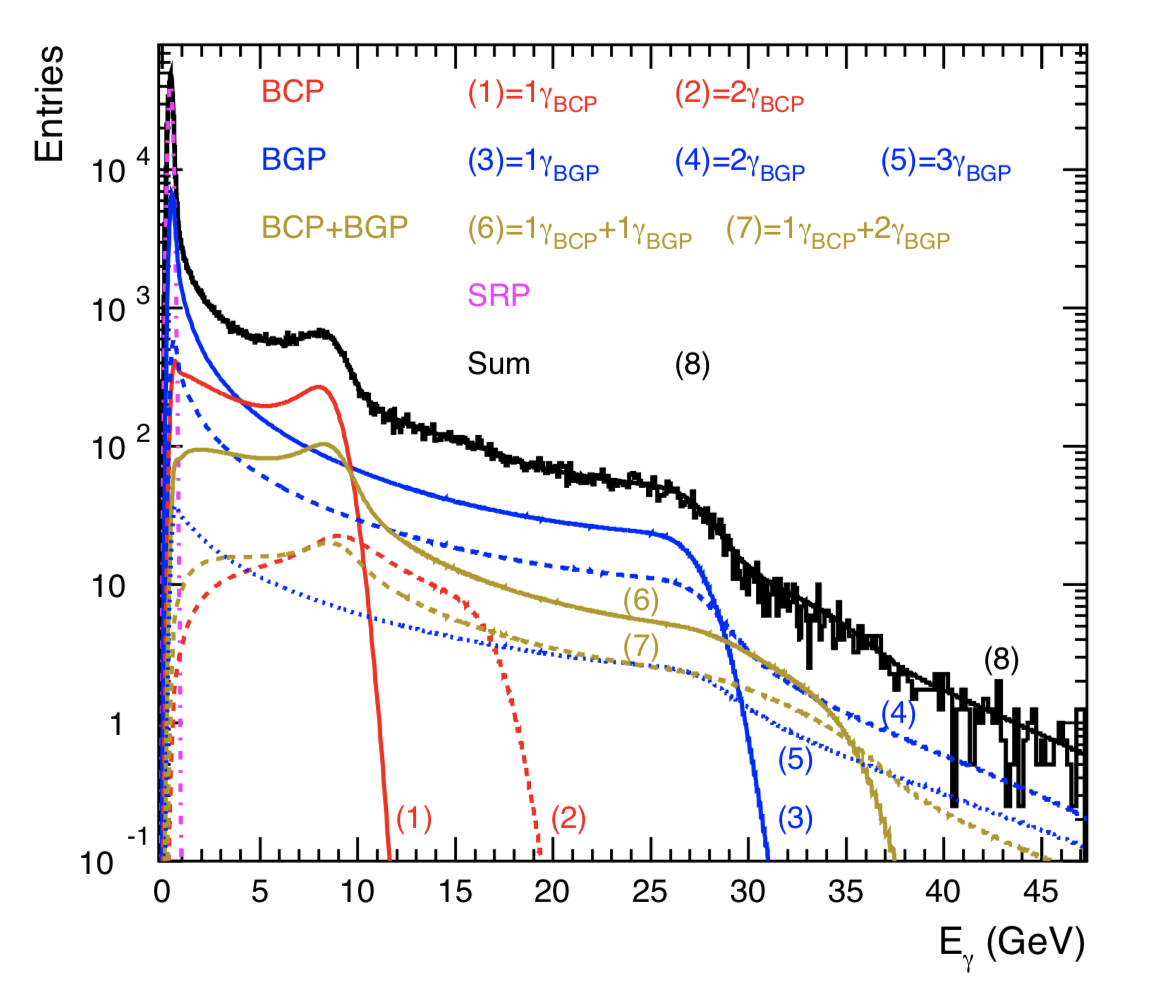}
\caption{A measured spectrum of photon calorimeter in HERA LPOL2 with data and a fit. Reproduced from Ref.~\cite{Baudrand_2010}.}
\label{fig:HERALPOL2spec}
\end{figure}

The photon calorimeter is designed to measure the energy of the photons scattered in the direction of the incident electron beam. Since it must measure photons at a rate of 250 MHz, a fast scintillating material must be used. BaF2 crystals are well suited for that goal, provided that the slow component is well filtered out by Yttrium doping and/or use of solar blind photodetectors \cite{Zhu_2019}. It it also compliant \cite{Zhu_2019} with the very large integrated dose delivered by the signal itself of about 0.6\,MGy assuming about 6 months of operation of the system. It is foreseen to implement only one channel for the measurement. This needs to be confirmed by simulation since the use of two or four detectors may allow to better align the detector from data itself. The resolution is not a critical aspect of the measurement and a resolution of several $\%/\sqrt{GeV}$ does not  significantly degrade the precision, according to preliminary estimates. The possible contribution from pile-up  will require careful estimation and subtraction \cite{Baudrand_2010}, with remaining fluctuations contributing to the detector  resolution with magnitudes depending mainly on the decay time of the measured signal. Dedicated test-beam experiment may be needed to validate the design of this detector and its associated electronics. A similar electronics to that implemented at HERA \cite{Baudrand_2010} but able to cope with the higher rate of SuperKEKB may be realized. At that time an online extraction of the polarization was made possible and was checked with offline fits of the measured spectra, an example of which is shown on Fig.~\ref{fig:HERALPOL2spec}. Estimates of background levels by dedicated simulations or measurements in the SuperKEKB environment will be needed to precisely assess the performance of this detector. 

An electron spectrometer employing a electron counter horizontally segmented after a dispersive magnetic element, ideally one already existing in the SuperKEKB lattice, will provide complementary information. The spectrum of the scattered electrons will be measured by a HVMAPS array. Despite the fact that this detector is slow compared to the 4~ns bunch separation, the pixel occupancy will be much smaller compared to that of the photon calorimeter, assuming only one scattering per bunch crossing. Thus each bunch crossing may still be well separated. Alternatively diamond strips could be used, which will be naturally radiation hard and fast. However there is at present no commercial solution for that technology. HVMAPS sensors for electron detection in Compton polarimetry is currently under development for the MOLLER experiment at Jefferson Lab (11\,GeV electron beam).

The integration of these detectors present some technical difficulties that need to be solved. The extraction of the photons requires that the existing BLA2LE dipole of the HER ring is modified. It is currently too long to allow the gamma beam to be extracted. One solution could consist in locally modifying the lattice to use a shorter dipole with larger rigidity or split it in two shorter dipoles with similar rigidity and a small straight section in between to locate the gamma-beam detector. This detector could be placed at  atmospheric pressure right after a window. This detector as well as the laser chamber are not expected to affect the beam impedance. The integration of the detector for the deflected electrons can be placed few meters after BLA2LE. However the detector, to be placed under vacuum, requires a relatively large chamber that may cause issues for the beam impedance (a strategy typically referred to as a ``Roman pot''). Alternatively, it may be possible to maintain the original beam pipe environment and place the electron detector in a secondary vacuum chamber separated by a thin beryllium window from the beam pipe vacuum (which would degrade the position resolution to an extent to be constrained by simulation studies).

\subsubsection{Expected performance}
Preliminary estimates suggest that the polarization may be estimated with less than 1\% statistical precision within a minute for each bunch, provided that about one scattering takes place at each crossing of electron and laser bunches. A major common source of systematic uncertainties for both detectors is the knowledge of the laser polarization. This aspect will need to be controlled at the per-mille level for SuperKEKB thanks to dedicated diagnostics as mentioned in the previous section. Expected systematic uncertainties have partially different sources for the two detectors. For the photon detection \cite{Baudrand_2010}, knowledge of backgrounds, precise knowledge of the calibration and alignment of the detector,  and associated electronics contribute with an overall systematic uncertainty well below 1\%. These will need to be re-assessed precisely in the present case, in particular in view of the potential impact of pile-up related to the short bunch separation in time. Data-driven calibration and alignment techniques may be investigated to reduce these systematic uncertainties.  For the electron spectrometric measurement \cite{PhysRevX.6.011013}, the knowledge of the magnetic field, the residual angular mis-alignment of the detector, knowledge of the beam  energy, knowledge of the detector response are expected to contribute within 0.6\%. Beam jitters and emittance growth related to the top-up injection of fresh beam inside the SuperKEKB ring as been estimated and found to contribute negligibly to the systematic uncertainties.

\subsection{Summary of required R\&D before Technical Design Report}
The studies performed at this stage give confidence that commercial laser systems can be used to provide the  photons that will interact with the electron beam of SuperKEKB to realize a measurement of electron beam polarization with Compton backscattering. Expected statistical precision of this device is better than 1\% per minute for each bunch. Residual systematic uncertainties are expected to lie well below 1\%. One of the key aspects related to the laser beam is the control and survey of its polarization components that enter into the systematic uncertainties for the extraction of the polarization of the electron beam. Some limited R\&D work needs to be performed on this topic to ensure a precise and accurate estimation in the SuperKEKB environment with the laser system under consideration. It is planned to use the existing/upgraded SuperKEKB lattice with minor modifications related to the insertion of the laser interaction chamber and the photon and electron detectors. Existing technology based on HVMAPS can be implemented for the counting of electrons, with no specific need for a dedicated R\&D program. A measurement of the scattered photon spectrum using a BaF2 crystal coupled to a PMT could complementarily be implemented. It is expected to be compliant with the short time-separation of bunches at SuperKEKB. This may be confirmed with some limited R\&D to ensure the performance of this system. 

%% file: TauPolarimetry.tex
\begin{figure}
	\includegraphics[width=0.3\linewidth]{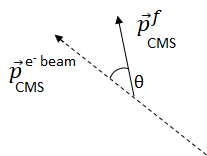}
	\includegraphics[width=0.3\linewidth]{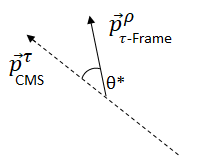}
	\includegraphics[width=0.3\linewidth]{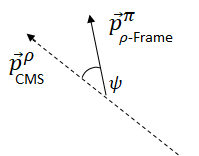}
	\caption{Cartoons illustrating $\cos\theta$ (left, $f$ represents a final state particle), $\cos\theta^\star$ (center), and $\cos\psi$ (right). }
	\label{fig:cartoons}
\end{figure}
\begin{figure}
	\centering
	\includegraphics[width=0.4\linewidth]{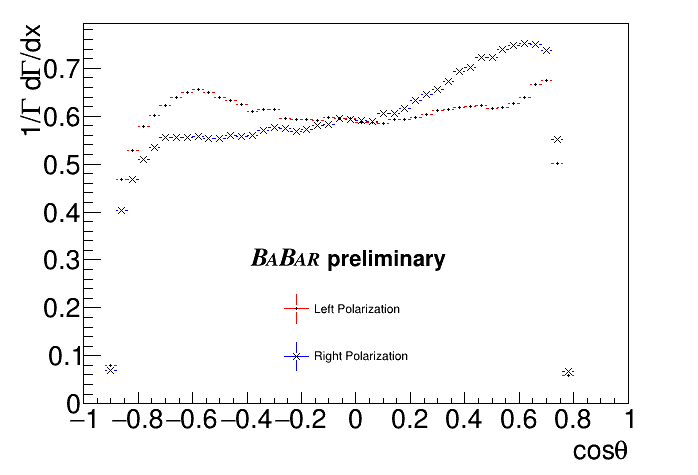}
	\caption{$\cos\theta$ distribution in MC for positively charged rho decay.}
	\label{fig:ctpolar}
\end{figure}
\begin{figure}
	\centering
	\includegraphics[width=0.4\linewidth]{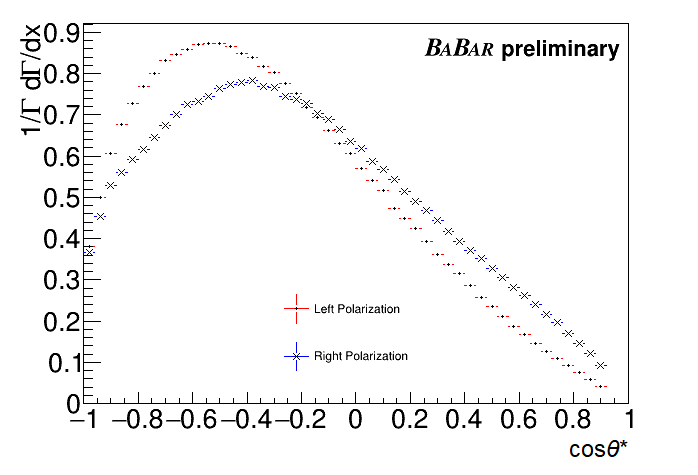}
	\includegraphics[width=0.4\linewidth]{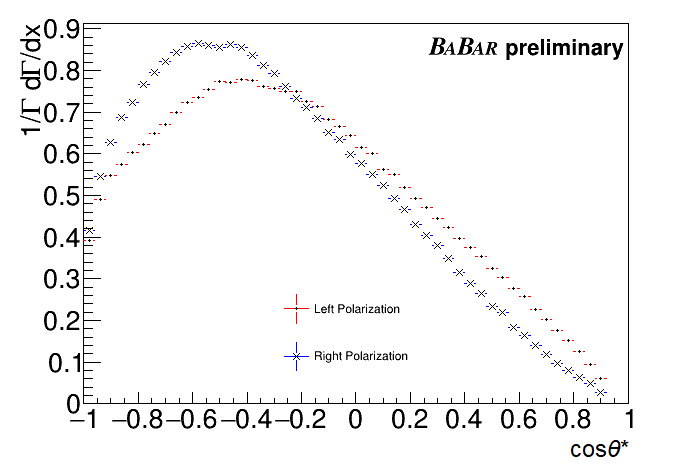}
	\caption{$\cos\theta^\star$ distribution in MC for positively charged rho decay. For $\cos\theta<0$ (left) and $\cos\theta>0$ (right).}
	\label{fig:zctpolar}
\end{figure}
\begin{figure}
	\centering
	\includegraphics[width=0.4\linewidth]{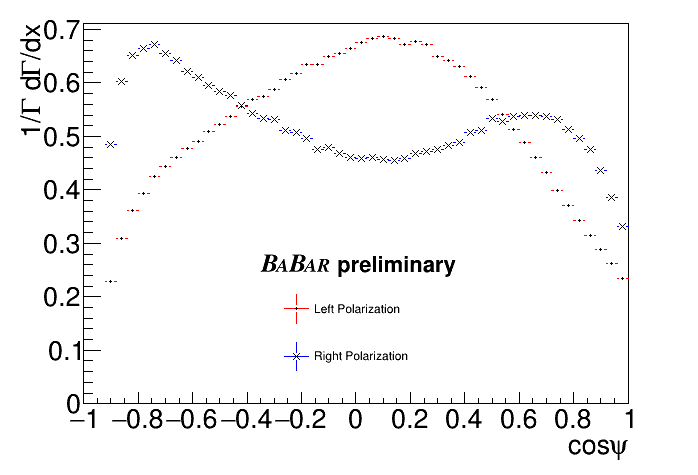}
	\includegraphics[width=0.4\linewidth]{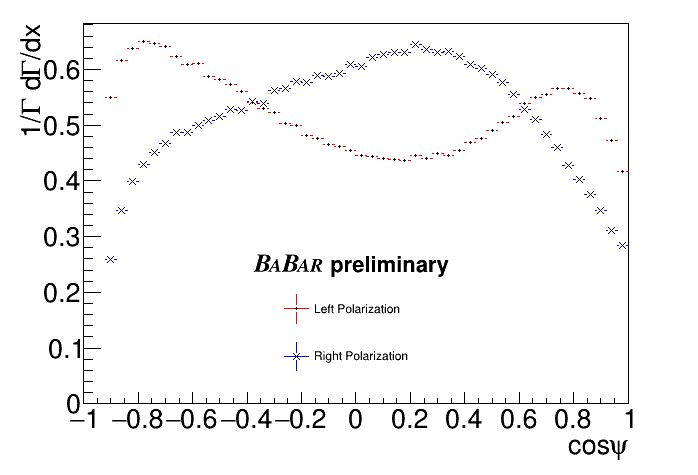}
	\caption{$\cos\psi$ distribution in MC for positively charged rho decay. For $\cos\theta<0$ (left) and $\cos\theta>0$ (right).}
	\label{fig:xctpolar}
\end{figure}

\label{TauPolarimetery}
As mentioned in the Physics Case, a precise determination of the average beam polarization is required to maximize the sensitivity of the planned measurements. Due to the left handed nature of the weak nuclear force the $\tau$ particle is uniquely suited for gaining access to beam polarization. This is due in part to the $\tau$ decaying while inside the detector and secondly the kinematics of the decay products are sensitive to the $\tau$ spin state. The spin state of the tau produced in an $e^+e^-\rightarrow \tau^+\tau^-$ directly couples to the helicity of the electron beam as shown in Equation \ref{eqn:ePoltoTauPol}.
\begin{equation}
    	P_\tau=P_e\frac{cos\theta}{1+cos^2\theta}-\frac{8G_Fs}{4\sqrt{2}\pi\alpha}g^\tau_V\left(g^\tau_A\frac{|\vec{p}|}{p^0}+2g^e_A\frac{cos\theta}{1+cos^2\theta}\right)
	\label{eqn:ePoltoTauPol}
\end{equation}
Where $P_\ell$ is the polarization of the tau or electron, $\cos\theta$ is the opening angle between the tau and the electron beam, $G_F$ and $\alpha$ are the Fermi constant and fine structure constant respectively, and $g^\ell_{V,A}$ are the vector and axial neutral current couplings for their respective leptons. 
The work we present in this section is focused on determining the polarization sensitivity in the $\tau^\pm\rightarrow(\rho^\pm\rightarrow\pi^\pm\pi^0)\nu_\tau$ decay mode and determining the associated statistical and systematic uncertainties. This decay mode was chosen for the large branching fraction of the tau decay, 25.93\%, as well as the high tau pair purity achievable with a lepton tag on the other tau decay. For this $\rho$ decay three angular variables are required to extract the beam polarization. The polarization sensitive variables are defined in equations in \ref{eqn:zct} and \ref{eqn:xct}~\cite{HAGIWARArho}. 
\begin{equation}
	\cos\theta^\star=\frac{2z-1-m^2_{\rho}/m^2_{\tau}}{1-m^2_{\rho}/m^2_{\tau}} \hspace{2cm} z\equiv\frac{E_\rho}{E_{\textrm{beam}}}
	\label{eqn:zct}
\end{equation}
\begin{equation}
	\cos\psi=\frac{2x-1}{\sqrt{1-m^2_{\pi}/m^2_{\rho}}} \hspace{2cm} x\equiv\frac{E_\pi}{E_{\rho}}
	\label{eqn:xct}
\end{equation}
For the mass of the pion and the tau we use the PDG values, while for the mass of the rho we used the event-by-event reconstructed $\pi\pi^0$ mass. $\cos\theta^\star$ is defined as the opening angle between the between the tau flight path in the center-of-mass frame and rho direction in the tau rest frame. Similarly $\cos\psi$ is the opening angle between the rho flight direction in the center-of-mass frame and the pion direction in the rho rest frame. Forward and backward regions are defined from the direction of the final state momentum with respect to the beam axis, $\cos\theta$, and the prior angular variables reverse their polarization behaviour switching between these regions. Figure \ref{fig:cartoons} illustrates the angular definitions. The distributions of these variables are depicted in Figures \ref{fig:ctpolar}, \ref{fig:zctpolar}, and \ref{fig:xctpolar} for different values of the electron beam polarization.
\subsection{Event Selection}
\begin{figure}
	\centering
	\includegraphics[width=0.3\linewidth]{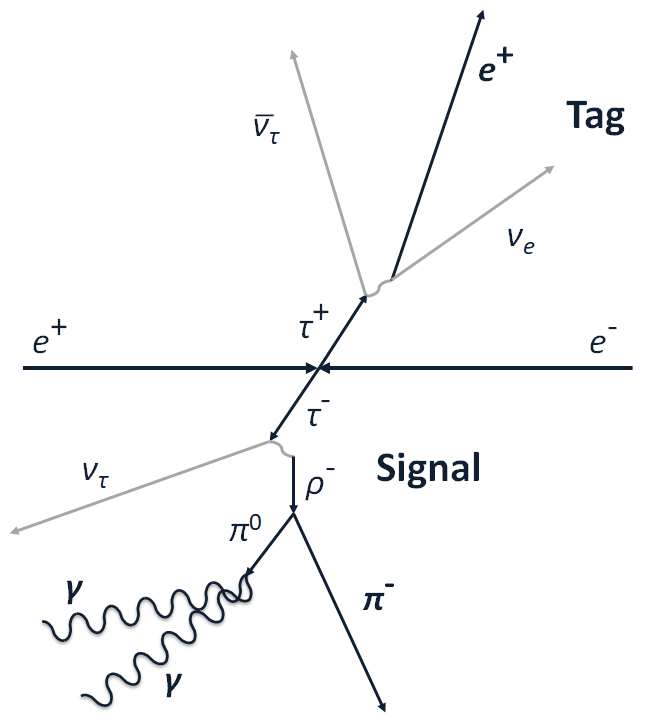}
	\caption{Example tau-pair decay. Signal \taurho decay tagged with a \taue decay}
	\label{fig:topo}
\end{figure}

A preliminary study of this technique has been implemented at \babar~in order to identify dominant uncertainties. At \babar~the event selection is designed to select tau pair events where each of the tau particles decay into a single charged particle. One tau lepton, labelled the as the tag, is required to decay leptonically into an electron while the other tau, the signal, decays into a charged and a neutral pion.  Figure \ref{fig:topo} shows this event topology. This requirement for a single charged hadron and a single charged lepton excludes nearly all Bhabha, $\mu\mu$, and $q\overline{q}$ events. The requirement of the neutral pion on the signal side of the event significantly reduces any remaining Bhabha events which contain one electron misidentified as a pion. 
The remaining Bhabha events are further reduced by a factor of three by slightly trimming $\cos\theta^\star$, $-1<\cos\theta^\star<0.9$, and $\cos\psi$, $-0.9<\cos\psi<1$. These requirements achieve a final tau selection which is 99.7\% pure and with a 0.70\% efficiency. The branching fraction for the tagging tau is 17.82\% and 25.49\% for the signal tau. Including these in our efficiency calculation, brings our efficiency to 15.40\% for selecting $\tau^\pm\tau^\mp\rightarrow\rho^\pm\nu_\tau+e^\mp\nu_e\overline{\nu_\tau}$ events. Our largest non-tau background is Bhabha events which make up 0.3\% of the final sample. The final event selection break-down as predicted by MC is shown in Table \ref{tab:events}.
\begin{table}
\centering
\begin{tabular}{lr}
MC source & \quad Fraction \\ \hline
Bhabha & 0.289\%\\
\mumu & 0.000\%\\
\uubar,\ddbar,\ssbar & 0.005\%\\
\ccbar & 0.002\%\\
\bbbar & 0.000\%\\
\tautau & 99.704\%\\
\end{tabular}
\hspace{2cm}
\begin{tabular}{lr}
	Tau Signal & Fraction \\ \hline
	\taue	&	0.012\%	\\
	\taumu	&	0.018\%	\\
	\taupi	&	0.033\%	\\
	\taurho	&	89.994\%	\\
	\taua	&	8.028\%	\\
	$\tau\rightarrow$ else	&	1.915\%	\\ 
\end{tabular}
\caption{Fraction of event types in MC in the final event selection. The tau pair events are further broken down to show the \taurho selection.}
\label{tab:events}
\end{table} 
\subsection{Fitting}
\begin{figure}
	\centering
	\includegraphics[width=0.3\textwidth]{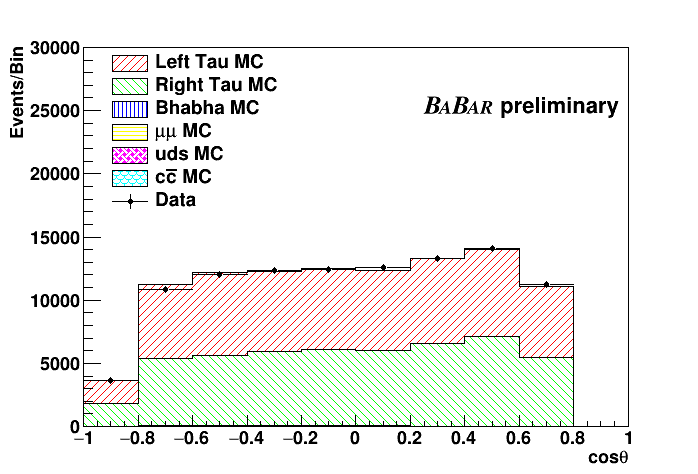}
	\includegraphics[width=0.3\textwidth]{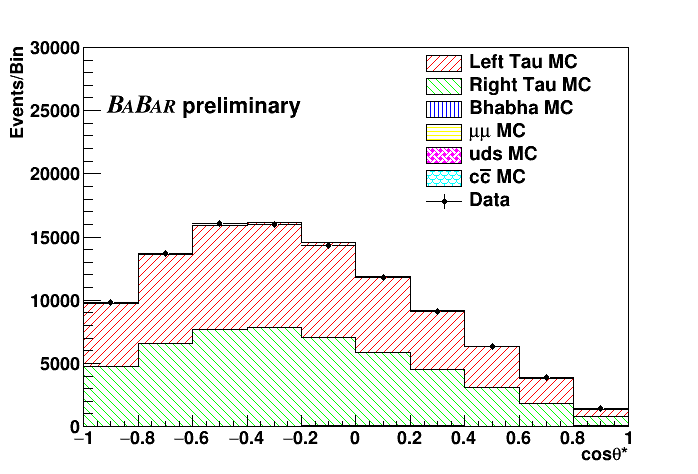}
	\includegraphics[width=0.3\textwidth]{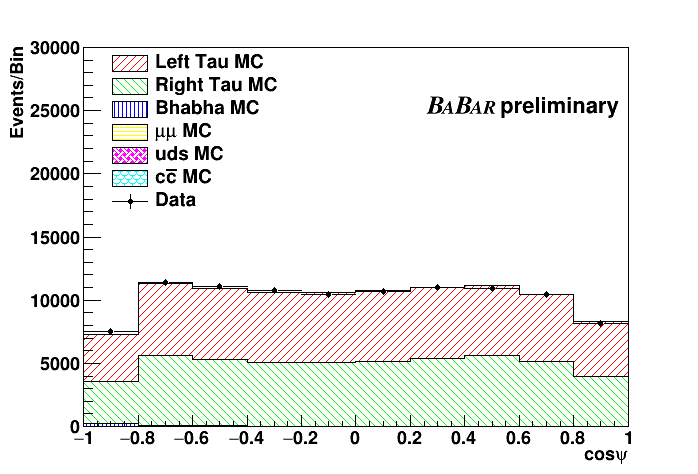}
	\caption{One dimensional projection of the fit result for $\cos\theta$ (left), $\cos\theta^\star$ (middle), and $\cos\psi$ (right) in positively charged candidates in Run 3.}
	\label{fig:fitproj1}
\end{figure}

To extract the average beam polarization we perform a binned likelihood fit as described by Barlow and Beeston~\cite{barlow}. We fill three dimensional histograms of $\cos\theta^\star$, $\cos\psi$, and $\cos\theta$, for each of the data and MC modes. The 3D histogram for data is then fit as a combination of the MC histograms, where the weights of the non-tau MC is fixed based on the MC efficiency studies. This leaves only the left and right polarized contribution to vary in the fit, and with a restraint that the contributions sum to 1 leaves only one parameter in the fit. We define this fitted parameter as $\langle$P$\rangle$=L-R, where L and R are the fitted fractions reported by the fit for the left and right polarized tau MC. As the polarization sign flips with electric charge we preform the fit separately for positively and negatively charged signal candidates, and extract the fit result, statistical uncertainty, the chi squared as defined by TFractionFitter, and the number of degrees of freedom (bins in fit with an event minus one fit parameter). A final fit result is then reported as the combined average of the two independent fits. In addition to the data fit we use unpolarized tau MC mixed with non-tau MC to produce three ``data-like" samples for approximating the level of statistical uncertainty present in small perturbations to the fit. Figure \ref{fig:fitproj1} shows the 1D projections of the run 3 fit, where the data points represent the data and the MC is stacked with relative proportions equal to the fit results.
As the \babar~data sets are split into multiple data collection periods, run numbers 1 to 6, and each set has it's own unique beam conditions we treat each sample independently. As such we obtain 6 measurements of the beam polarization and corresponding statistical and systematic uncertainties. A final measurement of the beam polarization is then obtained from combining these measurements and accounting for correlations in the systematic uncertainties.

\subsection{Extracted vs Input Beam Polarization Study}
\begin{figure}
	\centering
	\includegraphics[width=0.4\linewidth]{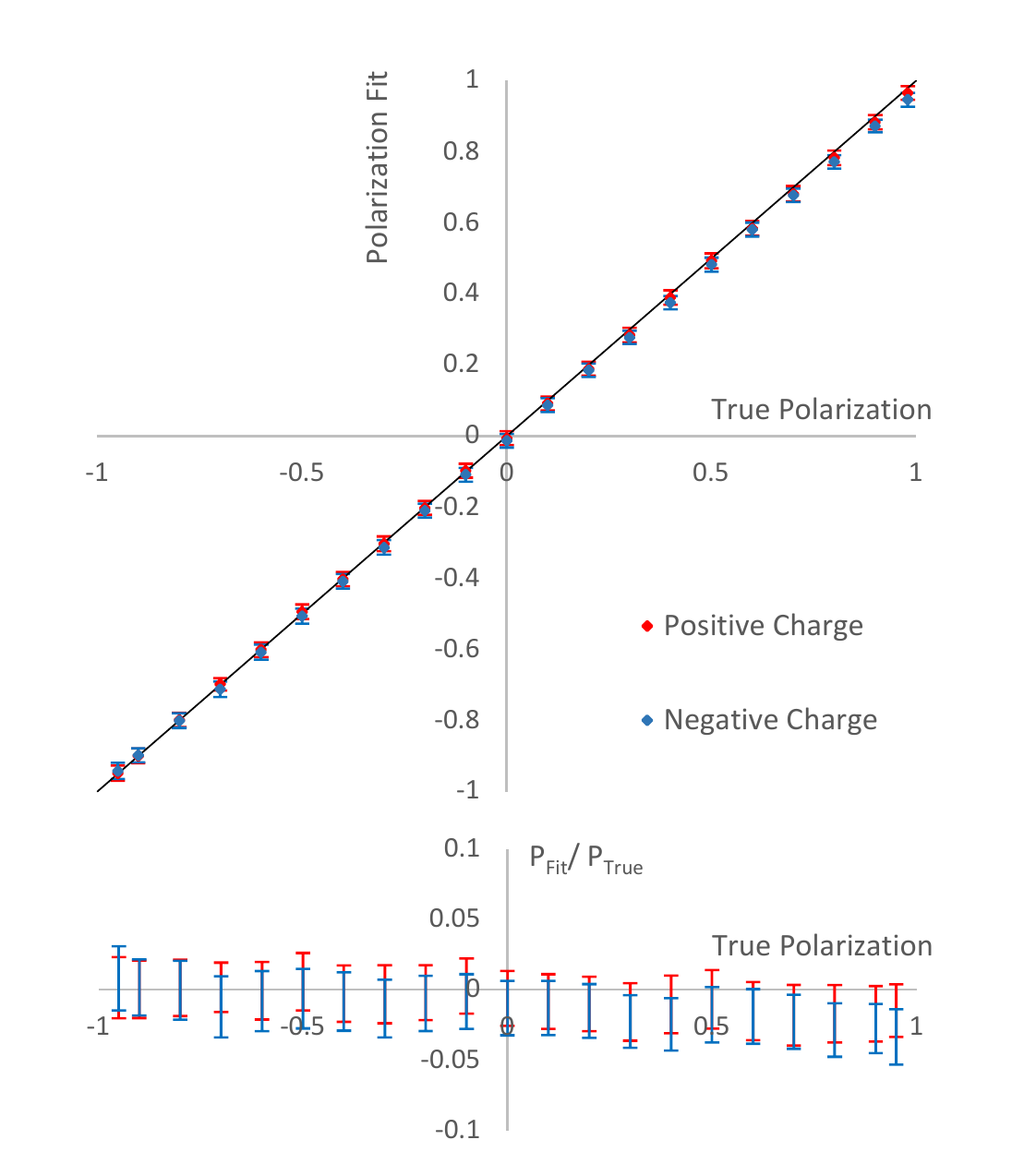}
	\caption{Measured beam polarization for a given input polarization mixed from polarized tau MC. The red points correspond to the measurements for positively charged signal candidates while the blue points correspond to the negatively charged candidates.}
	\label{fig:polarSense}
\end{figure}

In order to ensure the polarimetry measurement is valid at beam polarization states other than 0, as expected for PEP-II runs, we used polarized tau MC to study the extracted polarization from this analysis of the rho channel at multiple input beam polarization states. This was done by splitting each of the left and right polarized tau MC into two distinct samples, one reserved for fitting the beam polarization in MC ``measurements'' and the other for mixing beam polarization states. With the samples reserved for mixing beam polarization states specific beam polarization states can be created, ie. 70\% polarized is made with 85\% left polarized MC and 15\% right polarized MC. Using this technique we tested polarization states from -1 to 1 in steps of 0.1, the results of which can been seen in Figure \ref{fig:polarSense}. The fit results to the MC are within good agreement of the input MC beam polarization states, which demonstrates the measurement technique will yield the correct polarization for any beam polarization.
\subsection{Systematic Uncertainties}
Systematic uncertainties are evaluated by studying the response in the MC and data polarization fits as we systematically vary parameters in the fit templates. To evaluate the systematic uncertainty corresponding to a particular source, the fit is performed in a ``default" mode then repeated with some variation in the systematic source. The shift in data, as well as the average shift in MC is found, and the relative shift in the agreement between data and MC is taken as the systematic uncertainty. In order to approximate the level of statistical uncertainty present in the measurement we split the MC into 3 independent samples, each roughly equal in size to the data set being studied, and take the RMS of the shifts in the MC as a statistical uncertainty on the systematic uncertainty. This provides both a central value for the systematic shift and a statistical uncertainty on the systematic uncertainty. As a conservative measure, in the case the statistical component is larger than the central value of the systematic uncertainty we list the statistical part as the systematic uncertainty.
As systematic uncertainties are expected to be run dependent we treat each run independently and started with \babar~Run 3 as a study sample as the data set, which represents 32.3 fb$^{-1}$, is relatively small. The following systematic uncertainties discussed in this section are the result of the Run 3 studies.
In order to determine the level of variation required to accurately evaluate a potential systematic source, the mean of the distributions for the variable are compared between data and MC. The level of disagreement between data and MC is taken as the uncertainty in the variable and used in the systematic uncertainty tests. For the cuts applied to $\cos\psi$, the cut was varied from $\cos\psi>-0.9$ up and down to $\cos\psi>-0.91$ and $\cos\psi>-0.89$. This resulted in an overall shift in the level of agreement between data and MC in the polarization fit of -0.0007$\pm$0.0013. As the statistical component dominates, we list 0.0013 as the uncertainty seen in Table \ref{tab:systematicSummary}. A similar approach for $\cos\theta^\star$ results in a systematic uncertainty of -0.0002$\pm$0.0001, or 0.0002 in our final table.
To evaluate the uncertainties arising from our treatment of neutral clusters we vary the thresholds for the related cutoffs. In the case of the hadronic split-off modelling the relevant variable is the distance within we associate neutral clusters with charged particles, 40 cm by default. The MC suggests the resolution for the distance between a charged particle and a neutral cluster is better than 1cm. By varying the acceptance cut by 1 cm a systematic uncertainty of 0.27\% was estimated. 
For the minimum neutral energy, 50 MeV, as well as the photon energy cutoff for \piz reconstruction, 100 MeV, we vary the acceptance by $\pm$1 MeV, as dictated by the level of agreement between MC and data. Both of these variables resulted in small systematic uncertainties of 0.13\% and 0.11\% respectively. 
The systematic uncertainty for momentum resolution is evaluated by scaling the momentum up or down based on its resolution and results in a 0.02\% systematic uncertainty in the polarization fit. Similarly for $\cos\theta$ the angle is varied by it's resolution and results in a 0.10\% systematic uncertainty.
The \piz identification systematic uncertainty was evaluated by varying the mass window for the reconstructed \piz's by 1 MeV or by tweaking the \piz likelihood acceptance by 5, depending which selection criteria selected a particular \piz. The combined systematic uncertainty for the \piz identification is 0.19\%.
In addition to already mentioned variables, similar studies were performed on the $\tau$ trigger decision, the boost calculation, the charged track definition, the total event transverse momentum, uncertainties within the $\tau$ branching fraction, the weights of non-$\tau$ backgrounds, the luminosity weighting of the MC, the effects of re-binning in the fit, the electron PID requirements, and the cuts on $\cos\psi$ and $\cos\theta^\star$. None of these exhibited any notable effects.
After carrying out the full measurement on the remaining \babar~data-sets the run-by-run systematic uncertainties were combined in a process that accounts for correlations between the runs. Table \ref{tab:systematicSummary} shows the systematic uncertainties for each run as well as the final uncertainty for each systematic source once correlations are accounted for. The individual uncertainties are summed in quadrature to arrive at the total uncertainty shown in the final row.
\begin{table}[ht]
	\centering
	\begin{tabular}{l|r}	
		Source &\multicolumn{1}{c}{Final} \\ \hline	
		$\piz$ Likelihood		&	\textbf{	0.0015	}	\\
		Hadronic Split-off Modelling		&	\textbf{	0.0011	}	\\
		$\cos\psi$	&	\textbf{	0.0010	}	\\
		Angular Resolution		&	\textbf{	0.0009	}	\\
		Minimum Neutral Energy		&	\textbf{	0.0009	}	\\
		$\piz$ Mass	&	\textbf{	0.0009	}	\\
		$\cos\theta^\star$		&	\textbf{	0.0008	}	\\
		Electron PID	&	\textbf{	0.0007	}	\\
		Tau Branching Fraction	& \textbf{	0.0006	}	\\
		Event Transverse Momentum &	\textbf{	0.0005	}	\\
		Momentum Resolution	&	\textbf{	0.0005	}	\\
		$\piz$ Minimum Photon Energy	&	\textbf{	0.0004	}	\\
		Rho Mass	&	\textbf{	0.0003	}	\\
		Background Modelling	&	\textbf{	0.0003	}	\\
		Boost 		&	\textbf{	0.0002	}	\\ \hline
		Total	&	\textbf{	0.0030	}	\\
	\end{tabular}
	\caption{Summary of systematic uncertainties associated with polarization measurement.}
	\label{tab:systematicSummary}
\end{table}
\subsection{Preliminary results from \babar~Beam Polarization Fit}
The fit results for all of the \babar~data sets are shown in Table \ref{tab:fitresults}. 
Taking the weighted mean of these fit results gives the average beam polarization of PEP-II runs to be $\langle P\rangle=0.0001\pm 0.0035_{\textrm{stat}}\pm 0.0030_{\textrm{sys}}$ with a sample of 0.4~ab$^{-1}$. 
We can estimate that 56 fb$^{-1}$ of data is needed to achieve a total uncertainty of 1\%, where the statistical uncertainty scales with 1/$\sqrt{N_{\textrm{Events}}}$.
The absolute systematic uncertainty of $\pm 0.003$  demonstrates that a 0.4\% relative systematic uncertainty on a beam polarization of 70\%  can be achieved in \epem colliders, and with sufficient statistics  represents the achievable total uncertainty. Such a precise beam polarization measurement in SuperKEKB enables the high precision electroweak measurements. 
These preliminary results have been presented at the Lake Louise Winter Institute Conference\cite{LLMiller}.

\begin{table}
	\centering
	\begin{tabular}{l|c|cr|cr|c}
		Data  & Luminosity & Positive  & $\chi^2$/NDF & Negative  &$\chi^2$/NDF& Average  \\ 
			 Set & (fb$^{-1}$)&  Charge & &  Charge &  &  Polarization \\ \hline
		Run 1 & 20.37  &-0.0018$\pm$0.0222&934/881& 0.0143$\pm$0.0224&1022/882& 0.0062$\pm$0.0157 \\
		Run 2 & 61.32  &-0.0064$\pm$0.0127&785/884& 0.0056$\pm$0.0128&819/877&-0.0004$\pm$0.0090 \\
		Run 4 & 99.58  & 0.0054$\pm$0.0101&890/888&-0.0280$\pm$0.0100&832/883&-0.0114$\pm$0.0071 \\
		Run 5 & 132.33 & 0.0053$\pm$0.0092&914/886&-0.0124$\pm$0.0087&993/886&-0.0040$\pm$0.0063 \\
		Run 6 & 78.31  & 0.0256$\pm$0.0117&939/881& 0.0060$\pm$0.0116&1022/882& 0.0157$\pm$0.0082 \\ \hline
		Total & 424.18 & 0.0070$\pm$0.0052&&-0.0087$\pm$0.0051&& -0.0010$\pm$0.0036 \\
	\end{tabular}
	\caption{Average beam polarization measured in each data set. The average for each run is found from the weighted mean of the positive and negative fit results.}
	\label{tab:fitresults}
\end{table}